# Archimedes
# the Free Monte Carlo simulator

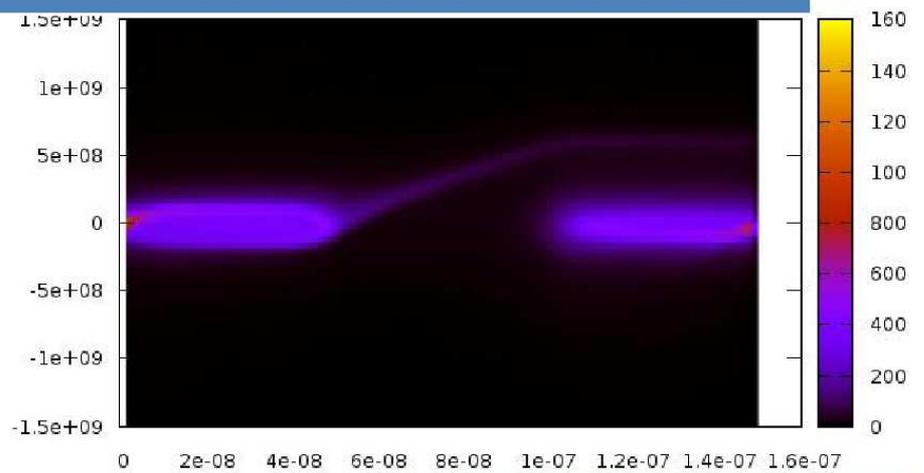

Jean Michel D. Sellier

**This document is a draft of the manual for the GNU package Archimedes.**

**<span style="color:red">Release 1.0</span>**

**This manual is released under GFDL v1.3 or later**

**GNU Archimedes is released under GPL v3.0 or later**

**Table of Contents DRAFT**

# Preface

First of all, thanks for reading this manual! By reading it and spreading the word around, you are contributing to the success of the GNU project Archimedes. It is important to support GNU projects and, thus, software freedom. As long as there will be someone that support these projects, we will keep on having very good packages that we can download, modify, compile, run and redistribute and that represents an extremely important value.

***The Ethical Motivations, a New Paradigma in Science***
Archimedes has been created after observing the situation of semiconductor simulations around the world. One easily observes that the all codes developed for simulation are usually not free and/or proprietary codes. That is a very bad situation, at least for accademic purposes, since it forces people to reinvent the wheel everytime a piece of code is needed. This surely slows down the progress of Science (immagine you had to rediscover the Newtonian laws every time you need them...).

The actual situation is that we have a huge amount of papers describing a lot of numerical methods for advanced simulations of semiconductor devices, but nobody can access single code on which to build new and even more advanced methods.

So, today, every university (and even every group in a university) has its own Monte Carlo simulator, its own NEGF simulator and so on.. Would not it be better if we could avoid this incredibile duplication of efforts all around the world?

That is why Archimedes has been created...

**Do you want to support GNU/Archimedes ?**
Please remember that the development of GNU archimedes is a volunteer effort, and you can also contribute to its development. For information about contributing to the GNU archimedes Project and/or request of enhancements and new features, please contact me at ***jeanmichel.sellier@gmail.com***.

Archimedes is a free software, which means that you can run, modify and redistribute the code (as long as the code is delivered under the same license of Archimedes). Archimedes is released under GPL version 3. If you have any suggestion to the book and/or Archimedes you are VERY welcome to contact the main developer and maintainer of this GNU package at ***jeanmichel.sellier@gmail.com***.

# About the Author

Jean Michel Sellier today is a Research Assistant Professor at Purdue University, member of Prof. Klimeck Group.

He is currently part of the NEMO5 team. His main interests are the Monte Carlo simulations of electron transport in semiconductor devices and the simulation of Schroedinger equation coupled to Poisson equation in both stationary and transient (time-dependent) regimes to study the feasibility of quantum computing. He is currently working on massive parallelization of Schroedinger-Poisson solutions using the Lanczos eigensolver in the aim to understand Single Impurity Devices (Quantum Computers) and Decoherence effects in very small devices.

Jean Michel D. Sellier studied mathematical physics at the University of Catania (Italy). His PhD tutor was one of the most influent mathematical physicist in Italy at that time (A.M. Anile). Jean Michel gained experience during his postdocs at Imperial College London (UK) in Plasma Simulations and at INRIA (Institut national de recherche en informatique et en automatique), Rocquencourt (France), in Semi-classical Hydrodynamical Electron Transport models. He has also been a Research Associate at Purdue University, IN, USA working with Prof. G. Klimeck.

He holds a "laurea in matematica" magna cum laude and a PhD in Mathematics (simulation of semiconductor devices), both from the University of Catania (Italy).

Jean Michel is the developer of Archimedes and Aeneas, GNU packages, two tools for the design and simulation of semi-classical and mesoscopic semiconductor devices in 2D and 3D respectively.
Jean Michel is the main maintainer of three nanoHUB tools, i.e. Archimedes, 1dhetero and RTDNEGF for Monte Carlo, quantum structures and quantum transport in nano devices.

He is also the expert for Monte Carlo simulations in the nextnano³ team.

In the following, a list of some simulators implemented and maintained by JM Sellier:

---

http://www.nanohub.org/tools/rtdnegf
http://www.nanohub.org/tools/1dhetero
http://www.nanohub.org/tools/archimedes
http://www.gnu.org/software/archimedes
http://www.gnu.org/software/aeneas

# Introduction

**What is Archimedes about?**

Archimedes is the GNU package for semiconductor devices simulations in semi-classical and quantum regime that can be used to simulate, respectively, submicron and nanoscale devices in a reliable and predictive way.

The first version of Archimedes has been implemented in 2004 and the first version to be released (0.0.1 in 2005) was just a very simple Monte Carlo simulator. Since then, many things have changed in Archimedes. Many new models have been implemented in it, like hydro-dynamical models for holes, effective potentials to mimic quantum effects due to non-zero dimensions of electrons, etc. Archimedes does not aim to be the best simulator around, but is certainly one of the most interesting since it is released under GNU General Public License (well-known as GPL) that give the user the freedom to download, use, modify and redistribute the sources (as long the original license is kept).

Archimedes can be, at least, considered a good starting point for students that want to understand electron transport in semiconductor devices, but also as a starting point for researchers that want to have a tested and working transport code to start from. This package is used today by many universities and semiconductor companies all around the world, and has been also used as a starting point to develop new interesting codes to simulate particular/experimental devices.

Archimedes is very easy to use. You can set up and run a simulation in just a few minutes. No need to know the physics or even the code behind. The only thing the user has to know is how to describe devices in a scripts and how to specify the transport model to be used. This is clearly shown in the following example where we simulate a Metal Semiconductor Field Effect Transistor (MESFET) is simulated.

MESFETs are usually constructed in compound semiconductor technologies lacking high quality surface passivation such as GaAs, InP, or SiC, and are faster but more expensive than silicon-based JFETs or MOSFETs. Production MESFETs are operated up to approximately 45 GHz and are commonly used for microwave frequency communications and radar. They are quite similar to a JFET in construction and terminology. The difference is that instead of using a p-n junction for a gate, a Schottky (metal-semiconductor) junction is used.

The MESFET simulated here is a structure like the one below.

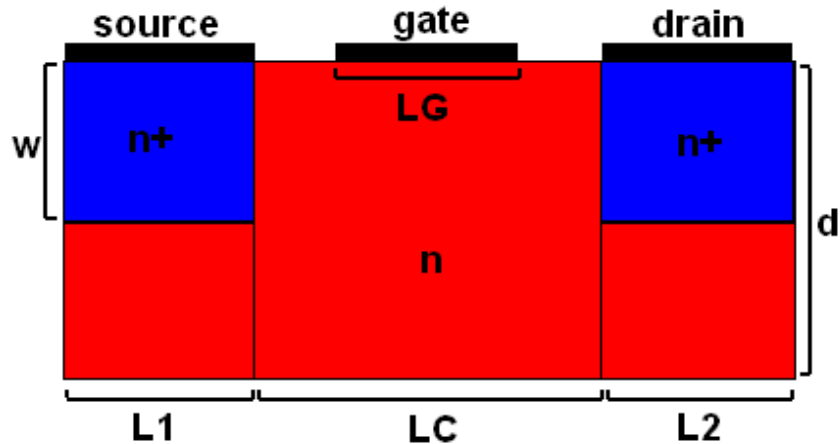

To simulate this device using Archimedes, one needs to describe the geometry and the doping of the device, and the applied bias. To do this the user has to write a short script. For example, the script below describe such a double-barrier structure.

```
TRANSPORT MC ELECTRONS

FINALTIME 6.0e-12
TIMESTEP 0.0015e-12

XLENGTH 0.6e-6
YLENGTH 0.2e-6

XSPATIALSTEP 120
YSPATIALSTEP 40

# definition of the material (all the device is made of Silicon)
MATERIAL X 0.0 0.6e-6    Y 0.0 0.2e-6  SILICON

# Definition of the doping concentration
# =======================================
DONORDENSITY    0.        0.         0.6e-6     0.2e-6     1.e23
DONORDENSITY    0.        0.15e-6    0.1e-6     0.2e-6     3.e23
DONORDENSITY    0.5e-6    0.15e-6    0.6e-6     0.2e-6     3.e23
ACCEPTORDENSITY 0.        0.         0.6e-6     0.2e-6     1.e20

# Definition of the various contacts
# ==================================
CONTACT DOWN  0.0     0.6e-6 INSULATOR 0.0
CONTACT LEFT  0.0     0.2e-6 INSULATOR 0.0
CONTACT RIGHT 0.0     0.2e-6 INSULATOR 0.0
CONTACT UP    0.1e-6 0.2e-6 INSULATOR 0.0
CONTACT UP    0.4e-6 0.5e-6 INSULATOR 0.0
CONTACT UP    0.0     0.1e-6 OHMIC      0.0 3.e23
CONTACT UP    0.2e-6 0.4e-6 SCHOTTKY -1.3
CONTACT UP    0.5e-6 0.6e-6 OHMIC      1.0 3.e23

NOQUANTUMEFFECTS
MAXIMINI
# SAVEEACHSTEP

LATTICETEMPERATURE 300.

STATISTICALWEIGHT 1000
```

The script is very easy to understand, even at a first glance, and no knowledge of the physics and/or of the code is required. Let us describe it shortly (not that every line that starts by a # symbol is a comment).

First, one must specify the kind of transport to simulate. This is specified in the first line of the script.

```
TRANSPORT MC ELECTRONS
```

In this case, the user wants to simulate electrons transport by means of NEGF model.

Then, one must specify the dimensions of the device. This is easily with the following two rows.

```
XLENGTH 0.6e-6
YLENGTH 0.2e-6
```

In this case, we want the device to be 600 nanometers in the x-direction and 200 nanometers in the y-direction.

Since, our simulator uses approximations to simulate a device, we must specify the grid used for the approximation. This is done as follows.

```
XSPATIALSTEP 120
YSPATIALSTEP 40
```

Here we want the space grid to have 120 points in the x-direction and 40 in the y-direction.

At this point, one can specify the materials the device is made of. This is done in the following lines.

```
MATERIAL X 0.0 0.6e-6   Y 0.0 0.2e-6  SILICON
```

Let us explain it quickly. Here we specify to Archimedes that for $0 \leq x \leq 600 \, nm$ and $0 \leq y \leq 200 \, nm$ the device is made of Silicon (i.e. the whole device). One could add more lines like that to simulate, for example, an heterostructure.

The donor and acceptor density of the device is then specified.

```
DONORDENSITY     0.        0.        0.6e-6    0.2e-6    1.e23
DONORDENSITY     0.        0.15e-6   0.1e-6    0.2e-6    3.e23
DONORDENSITY     0.5e-6    0.15e-6   0.6e-6    0.2e-6    3.e23
ACCEPTORDENSITY 0.        0.        0.6e-6    0.2e-6    1.e20
```

The meaning of those lines is easy. For example, in the first line we say that for $0 \leq x \leq 600 \, nm$ and $0 \leq y \leq 200 \, nm$ the assigned donor density is $10^{23}/cm^3$. The other lines are to be interpreted in the same way and supersede the previous ones in the overlapped areas.

Then we define the contacts and edge.

```
CONTACT DOWN  0.0      0.6e-6 INSULATOR 0.0
CONTACT LEFT  0.0      0.2e-6 INSULATOR 0.0
CONTACT RIGHT 0.0      0.2e-6 INSULATOR 0.0
CONTACT UP     0.1e-6 0.2e-6 INSULATOR 0.0
CONTACT UP     0.4e-6 0.5e-6 INSULATOR 0.0
CONTACT UP     0.0    0.1e-6 OHMIC       0.0 3.e23
CONTACT UP     0.2e-6 0.4e-6 SCHOTTKY -1.3
CONTACT UP     0.5e-6 0.6e-6 OHMIC       1.0 3.e23
```

It is easy to understand how these lines work at this point.

Some results obtained using Archimedes are shown below.

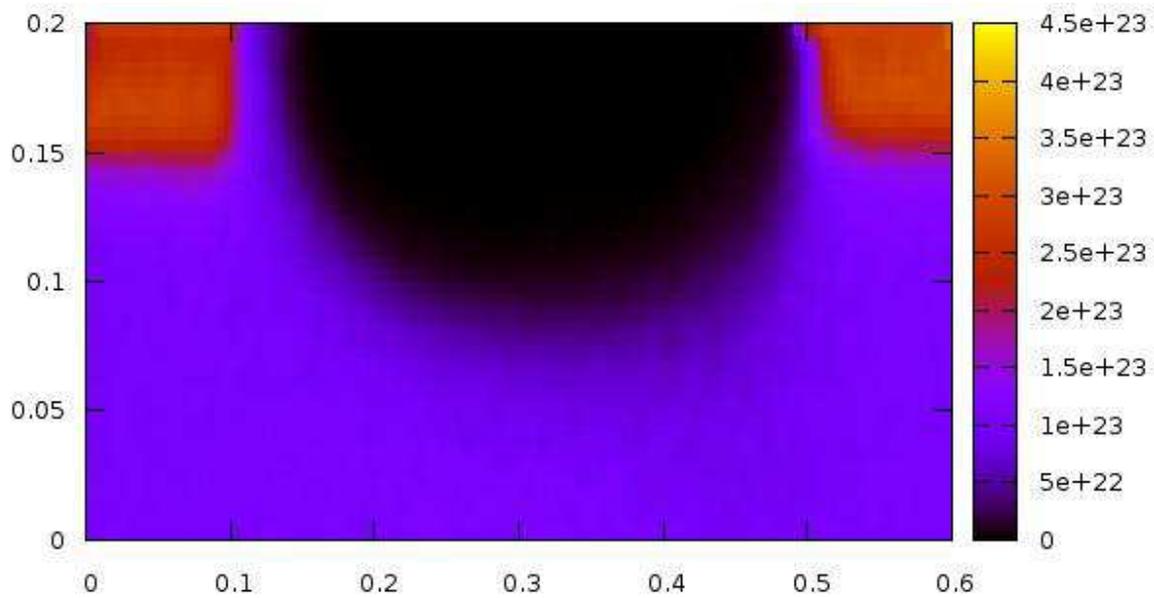

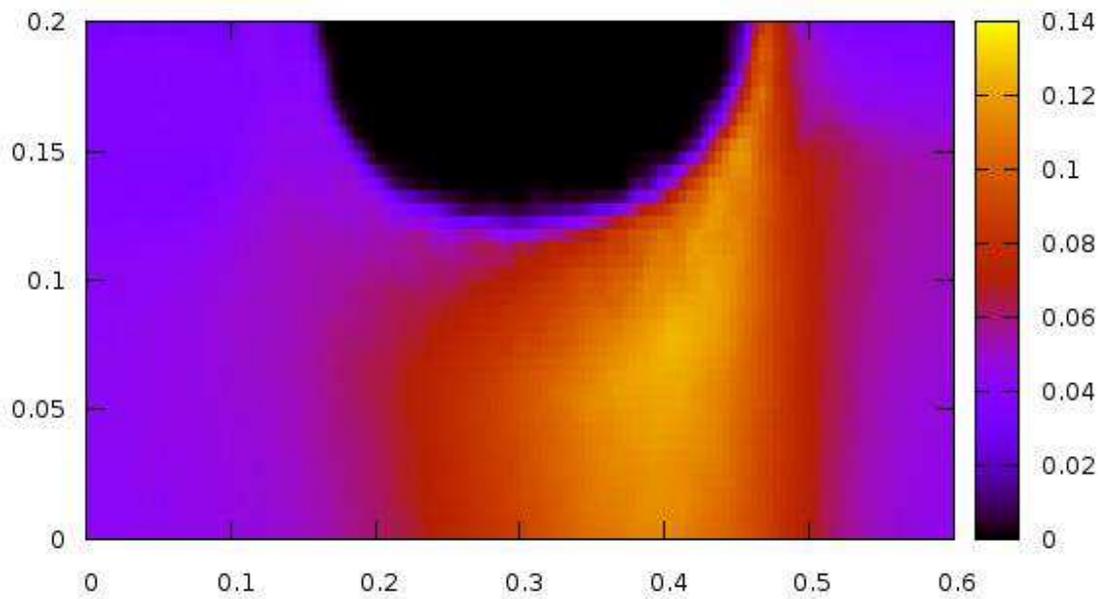

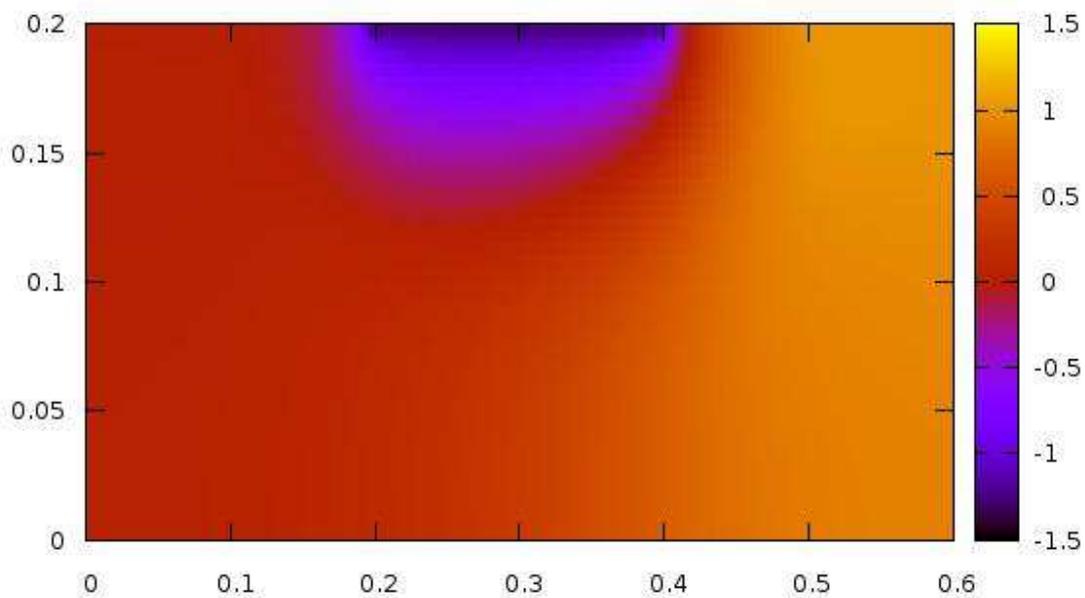

There is plenty of parameters that can be specified in scripts. We briefly have shown how to implement a script for a MESFET, just to give an idea of what can be done in Archimedes. We will see (and we have the whole book for this) other commands in other chapters.

**What is this book about?**

The aim of this book is two-fold. On one hand, I'd like to provide a manual for Archimedes, the free GNU package for submicron and nanoscale semiconductor device simulations. On the other hand, I'd like to provide an introduction to the Monte Carlo method for students, engineers and scientists to the field of semiconductor device simulations.

At the time this book is being written, the field of semiconductor device simulations is going through a very exciting period. Devices of the order of a few nanometers have been successfully manufactured that show very interesting effects. We have reached incredibly small scales that could even not be thought of just a one decade ago. To understand the behavior of such devices, it is very important to understand the mathematical tools scientists and engineers have developed in the last decades.

TCAD (Technology Computer Aided Design) is a well-developed branch that has demonstrated to be of high importance in both Physics and Engineering. It has helped us to understand the extremely rich physics behind electrons transport in semiconductor devices, but it has been extremely useful as a tool to predict the behavior of novel devices even before they have been manufactured (allowing industries to save time and money).

Today, we are in front of two well-developed and understood formalisms i.e. the semi-classical and the quantum theories to describe electron transport in semiconductor devices. These two methods are well represented by, respectively, the well-known Monte Carlo method (for semi-classical transport) and the Wigner equation formalism (for nanoscale devices). At a first glance, the two methods seem to be very different but, as we will see through the whole book, they have many points in commons and, actually, describes the very same problem (the dynamics of an electron in a semiconductor device) from two different point of views. Obviously, the two methods do not give the same results when applied to a very small device, but we will also see that this is due to the fact that the two models actually give answers not to the same questions. One method is well suited to the description of electrons considered as particles and the other is well suited for electrons that behaves like waves.

In Archimedes, the models developed are the semi-classical Monte Carlo method and the Monte Carlo Wigner method. These models, today, represent the state-of-the-art of semiconductor device simulations respectively for semi-classical and quantum regimes. The user can specify the model he/she wants to use in a script that will be parsed by Archimedes. There is, actually, no need to know all details of the physics to run a simulation since everything is taken in charge by Archimedes. Even so, it is useful to know the theory, at least, to interpret the results obtained running Archimedes.

This book is, thus, organized as follows:

**In the first chapter**, we introduce the physics used to describe an electron and its dynamics in a semiconductor device. We first try to understand what an electron is, and then we develop the formalism needed to describe the dynamics of such a particle in the semi-classical and the quantum regimes. When talking about quantum mechanics, we will follow, in this book, the well-known Copenhagen interpretation which is the best interpretation that has been presented since now. Still in chapter one, we introduce the formalism used to describe the scattering phenomena that occurs in semiconductor devices, especially in the semi-classical regime since these devices are the ones that experience important scattering effects (they also occur, in a minor quantity, in quantum regime as we will see). We also show the differences and the similitudes of the two approaches and how to get a model that is, somehow, between the two regimes by means of corrections like quantum effective potentials, Wigner corrections, etc. Finally, we shortly introduce the quantum de-coherence and the emergence of classical behavior from the quantum regime.

**In chapter two**, we specifically introduce the Monte Carlo method and the numerical approximations done to calculate the position and pseudo-wave vectors of an electron. This chapter will often refer to Archimedes code snippets to show how things are coded in this package. This is two-fold: it is propedeutic to students that want to develop their own code and it is useful to who wants to understand how the Monte Carlo method is coded inside Archimedes.

**In chapter three**, we present the syntax of the scripts used to describe a device to be simulated. We will present a list of the commands parsed by Archimedes along with their description. Some examples will be also presented to clarify the use of those commands.

**In chapter four**, we present the GUI that has been developed to run Archimedes in a simple way without having to type any scripts. We will show how to use it and what tools have been implemented to analyze the results. This GUI runs on line and locally and we will show how to install it on a local machine.

The book is intended to be a self-contained explanation of Archimedes and its implemented physics, and try to reduce constant references to outside material, in the spirit of making things clear. It has, anyway, at the end of every chapter a section of references for the motivated readers that want to study the models in further details. A previous knowledge of solid state physics is recommended though many results of this field are reported in this book. Users are strongly encouraged to run the examples reported in this book, downloading, modifying and compiling Archimedes sources.

# Chapter 1

# Basics of semiconductor physics

# and the Monte Carlo method

## 1.1 What is an electron?

In every simulation of semiconductor devices, whatever the transport regime is, semi-classical or quantum, whatever the device is, the basic entity is always what is called an electron. But what is an electron? This is a basic question that needs an answer. Understanding such an entity is of paramount importance. Even if we do not have a definitive answer to this question (and we will see why), we will still try to give an answer. In doing so, we will follow the approach developed by Werner Heisenberg, one of the father of quantum mechanics, in his lessons given at Chicago in Spring 1929. To fully understand what an electron is, we need to understand the experiments that leaded to quantum physics. Those experiments give us a clear description of how an electron behaves in the real world and help us to understand the limits of the two formalisms that have been developed to described this entity. Note that we are using the word entity in a broad sense, without referring to a particle or a wave. The meaning of that will be clear once we understand the experiments.

### 1.1.1   The Wilson chamber

The cloud chamber, known also as the Wilson chamber, is a sealed environment containing highly saturated vapor (that can be water or alcohol). It is used to detect particles and/or radiation. When a particle enters the chamber, it leaves a pattern like the ones shown in figure 1.

It is interesting that the tracks observed in a Wilson chamber can be predicted by means of classical mechanics using the concept of a particle, i.e. an entity which can be considered as a point having a mass and that follows the Newtonian mechanics (classical mechanics). It is indeed possible, applying known electric and magnetic fields inside the chamber, to calculate the mass of the particles using the Newtonian laws, according to the lines observed in the chapter.

What is interesting to note in this experiment is the fact that if radiations are introduced in the chamber by means of radioactive substance one observe tracks that are similar to the ones in figure 1. In other words, this experiment explicitly proofs the discontinuous nature of radiations. Even in this case, indeed, it is possible to predict the shape of the tracks by means of classical mechanics.

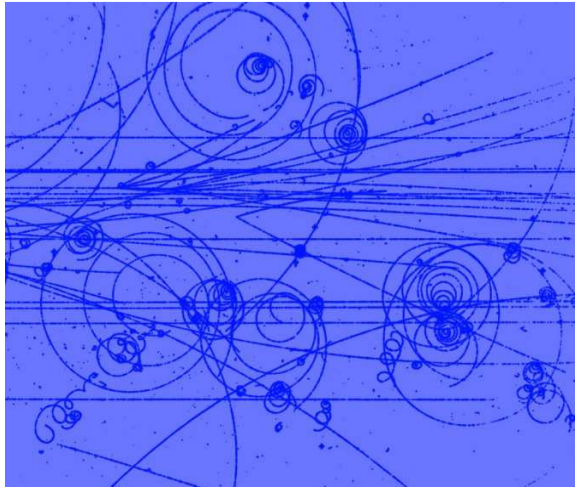

Figure 1 - Particle Tracks in a Wilson Chamber

### 1.1.2  Single-slit Diffraction

If radiations are projected onto a screen by a source of whatever nature nothing actually happens. If the same radiation is projected on a slit of infinitesimal width made on the same screen, one obtains the patterns observed in figure 2. This clearly shows the wave nature of radiations. The patterns reported in figure 2 can be explained using the Maxwell laws of classical mechanics that describe electromagnetic field described as waves.

The incredible fact about this experiment is that one obtains the very same pattern if, instead of electromagnetic waves, one projects a beam of electrons on the slit. This is a very surprising fact that cannot be explained using classical laws. This clearly shows one incredible fact: in certain conditions, electrons show a wave behavior. The only way to explain the patterns of figure 2 is by using the laws of quantum mechanics. Quantum mechanics can explain wave patterns of electrons, classical mechanics cannot.

### 1.1.3  Conclusions

The experiments presented in the previous two sections clearly show the dual nature of radiations and, surprisingly enough, of particles. In one experiment (the Wilson chamber) both waves and particles show their particle nature, in the other (single-slit experiment) both show their wave nature. It is like particles and waves were two different aspects of the same entity. According to these two experiments, we could think of an electron as an object that, in some cases, behaves like a classical particle while, in other cases, behaves like a wave. How to explain this fact?

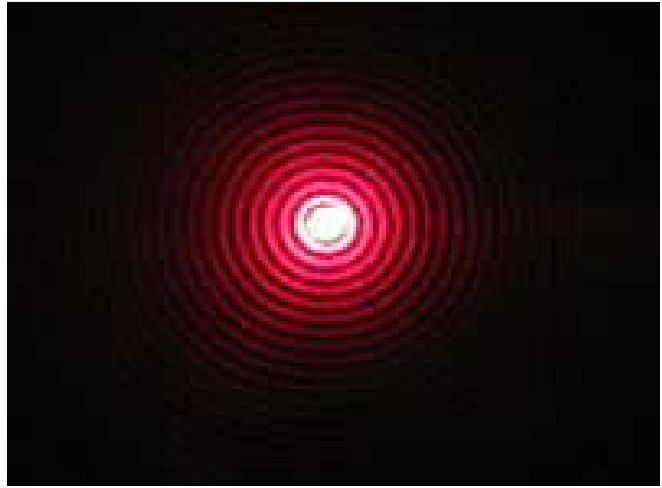

To make things short and give a useful answer to the question, we will use the same words of W. Heisenberg:

"*From experiments…it is clear that matter and radiation present a double nature: their behavior is, on one hand, similar the behavior of waves and, on the other, similar to the behavior of particles… Now, it is clear that matter cannot be made of waves and particles at the same time and that the two representations are profoundly different… The solution of this difficult problem is in the fact that both models (waves and particles) should be considered only as analogies that are valid in some situations and not in others…For example, the electron behaves as a particle in some experiences. But this does not proof at all that the electron holds all properties of a particle. The same can be said, mutatis mutandis, for the wave model. Both representations can be considered valid as analogies only in specific limit cases; but, as a matter of fact, the atomic phenomena cannot be described directly using our usual language. Light and matter are unitary physical phenomena; their apparent double nature comes from the intrinsic limits of our language.*" (translated from Die physikalischen Prinzipien der Quantentheorie, Leipzig 1930)

In other words, particles and waves are just mathematical models that we have created and that are based on our every-day experiences. We created a language that is well-suited to objects of the order of human experiences. Though these models are adapt to describe classical objects, they show their intrinsic limits when applied to the atomic world. We can still use the words "*particles*" and "*waves*", but we should always remember that they have been created for classical objects. They can be used only as analogies that help us to "*visualize*" the physics when applied to the atomic world.

We show now how the two models look like mathematically.

In classical physics, also known as Newtonian physics, an electron is represented by a particle, i.e. an extremely small object having a mass and which dimensions are negligible. A particle is described, in any time, by two vectors, i.e. the position

$$\bar{x} = (x, y, z)$$

and the velocity

$$\bar{v} = \left(v_x, v_y, v_z\right)$$

The dynamics of a particle in this regime (and so the dynamics for an electron) is described by the following equation (the Newton law):

$$m\frac{d^2\bar{x}}{dt^2} = -\nabla U \qquad (1.1)$$

Where $m$ is the mass of the particle and $U = U(x, y, z)$ is a function known as the potential, that represents the forces acting on the particle. From equation (1.1) it is possible to recover an equation describing the evolution in time of the velocity vector by considering the fact that :

$$\bar{v} = \frac{d\bar{x}}{dt}$$

In the quantum regime, things are different. An electron is no more described by means of position and velocity vectors. An electron is now described by a mathematical object called a *wave function* $\Psi = \Psi(x, y, z)$ which is a complex function similar to a wave and which square module tell us the probability of finding an electron in that point. The equation that describes the dynamics of this wave function is (the Schroedinger equation):

$$i\hbar \frac{\partial \Psi}{\partial t} = \hat{H}\Psi \qquad (1.2)$$

Even if equations (1.1) and (1.2) look very different from each other, they are both deterministic i.e. they describe the state of a system at a time, basing itself on the past state of the system, but while one describes the evolution of position and velocity of an electron, the other describes the evolution of the probability of finding an electron in a given point. So, the approaches have some analogies (being both deterministic) but differ for the evolving quantities they describe.

## 1.2 Crystal lattices, Energy bands and Bloch theorem

A semiconductor device, in simple words, can be considered as a piece of semiconductor material which, at the atomic level, is simply an highly ordered arrangement of atoms (or molecules).

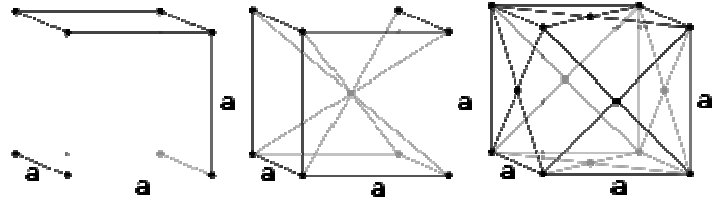

All crystal structures are described in terms of a *unit cell*, which look like one of the ones reported in Figure 3. The unit cell is the basic element that is repeated in all directions and that constitutes the crystal. The points in the cell represents the atoms or molecules that the crystal is made of. The points can be all of the same atomic element or not. Usually, a unit cell is given by the *lattice constants*, which are the lengths of the cell edges, and the atomic positions, which are the positions of the atoms inside the cell.

Mathematically, a crystal is described by what is known as a *Bravais lattice*. A Bravais lattice is an infinite discrete set of points generated by the application of translation operations. The translation reads:

$$\bar{R} = n_1 \bar{a}_1 + n_2 \bar{a}_2 + n_3 \bar{a}_3 \qquad (1.3)$$

where $n_i \in \mathbb{N}$, and $\bar{a}_i$ are three-dimensional vectors (which length is equal to the lattice constants). Applying this definition, a crystal is, thus, described as an arrangement of one or more atoms (known as the *basis*) repeated at each lattice point.

For example, in Figure 4 we report the five fundamental two-dimensional Bravais lattice to show how they look like. From that picture, one clearly see how to use a Bravais lattice to describe a real crystal lattice. It is important to have a mathematical description of the crystal structure since, as we will see it is of paramount importance to calculate the allowed energies an electron can have in that crystal. For this, we introduce a further very useful concept, the reciprocal lattice.

Mathematically a *reciprocal lattice* of a given Bravais lattice, as the one in (1.3), is a lattice that is described by the following vectors

$$\bar{A}_1 = \frac{2\pi(\bar{a}_2 \times \bar{a}_3)}{\bar{a}_1 \cdot (\bar{a}_2 \times \bar{a}_3)}, \qquad \bar{A}_2 = \frac{2\pi(\bar{a}_3 \times \bar{a}_1)}{\bar{a}_2 \cdot (\bar{a}_3 \times \bar{a}_1)}, \qquad \bar{A}_3 = \frac{2\pi(\bar{a}_1 \times \bar{a}_2)}{\bar{a}_3 \cdot (\bar{a}_1 \times \bar{a}_2)}$$

Reciprocal lattices are useful for the calculation of the allowed energies.

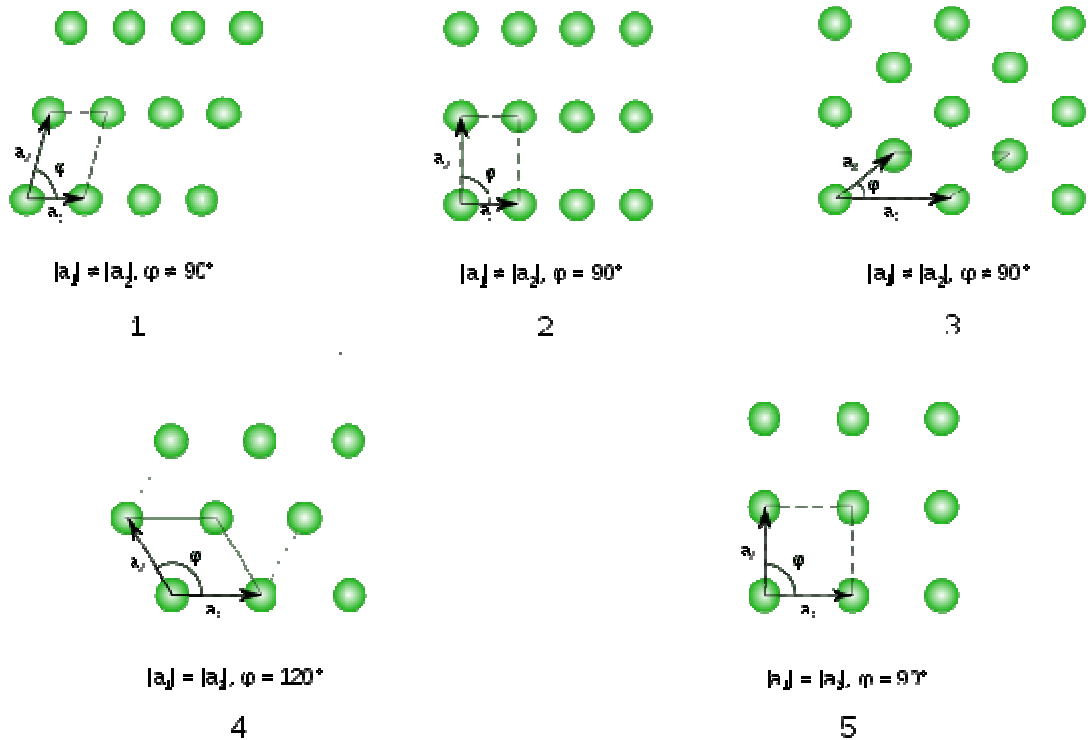

Figure 4 - The five fundamental two-dimensional Bravais lattices: 1 oblique, 2 rectangular, 3 centered rectangular (rhombic), 4 hexagonal, and 5 square

We introduce now the *Bloch theorem*, one of the most important theorem of solid state physics. We will not provide a proof of it. Instead we will see how to utilize the results of this theorem for the purpose of describing electron dynamics in crystal structures.

*Bloch Theorem*

Let us suppose we have an electron in periodic crystal lattice (equivalently a periodic potential). The wave-function of the electron is such structure is given by:

$$\Psi_{n\bar{k}}(\bar{x}) = e^{i\bar{k}\cdot\bar{x}}u_{n\bar{k}}(\bar{x})$$

i.e. as the product of a plane wave envelope function and a periodic function $u_{n\bar{k}}$ that has the same periodicity of the crystal lattice. The corresponding energies have the periodicity of the corresponding reciprocal lattice, i.e. $\varepsilon_n(\bar{k}) = \varepsilon_n(\bar{k} + \bar{K})$, where $\bar{K} = n_1\bar{A}_1 + n_2\bar{A}_2 + n_3\bar{A}_3$ and $n_i$ is an integer.

The vector $\bar{k}$ is called the pseudo-wave vector and it can be proved (for a free electron) that it is directly proportional to the electron velocity, i.e.

$$\hbar\bar{k} = m\bar{v}$$

*Energy levels and Energy bands*

As we saw earlier, an electron in crystal lattice is subject to a periodic potential due to the lattice of positive ions that constitutes the crystal. Mathematically this potential reads:

$$V(\bar{x}) = V(\bar{x} + \bar{X})$$

where $\bar{X} = n_1\bar{a}_1 + n_2\bar{a}_2 + n_3\bar{a}_3$, as usual.

The calculation of the energies an electron should have in that potential should be done by solving the Schroedinger equation (1.2), where the Hamiltonian should include that potential, i.e.

$$\hat{H} = -\frac{\hbar^2\nabla^2}{2m} + V(x, y, z) \tag{1.4}$$

The calculated eigenvalues and eigenvectors would represent, respectively, the allowed energies and the wavefunctions of an electron in that crystal.

**The Bloch theorem gives the solution to the problem without having to use any sophisticated mathematical method.**

Furthermore, the Bloch theorem give us a way to classify the solutions since the energy levels depend of the pseudo-wave vector. The explicit mathematical relation that give us the energy in function of the pseudo-wave vector depend on the crystal lattice and for each lattice we have a corresponding relation. This relation is called the *energy band* of the material. In Figure 5, we show the energy bands for several relevant semiconductor materials, i.e. Silicon, Germanium and Gallium Arsenide.

As one see from the figure, more than one energy levels can correspond to the same pseudo-wave vector.

It is interesting that, for practical problems like devices simulations, one can use approximations of the energy band. In Archimedes, for example, two approximations are implemented (along with the full-band approach), i.e. the parabolic band and the Kane approximation.

The parabolic band, as the name itself suggests, approximate the energy band in the minimum energy valley (for GaAs, for example, it is the point named Γ in figure 5) where the vast majority of electrons is (though this approximation is not valid in some cases, as we will see latter on).

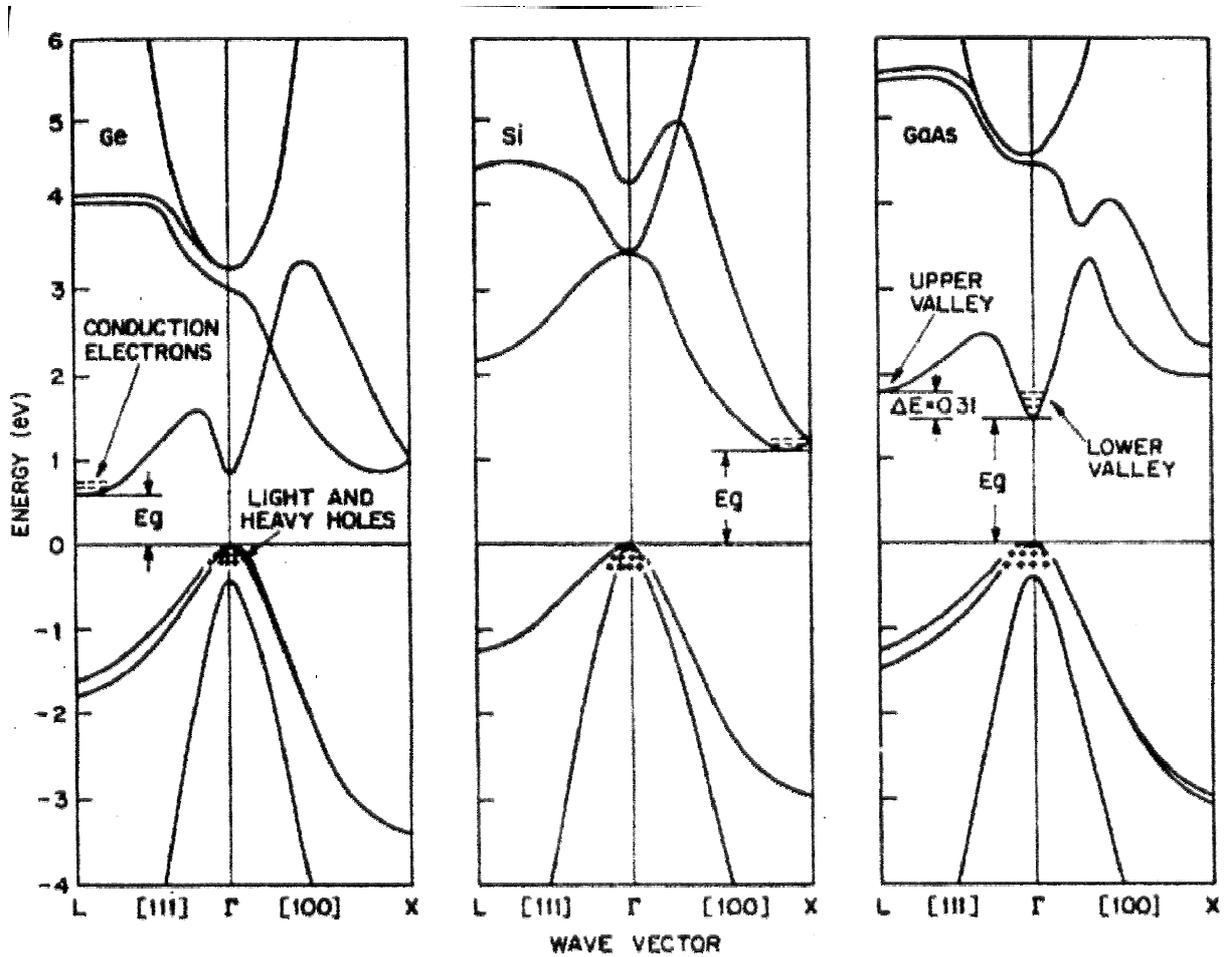

The parabolic band reads:

$$\varepsilon\left(\bar{k}\right) = \frac{\hbar k^2}{2m^*} \tag{1.5}$$

where $m^*$ is called the *effective mass* of the electron. This approximation basically treats the electron in the lattice as it were free and had a mass equal to $m^*$.

The most used non-parabolic approximation of the energy band is the so called Kane approximation. In this case the energy band reads:

$$\mathcal{E}\left(\bar{k}\right) * \left(1 + \alpha \mathcal{E}\left(\bar{k}\right)\right) = \frac{\hbar k^2}{2m^*} \tag{1.6}$$

Finally, once the energy band is fixed, an important relation is the following (that is extremely useful for calculating the electron velocity)

$$\bar{v} = \frac{\nabla_{\bar{k}}\varepsilon(\bar{k})}{\hbar} \tag{1.7}$$

This relation will be very useful when we will need to find the velocity of an electron in the Monte Carlo method.

## 1.3 Scattering

### 1.3.1 Phonon Scattering

A crystal lattice is not a fixed object. The atoms (or molecules) that constitute the nodes of the lattice are actually in motion, they vibrate around an equilibrium position. If one should take into account the exact dynamics of that vibration, the mathematical problem would be impossible to solve (even with numerical approximation techniques). We approximate, thus, the problem to a simpler one.

The lattice vibrations are quantized and a new "particle" is introduced, the *phonon*. The phonon can be seen as the quantization of the vibration exactly in the same way the photon is the quantization of the electromagnetic field.

The lattice vibrates because of the non-zero temperature of it. The highest the temperature, the more the vibration of the lattice. The lattice vibration is nor a random phenomenon. The lattice uniformly oscillates since the nodes are all connected by forces (two positive ions cannot approach to each other indefinitely).

In classical mechanics, these lattice vibrations are called *normal modes* and every vibration mode of a lattice can be described as a superposition of normal modes. Every solid (and in particular semiconductor materials) that has more than one atomic element (or different masses or bonding strengths) in its unit cell can produce two types of vibrations, and thus two types of phonons:

- *Acoustic* phonons
- *Optical* phonons

These two types differs by their dispersion relations, i.e. the law that connects the frequency of the phonon with its pseudo-wave vector (there is no need for us to know this equation). The one for acoustic phonon is linear. More simply, to have a physical idea of the acoustic and optical phonons, we can think of acoustic phonons as quantization of the lattice vibrations where neighboring atom displace all in the same direction while for optical phonons they displace in opposite directions.

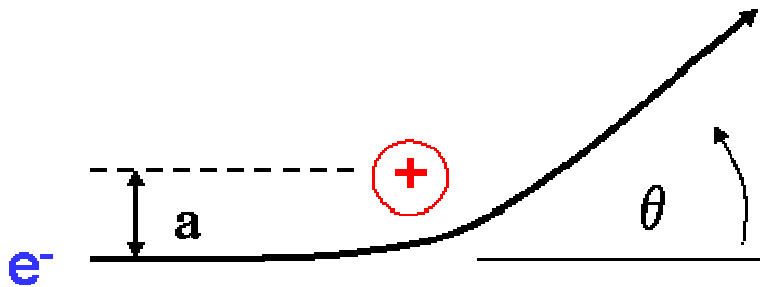

Acoustic phonons are the quantization of acoustic waves in the crystal. They come in two flavor :

- *Longitudinal*
- *Transversal*

according to the direction of the acoustic wave.

Optical phonons are called so because in ionic crystals they are excited by infrared radiation. Also optical phonons can be longitudinal or transversal.

An electron in a crystal can experience frequent collisions with phonons and we define this phenomenon as *phonon scattering*. Scattering is an important phenomenon that has to be taken into account if a realistic (see reliable) simulation have to be done for a device. Scattering depends on the temperature and dimensions of the device. If an electron scatters with an acoustic phonon, we say that the scattering is acoustic. When the electron scatters with an optical phonon, we say that the scattering is optical.

In this book, for the sake of simplicity, we will not go through all details of scattering theory. We will simply report the way scattering probabilities are calculated and how it is included in Archimedes.

### 1.3.2 Impurity Scattering

Electrons in a semiconductor crystal can experience an other type of scattering, the one generated by the presence of ionized impurities in the crystal. This is a well-known phenomenon that can be demonstrated experimentally that happens in highly doped devices.

Impurities can be considered to be localized positive point charges in the crystal. If this positive point charge were in the vacuum it would be easy to calculate the Coulomb potential generated by it. The electron would be deviated as reported in Figure 6. But we have to take into account the fact that it is located in a crystal and, thus, we have to deal with the electrons that tend to screen this potential. To do this, we need to find a way to calculate the screened potential of the impurity.

Two methods exist to calculate the screened potential of an impurity known as the Brooks-Herring and the Conwell-Weisskopf approaches. The one implemented in Archimedes is the Brooks-Herring method and this is the one we will describe in the next section and use for our simulations.

### 1.4 Scattering formulae

It is possible, by means of quantum mechanics calculations, to obtain a formula that gives the probability an electron has to scatter with a given scatter center (an impurity or a phonon). The probability is calculated by means of the *Fermi Golden Rule*, that is a method to express probability as the result of a first order perturbation problem.

In general, the probability that an electron of pseudo-wave $\bar{k}$ scatters is given by the following formula:

$$W(\bar{k}) = \frac{\Omega}{(2\pi)^3} \int_0^{2\pi} \int_0^{\pi} \int_0^{\infty} S(\bar{k}, \bar{k}') dk' d\vartheta d\phi \qquad (1.8)$$

where $S(\bar{k}, \bar{k}')$ is the transition rate, the transition probability per unit time that an electron scatter from the initial state $\bar{k}$ to the state $\bar{k}'$, $\phi$ is the azimuthal angle, $\theta$ is the polar angle of the pseudo-wave vector and $\Omega$ is the volume of the crystal. This is the formula will use as starting point for the probability of all scattering phenomena implemented in Archimedes.

In this section, we do not go in all details but simply report the formulae to calculate the probability an electron scatters with an impurity or a phonon. Along with the probability formula to obtain a scattering, we report also how to calculate the final state of an electron after the scattering has occurred. This will be useful for the implementation of scattering phenomena in the Monte Carlo calculations of electron transport.

### 1.4.1  Phonon scattering formulae

The transition probability for phonons depend on the type of phonon. We have to distinguish, thus, between optical and acoustic transitions. Furthermore, there is another distinction that has to be done with phonons. They can be *polar* and *non-polar*. Polar phonons are usually present in compound semiconductors. This scattering effect can be very strong. Polar phonons can be acoustic and optical. In the first case, the scattering is termed piezoelectric scattering. This phenomenon is very strong in extremely pure semiconductor crystals at very low temperature so we will consider it to be negligible in practical applications.

We now present the phonons scattering rates implemented in Archimedes, i.e.:

- Acoustic phonons (non-polar)
- Non-polar optical phonons
- Polar optical phonons

*Acoustic phonons*

Acoustic phonon scattering occurs at energies that are much smaller than the average thermal energy of electrons at room temperature $k_B T_L$, so it is considered to be elastic (momentum and energy conservation of the electron). The transition probability for acoustic phonons is given by:

$$S(\bar{k}, \bar{k}') = \frac{\Xi_d^2 k_B T_L}{8\pi^2 \hbar c_L} \frac{k}{E_k} \int \frac{1}{q} \delta\left(\frac{q}{2k} \mp \cos\theta'\right) d\bar{q} \qquad (1.9)$$

where $\Xi$ is the deformation potential, $\bar{q}$ is the phonon wave vector, $\theta'$ is the angle between the electron pseudo-wave vector and the phonon wave vector, and $c_L$ is the elastic constant of the material. Putting equation (1.9) into (1.8), one get the probability for an electron to scatter with an acoustic phonon:

$$W(\bar{k}) = \frac{2\pi \Xi_d^2 k_B T_L}{\hbar c_L} N(E_k) \qquad (1.10)$$

where $N(E_k)$ is the density of states given by:

$$N(E_k) = \frac{(2m^*)^{3/2} \sqrt{E_k}}{4\pi \hbar^3} \qquad (1.11)$$

The final state of an electron is then selected as follows:

- The azimuthal angle $\phi$ is determined by a random number between $0$ and $2\pi$ since the transition rate does not depend on $\phi$.
- $\cos\theta$ is a random number between $-1$ and $+1$.

*Non-polar Optical phonons*

Optical phonons (polar and non-polar) scattering occurs at energies $\hbar\omega_0$ that are comparable to the average thermal energy of electrons at room temperature and, thus, is considered to be an inelastic phenomenon. The transition probability for non-polar optical phonons reads as follows:

$$S(\bar{k}, \bar{k}') = \frac{\pi D_0^2}{\rho \omega_0 \Omega} \left(n_0 + \frac{1}{2} \mp \frac{1}{2}\right) \delta\left(\frac{\hbar^2 q^2}{2m^*} \pm \frac{\hbar^2 k q \cos\theta'}{m^*} \mp \hbar\omega_0\right) \qquad (1.12)$$

where $D_0$ is the optical deformation potential, $\omega_0$ is the angular frequency and $n_0$ is the number of polar phonons. Substituting (1.12) into (1.8) one obtains the scattering rate, i.e.:

$$W(\bar{k}) = \frac{\pi D_0^2}{\rho \omega_0}\left(n_0 + \frac{1}{2} \mp \frac{1}{2}\right)N(E_k \pm \hbar\omega_0) \tag{1.13}$$

Concerning the choice of the final state, the azimuthal angle $\phi$ and the polar angle $\theta$ are selected by two random numbers distributed uniformly. To obtain the electron energy after the collision, an energy equal to the optical phonon energy $\hbar\omega_0$ is added or subtracted to the initial energy of the electron, according the that a phonon has been, respectively, absorbed or emitted.

*Polar Optical phonons*

The transition probability for the polar optical phonon scattering is given by:

$$S(\bar{k}, \bar{k}') = \frac{\pi e^2 \omega_0}{\epsilon_p \Omega}\left(n(\omega_0) + \frac{1}{2} \mp \frac{1}{2}\right) \times \frac{\Omega}{(2\pi)^3}\int \frac{1}{q^2}\delta\left(\frac{\hbar^2 q^2}{2m^*} \pm \frac{\hbar^2 kq\cos\theta\prime}{m^*} \mp \hbar\omega_0\right) \tag{1.14}$$

Substituting, as usual, (1.14) into (1.8) one obtains the scattering rate for polar optical phonons:

$$W(\bar{k}) = \frac{e^2 \omega_0}{8\pi\epsilon_p}\frac{k}{E_k}\left[n(\omega_0) + \frac{1}{2} \mp \frac{1}{2}\right]ln\left(\frac{q_{max}}{q_{min}}\right) \tag{1.15}$$

where

$$q_{min} = k\left|1 - \left(1 \pm \frac{\hbar\omega_0}{E_k}\right)^{1/2}\right|$$

and

$$q_{max} = k\left[1 + \left(1 \pm \frac{\hbar\omega_0}{E_k}\right)^{1/2}\right]$$

Concerning the final electron state after scattering occurred, it is selected by a random azimuthal angle $\phi$ while the polar angle $\theta$ is determined by the following formula:

$$\cos\theta = \frac{1 + f - (1 + 2f)^r}{f}$$

where $r$ is a random number between 0 and 1, and

$$f = \frac{2\sqrt{E_k E_{k'}}}{\left(\sqrt{E_k} - \sqrt{E_{k'}}\right)^2}$$

### 1.4.2  Impurity scattering formulae

The transition probability for the impurity scattering is given as follows:

$$S(\bar{k}, \bar{k}') = \frac{2\pi}{\hbar} \frac{N_I Z^2 e^4}{\Omega \epsilon_S^2} \frac{\delta(E_{k'} - E_k)}{\left[2k^2(1 - \cos\theta) + q_D^2\right]^2} \tag{1.16}$$

where $\hbar$ is the reduced Planck constant, $N_I$ is the density of impurities, $e$ is the elementary charge, $\epsilon_S$ is the dielectric constant of the material, $E_k$ is the energy of the corresponding pseudo-wave vector, and finally :

$$q_D = \sqrt{\frac{e^2 n_0}{\epsilon_S k_B T_L}}$$

with $1/q_D$ the Debye length, $n_0$ the equilibrium electron density at temperature $T_L$ of the lattice, $k_B$ the Boltzmann constant.

The final state of an electron after an impurity scattering is calculated as follows:

- The azimuthal angle $\phi$ is determined by a random number between $0$ and $2\pi$ since the transition rate does not depend on $\phi$.
- The polar angle is calculated by the following formula where $r$ is a random number between 0 and 1:

$$\cos\theta = 1 - \frac{2r}{1 + (1 - r)\left(\frac{2k}{q_D}\right)^2}$$

Putting formula (1.16) into (1.8) one get the probability for an electron to scatter with an impurity center:

$$W(\bar{k}) = \frac{2\pi N_I Z^2 e^4 N(E_k)}{\hbar \epsilon_S^2} \frac{1}{q_D^2(4k^2 + q_D^2)} \tag{1.17}$$

## 1.5 Scattering rates for non-parabolic bands

All scattering rates reported in the previous sections have been calculated assuming that the energy band is parabolic and spherical. In this section, we show how to modify the rates in case of non-parabolic and/or elliptical bands. In case of non-parabolic and non-spherical bands, the analytical calculations are extremely difficult to carry on and some assumptions have to be assumed. We will not present such details here but briefly show the "corrections" to apply to the rates to take into account energy bands different than the parabolic and spherical ones.

In case of non-parabolicity, for example the Kane band in (1.6), it can be proved that the acoustic scattering rates remains the same as in (1.10) while the density of states (1.11) become

$$N(E_k) = \frac{(2m^*)^{3/2}\sqrt{\gamma(E_k)}}{4\pi^2\hbar^3}\frac{d\gamma(E_k)}{dE_k} \qquad (1.18)$$

where

$$\gamma(E_k) = E_k(1 + \alpha E_k)$$

When non-spherical bands are taken into account, the effective mass is represented by three values, each corresponding to a direction. In this case, the energy band reads:

$$E(\bar{k}) = \frac{\hbar^2}{2}\left(\frac{k_x^2}{m_x^*} + \frac{k_y^2}{m_y^*} + \frac{k_z^2}{m_z^*}\right) = \frac{(\hbar k^*)^2}{2m_d^*}$$

where

$$m_d^* = \left(m_x^* m_y^* m_z^*\right)^{1/3}$$

The acoustic scattering remains the same as in (1.10) and the density of states (1.11) is now:

$$N(E_k) = \frac{(2m_d^*)^{3/2}\sqrt{E_k}}{4\pi^2\hbar^3} \qquad (1.19)$$

## 1.6 Electrostatic Potential

The electrostatic potential is a quantity required if self-consistent device simulations are our goal. The equation that describes the electrostatic potential (or just potential for short) is known as the *Poisson equation*, which is part of a more general set of equations known as Maxwell equations that describes the electromagnetic field dynamics. This equation describes the evolution of the potential as a function of the charge density. Mathematically Poisson equation reads:

$$\nabla \cdot (\epsilon_S \nabla \varphi) = -\rho \qquad (1.20)$$

where $\epsilon_S$ is the dielectric constant of the semiconductor material, $\varphi$ is the potential and $\rho$ is the charge density which, in turn, reads:

$$\rho(\bar{x}) = e[n(\bar{x}) - N_D(\bar{x}) - p(\bar{x}) + N_A(\bar{x})]$$

with $n(\bar{x})$ the electron density, $p(\bar{x})$ the hole density, $N_D(\bar{x})$ the donor density and $N_A(\bar{x})$ the acceptor density.

Equation (1.20) has to be solved self-consistently with the selected transport model to obtain realistic and predictive results. Basically that means that the potential depends, through equation (1.20), on charge density and that, in turn, charge density depends on the potential, through the selected transport model. So at every time step, the two quantities have to be updated and calculated. We will see in the next chapter what self-consistency means in more details.

Whatever the electron transport model is, equation (1.20) has always to be solved. This is a common part of every semiconductor devices simulator.

## 1.7 Quantum Effective Potential

Today semiconductor devices are so small that their characteristic length starts to be comparable to the size of an electron wave packet. *Effective potential* models have been created to simulate some the quantum effects arising from the non-zero size of electron wave packets.

These models have the great advantage to be simple to implement. They are easy to include in semi-classical transport models but, unfortunately, they are not able to include important quantum effects like barrier tunneling and/or source-drain tunneling.

The main idea of Quantum Effective Potentials is well represented in the following picture (Fig. 7). The quantum corrections are incorporated into a Monte Carlo simulator by simply introducing a quantum potential term which is superimposed onto the classical electrostatic potential seen by the simulated particles (Poisson equation).

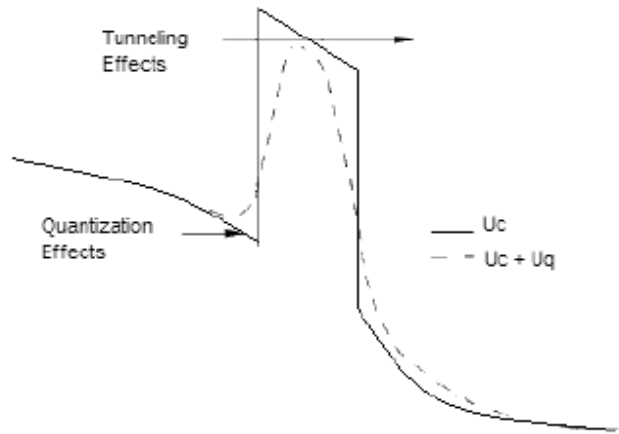

Quantum effective potential models differ from each other for the way quantum potentials are calculated. Today this kind of models is implemented in commercial codes with success.

The models implemented in Archimedes and are the following ones:

- *Bohm potential*

- *Calibrated Bohm potential*

- *Full effective potential*

- *Density gradient potential*

Mathematically the Bohm potential reads:

$$Q_{Bohm} = -\frac{\hbar^2}{2\,m^*} \frac{\nabla^2 n}{n} \qquad (1.21)$$

where $n$ is the electron density.

The Calibrated Bohm potential can be considered as a generalization of (1.21). The potential reads:

$$Q_{W.Bohm} = -\frac{\hbar^2}{2} \gamma \frac{\nabla[1/m^* \nabla n^\alpha]}{n^\alpha} \qquad (1.22)$$

where $\alpha$ and $\gamma$ are two fitting parameters that can be calibrated by means of more sophisticated (and thus more computationally demanding) quantum models.

The full effective potential reads:

$$Q_{full} = \alpha^2 \frac{\partial^2 V}{\partial x^2} \qquad (1.23)$$

Finally, the density gradient potential reads:

$$Q_{grad} = -\frac{\gamma \hbar^2}{6m} \frac{\nabla^2 \sqrt{n}}{\sqrt{n}} \qquad (1.24)$$

where $\gamma$ is a fit factor.

To use these effective potentials, basically, one calculates the classical potential by means of the Poisson equation (1.20) and add the selected effective potential to it. Then the electron transport is obtained based on the potential just calculated, i.e.:

$$V(\bar{x}) = \varphi(\bar{x}) + Q(\bar{x}) \qquad (1.25)$$

## 1.8 Many electrons problem: The Boltzmann equation

The problem of electrons transport in a semiconductor crystal lattice (or a device) with a applied and a self-consistent potential is a quantum multi-body problem that, in principle, should be resolved by means of multi-body Schroedinger equation. In this case the Schroedinger equations would read:

$$\hat{H}\Psi(\bar{x}_1, \bar{x}_2, \ldots, \bar{x}_N, \bar{X}_1, \bar{X}_2, \ldots, \bar{X}_N) = i\hbar \frac{\partial \Psi(\bar{x}_1, \bar{x}_2, \ldots, \bar{x}_N, \bar{X}_1, \bar{X}_2, \ldots, \bar{X}_N)}{\partial t}$$

with

$$\hat{H} = \sum_l \frac{\bar{P}_l}{2\,M} + \frac{1}{2}\sum_{l,m} \frac{Z^2 q^2}{4\pi\varepsilon_0 |\bar{X}_l - \bar{X}_m|} + \sum_i \frac{\bar{p}_i}{2m} + \frac{1}{2}\sum_{i,j} \frac{q^2}{4\pi\varepsilon_0 |\bar{x}_i - x_j|} + \sum_{i,l} \frac{Zq^2}{4\pi\varepsilon_0 |\bar{X}_i - \bar{X}_l|}$$

where the index $l$ is for the nuclei of the crystal, the $i, j$ is for the electrons in the crystal. $M$ is the mass of the nuclei, $Z$ the charge of the nuclei, $m$ the mass of the free electrons. This equation is, obviously, an incredibly difficult problem to solve, even from a purely numerical point of view. Approximations have to be introduced to simplify the problem.

The first approximation we introduce concerns the nuclei and the electrons that surround the nuclei. We can consider the whole set of such particles as a single entity that we keep on calling the nuclei.

The second approximation to use to simplify the problem is the well-known Born-Oppenheimer approximation that consists in separating the dynamics of nuclei and the dynamics of electrons. We consider the nuclei much more heavy and slow than the electrons and, as so, they can are treated as a separated classical problem.

We also suppose that the quantum nature of electrons can be neglected which is equivalent to say that an electron behave as a classical object (a billiard ball).

Finally, we suppose that the overall electrostatic field can be treated by means of the Poisson equation.

This set of approximations, without going through all details, brings us to a very powerful tool known as the Boltzmann Transport Equation (BTE). The BTE is given by:

$$\frac{\partial f(\bar{x},\bar{k},t)}{\partial t} + \frac{1}{\hbar}\nabla_{\bar{k}}\mathcal{E}(\bar{k})\nabla_{\bar{x}}f(\bar{x},\bar{k},t) + \frac{qE(\bar{x},t)}{\hbar}\nabla_{\bar{k}}f(\bar{x},\bar{k},t) = \left[\frac{\partial f}{\partial t}\right]_{collision} \qquad (1.26)$$

coupled to the Poisson equation (1.20) with

$$n(\bar{x}) = \int f(\bar{x},\bar{k},t)d\bar{k}$$

The function $f = f(\bar{x},\bar{k},t)$ is a known as the Boltzmann distribution function and it is a dimensionless function which is used to extract all observables of interest. Mathematically speaking, the distribution function represents the probability for an electron to have a position $\bar{x}$, a pseudo-wave vector $\bar{k}$ at time t.

Unfortunately, even the BTE is a daunting task to solve, even from a numerical point of view. That is why the Monte Carlo method has been created, to solve this difficult problem.

The Monte Carlo method can be considered a stochastic method that actually solves the Boltzmann Transport Equation without directly dealing with it.

## 1.9 The Self-consistent Ensemble Monte Carlo method

The ensemble Monte Carlo method is based on the simultaneous calculation of many particles (ensemble) dynamics during a small interval of time $dt$ (the time step) till a final time $T_f$ is reached. This method can be used to get both the stationary and transient solutions of BTE applied to a semiconductor device.

The Monte Carlo method can be seen as a semi-classical statistical method to get numerical solutions to the BTE problem. It provides accurate solutions to very complex problems that need sophisticated things like full band energy bands and scattering terms. This method is said to be semi-classical. It is *classical* because particles are treated as classical particles that scatters with scattering centers. It is *semi* (-classical) because it uses scattering probabilities that are actually calculated quantum mechanically (as we saw, using the Fermi golden rule). Finally, it is *statistical* since the simulation of scattering effects is obtained by generating random numbers.

A flow chart of the ensemble Monte Carlo method is reported in figure 8.

It is easy to see, from this flow chart, how easy it is to implement this method and how this method actually works. We will see, in more details, how to implement this method in the next chapter, when we will describe the sources of Archimedes. For now, let us focus on the equations used in such method.

Despite the big run-times needed to simulate a BTE, Monte Carlo has some great advantages that make it a irreplaceable tool when accurate physics is needed. First, it easily includes scattering mechanisms, second it can include very easily any kind of energy band model, even sophisticated ones obtained by previous calculations (like tight-binding or pseudo-potential methods). The main drawback of the semi-classical Monte Carlo method is its inadequacy to cope pure quantum phenomena such as barriers tunneling, short channel quantization effects, etc (despite methods exist to include those effects, but in that case we solve the Wigner equation and not the Boltzmann equation anymore).

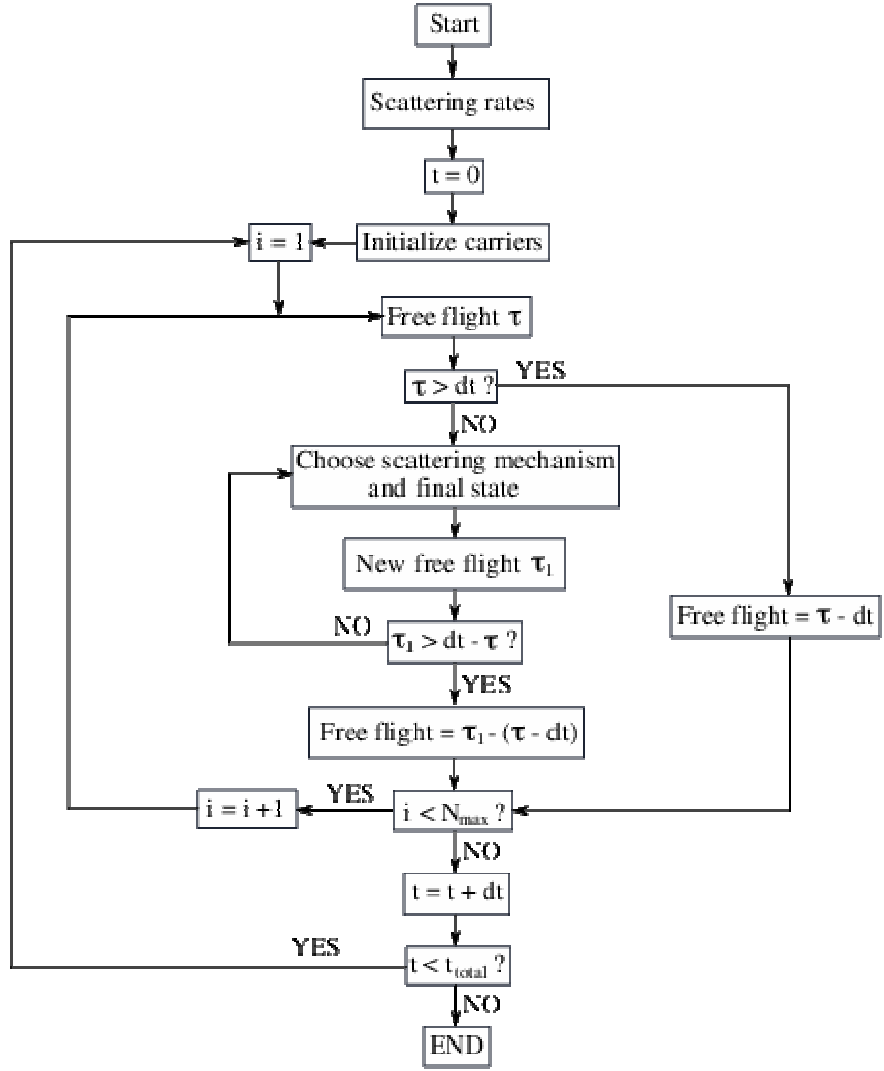

Let us see, now, how the transport of a particle is carried out in a Monte Carlo simulation. First of all, we consider that the motion of an electron is described by two quantities that are continuous variables, i.e. the position vector $\bar{x}$ and pseudo-wave vector $\bar{k}$. The equations describing the evolution of those quantities in space and time read:

$$\frac{d\bar{x}}{dt} = \frac{1}{\hbar} \nabla_{\bar{k}} \mathcal{E}(\bar{k}) \tag{1.27}$$

$$\frac{d\bar{k}}{dt} = \frac{q\bar{E}}{\hbar} \tag{1.28}$$

where $\bar{E}$ is the electric field, and the other quantities are well-known. This part of the Monte Carlo method is known as the *Drift Process*.

Electrons are scattered by centers in the semiconductor material that represents the vibration of the lattice due to the non-zero temperature. Equation (1.27) and (1.28) apply between one scattering and another. The motion of an electron between two scattering events is known as *Free Flight*. Let us see, now, how the time spent in a free flight is calculated in the Monte Carlo method.

## 1.10    Selection of the Random Flight

In a previous section of this chapter we saw that it is possible to calculate the probability that an electron in a state $\bar{k}$ collides with a scattering centers and this probability can be written a function $W(\bar{k})$. Without getting in all details, let us report an important result that is widely used in every Monte Carlo simulator.

The probability, per unit time, that an electron in free flight scatters is equal to:

$$P(\tau) = W_T(E_k)exp\left[-\int_0^\tau W_T(E_k)dt\right] \tag{1.29}$$

where $W_T(E_k)dt$ is the probability that an electron in the state $\bar{k}$ collides during the time $\tau$ and $W_T(E_k) = \sum_{j=1}^N W_j(E_k)$ is the total scattering rate (the sum runs over every possible scattering mechanism). The function in the integral sign is unfortunately impractical to generate stochastic free flights and a simplification of the formula must be used for practical cases.

To overcome such cumbersome formula, a fictitious *self-scattering* process is introduced. If a particle suffers such a scattering, basically nothing changes and the particle keep its initial pseudo-wave vector $\bar{k}$. We can, then, introduce the following constant:

$$\Gamma = \sum_{j=0}^N W_j(E_k)$$

where $W_0(E_k)$ is the probability that an electron suffers a self-scattering. It is possible to show that, by doing this, the probability (1.29) now reads much more simple, i.e.

$$P(\tau) = \Gamma e^{-\Gamma\tau} \tag{1.30}$$

It is now possible to obtain, from (1.30), the free flight time of a particle by generating a random number $r_1$ between 0 and 1, and using the following formula

$$\tau = -\frac{ln(r_1)}{\Gamma} \tag{1.31}$$

This is the formula implemented in every Monte Carlo simulator that aims to be fast and reliable while calculating the free flight time of an electron.

Random numbers can be used very simply to generate stochastic free flights. The computer time used for self-scattering is more than compensated for by the use of formula (1.31). Finally, it interesting to notice that, to enhance the speed of free flight time calculation, several schemes such as the *Constant Technique*, and the *Piecewise Technique* have been developed to minimize the self-scattering events.

## 1.11   Selection of scattering mechanism and final electron state

Two questions still remain open at this point. How to select the scattering mechanism at the end of an electron free flight? And what should be the final electron pseudo-wave vector after the collision? This section show how to get the answers to these questions in a very simple way.

The selection of the scattering mechanism after a free flight, can be made easily introducing the following functions $\lambda_1(E_k), \lambda_2(E_k), \ldots, \lambda_N(E_k)$ defined as follow ($N$ being the number of scattering mechanisms relevant in the simulation):

$$\lambda_i(E_k) = \frac{\sum_{j=1}^{N} W_j(E_k)}{\Gamma}$$

If an electron has energy an $E_k$, the *n-th* scattering mechanism is chosen, after generating a random number $r_2$ between 0 and 1, if the following condition holds

$$\lambda_{n-1}(E_k) < r_2 \leq \lambda_n(E_k) \tag{1.32}$$

## 1.12   Conclusions

We saw that the main physical effects to take into account to describe the dynamics of an electron in a semiconductor crystal are:

- Energy band
- Scattering
- Applied Potential

# Chapter 2

# Archimedes source code

The main feature of Archimedes is a not fancy or sophisticated model (even if the models implemented in it are correct and reliable) but that it comes with the source code. Everyone is free to download it from Internet, to study, modify, run and distribute the sources, as long as the original license, GPL, is maintained **[1]**. This is why we want to talk about the code in this chapter.

We will give, here, a short description of Archimedes sources trying to discuss some meaningful detail and leaving out everything that is not needed to understand the data flow and organization of the code. This chapter should be read by every user that wants to really understand how Archimedes works internally, by users that wants to implement a Monte Carlo simulator, or by users that want to use Archimedes as a starting point to develop their own simulator.

To understand this chapter, some knowledge of C programming language is required. Furthermore, since this code is strongly based on what presented in Tomizawa's book **[3]**, it is strongly advised to read this reference before reading this chapter.

Of course, if you are not interested in understanding the source code of Archimedes, and you want to use it only as a black box tool, you can skip this chapter and read the next one.

## 2.1 A little help: Cscope

In the previous chapter, we saw how the Monte Carlo method works, what the used equations are and what are the principles in it. We have not seen yet how to implement it. To understand how things works in Archimedes, some code spelunking has to be done (spelunk is an english verb that means to go exploring, usually in a cave). Code Spelunking is a set of tools and techniques for working with and comprehending large code bases, i.e. exactly what we need in this case.

A very good way to do code spelunking is to use a dedicated package for that, instead of using Unix commands like *grep*, *find*, etc (which are useful but makes your life really difficult if you have to study/understand a big code like Archimedes). There is plenty of packages around, but the best one is certainly Cscope **[2]**. Cscope is a package originally developed at Bell Labs (back in the day of PDP-11!!) and, so, has a perfect Unix pedigree. The package is released under BSD license (i.e. it is Free Software).

Cscope allows users to go through the code of a program in a very practical and simple way, making people in the position of effectively study the code. As the man page of the packages says, C*scope is an interactive, screen-oriented tool that allows the user to browse through C source files for specified elements of code.*

Some useful features of Cscope are (from website):

- Allows searching code for:
  - all references to a symbol
  - global definitions
  - functions called by a function
  - functions calling a function
  - text string
  - regular expression pattern
  - a file
  - files including a file
- Curses based (text screen)
- An information database is generated for faster searches and later reference
- The fuzzy parser supports C, but is flexible enough to be useful for C++ and Java, and for use as a generalized 'grep database' (use it to browse large text documents!)
- Has a command line mode for inclusion in scripts or as a backend to a GUI/frontend
- Runs on all flavors of Unix, plus most monopoly-controlled operating systems.

How to use Cscope to study Archimedes?

First thing to do is to go in the *src* directory:

```
# cd src
```

Then call Cscope (we suppose Cscope is already installed on your machine):

```
# cscope -R
```

(the –R option means that Cscope will get recursively into all sub directories). You should obtain something similar to figure 1. Now you can start to explore the sources in an effective way. For example, try a couple of searches (use the arrow keys to move around between search types, and 'tab' to switch between the search types and your search results). Hit the number at the far left of a search result, and Cscope will open Vim right to that location. Very useful indeed!

Also, try to put the cursor over a C symbol that is used in several places in your program. Type "CTRL-\ s" (Control-backslash, then just 's') in quick succession, and you should see a menu at the bottom of your Vim window showing you all the uses of the symbol in the program. Select one of them and hit enter, and you'll jump to that use.

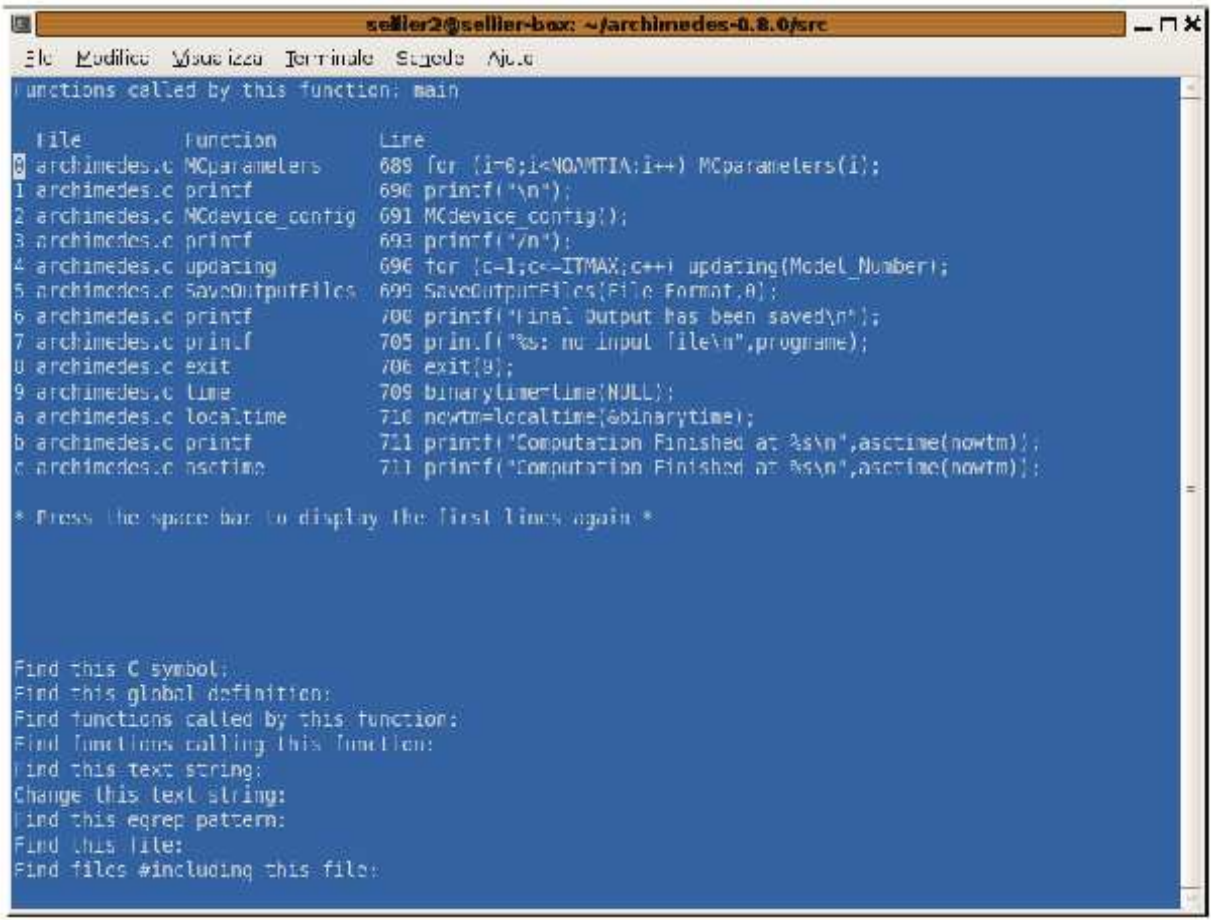

Figure 8 - Cscope applied to Archimedes

## 2.2 Archimedes sources

Archimedes source code consists of several files (the exact number depending on the release you download) with *c* and *h* extensions. All files can be found in the same directory *src* of your Archimedes repository. In the present release (2.0.0) the files are the ones reported in figure 2. The files are written in C, with no special C++ syntax in it. The code has been written in C to avoid any problem developers used to have with C++ compilers available back to the beginning of Archimedes (2004). At that time, C++ compilers tended to be slower and less reliable that the C ones. Today Archimedes is still developed in C for historical reasons but also to obtain a code that is as faster as possible (we want to avoid any run-time complications that could arise from the use of C++, we are doing numerical calculations after all..).

```
AlAs.h                archimedes.c                         mesher.h
AlP.h                 charge.h                             mm.h
AlSb.h                computecurrents.h                    mm2.h
GaAs.h                constants.h                          particlecreation.h
GaP.h                 deviceconfig.h                       poissonbcs.h
GaSb.h                diode_density_gradient_potential.input  quantumeffectivepotential.h
Germanium.h           drift.h                              random.h
HMEPbcs.h             electric_field.h                     readinputfile.h
Hole_bcs.h            electron_relaxation.h                saveoutput2dgnuplot.h
InAs.h                ensemblemontecarlo.h                 saveoutput2dholegnuplot.h
InP.h                 extrema.h                            saveoutput2dholemeshformat.h
InSb.h                faraday.h                            saveoutput2dmeshformat.h
MEP_interpolation.h   faradaybcs.h                         saveoutputfiles.h
Makefile.am           hole_relaxation.h                    scattering.h
Makefile.in           holemep2d.h                          sign.h
ParabMEP2D.h          mcparameters.h                       updating.h
Silicon.h             media.h
```

The main() of the code is in *archimedes.c.* This is the first file that we will analyze to understand the data flow in Archimedes. Concerning the other files, we give a description of what they do but the user should have a careful look to the sources to understand them. They should read the comments in the sources and use Cscope.

## 2.2.1 archimedes.c

After some comments on the Copyright and License of the code, and after including all standard headers required, some preprocessor definitions are done. The code is reported below.

```
// Preprocessor Definitions
#define real double
#define ON 1
#define OFF 0
#define KANE 0
#define PARABOLIC 1
#define FULL 2
#define QEP_BOHM 0
#define QEP_CALIBRATED_BOHM 1
#define QEP_FULL 2
#define QEP_DENSITY_GRADIENT 3
#define MN3 4
#define NXM 308
#define NYM 308
#define DIME 1003
#define ITMAX 10000000
#define POISSONITMAX 1500
```

```
#define SMALL 1.e-5
#define VMAX 1000000
#define NPMAX 10000000 // maximum number of super-particles
#define MCE 0  // MCE stands for MC for electrons only
#define MCH 1  // MCH stands for MC for holes only
#define MCEH 2 // MCEH stands for MC both for electrons and holes
#define MEPE 3 // MEPE stands for MEP model for electrons only
#define MEPH 4 // MEPH stands for MEP model for holes only
#define MEPEH 5 // MEPEH stands for MEP model for electrons and holes
#define GNUPLOTFORMAT 0 // output file in GNUPLOT format
#define MESHFORMAT 1 // output file in Mesh format
```

It is very useful to describe, at least, some of those definitions especially for those that want to modify the code.

First of all:

```
#define real double
```
This line defines the *real* type which is an internal type of Archimedes (it is not a C type obviously). This line is useful when precision needs to be modified. Every numeric variable and array in Archimedes is defined as a real. In our case, we say that a real is equivalent to a *double*. So, every time the compiler will find a real declaration it will actually perform a double declaration. Suppose now, that for some reason, we need to declare all variables and arrays as *float*. No need to go through the whole code and modify all declarations. The only we need to do is to modify the line above as:

```
#define real float
```
and then recompile. No more that that!

Two other important definitions are the following ones:

```
#define NXM 308
#define NYM 308
```
Basically, every array defined on a spatial grid will have those dimensions, which means that any spatial mesh cannot have more than NXM cells in the x-direction and NYM cells in the y-direction. If, for any reason, these limits are too small or too big, they can be modified. The code, obviously, needs to be recompiled.

In Monte Carlo simulations, a grid has to be defined for the energy. The maximum number of points for such energy mesh is defined by the following line:

```
#define DIME 1003
```
If, for any reason, the energy mesh has to be modified the code needs to be compiled again.

```
#define POISSONITMAX 1500
```

This line defines the number of iterations that the code has to do for the resolution of the (modified pseudo-time-dependent) Poisson equation.

Finally, a VERY important definition is reported below:

```
#define NPMAX 10000000 // maximum number of super-particles
```

This represents the maximum number of particles (pseudo-electrons) that can be accounted for during a simulation. During a Monte Carlo simulation, the number of particles in the device varies, due to the fact the some particles enter the contacts and some other leave. NPMAX is the maximum number of particles that can be contemporary simulated. Usually, 500,000 particles is enough for a standard device simulation but if, for any reason, NPMAX is too much limiting it can be modified. Be aware that if NPMAX is increased then the memory amount needed for a simulation will increase as well. One must be very cautious when modifying this integer. Obviously, once modified, the code needs to be re-compiled.

After preprocessor declarations are done, some definitions of global variables are done and inclusions of all functions/routines is implemented. This is done in the (partially) reported code below:

```
#include "mesher.h"
#include "poissonbcs.h"
#include "faradaybcs.h"
#include "media.h"
...
...
#include "hole_relaxation.h"
#include "updating.h"
#include "readinputfile.h"
//#include "SaveRappture.h"
```

If the user wants to add a new function to Archimedes, the first thing to do is to implement the code as a function and put it in a separate file. The file, then, needs to be included into Archimedes, and this is done by including it in this part of the code.

Then the main**(int** argc,**char\*** argv**[])** is started. The first part of the main() basically deal with the interpretation/check of the command line arguments. This part checks if the number of arguments is correct and decide what to do according to the given arguments. For example, if one launches Archimedes like this:

```
# archimedes –help
```

the output will be something like this:

```
GNU archimedes, a simulator for submicron and nanoscale semiconductor devices.
Copyright (C) 2004-2011 Jean Michel D. Sellier.
```

Usage: archimedes [OPTION] file...

 -h, --help        display this help and exit
 -v, --version     display version information and exit

Report bugs to jeanmichel.sellier@gmail.com or jsellier@purdue.edu

while if Archimedes is launched this way:

# archimedes input_file.inp

it will trigger a simulation run. The behavior is very different.

Once, the arguments are parsed, and the user wants to run a simulation, the semiconductor material parameters are defined (some of them calculated). Then the input file parser is invoked:

```
// Read the geometrical and physical description of the MESFET
// ========================================================
     Read_Input_File();
// ========================================================
```

This routine is the parser of the input file. Since this routine is very important, we report the details in the next chapter (where we discuss the scripting language to describe semiconductor devices) and we go ahead.

After parsing the input file, some further calculations are carried out on the material parameters (in particular the ones that depends on user provided values like, stechiometric concentration, lattice temperature, etc.).

At this point, the simulation starts. First, we save the initial time in a variable (that will be useful to understand how long a simulation takes) and we print this time on the screen. This is done in the following lines:

```
     binarytime=time(NULL);
     nowtm=localtime(&binarytime);

     printf("\n\nComputation Started at %s\n",asctime(nowtm));
```

The boundary conditions specified by the user in the input script file are then set for the Poisson and Faraday equations (respectively the equations for electrostatic and magnetic fields):

```
     PoissonBCs();
     if(FARADAYFLAG) FaradayBCs();
```

Some further settings are required for the Monte Carlo simulation and are done in the following lines:

```
// Initialisation for Monte Carlo
// ==============================
    if(Model_Number==MCE || Model_Number==MCEH){
     int i;
     for(i=0;i<NOAMTIA;i++) MCparameters(i);
     printf("\n");
     MCdevice_config();
    }
    printf("\n");
```

then the simulation starts according to the selected transport model in the line:

```
    for(c=1;c<=ITMAX;c++) updating(Model_Number);
```

Let us, now, in the following give a description of the other files of Archimedes. A short description is given and the user is strongly invited to have a look to the comments in the files and study the code.

### 2.2.5 charge.h

This part calculates the total electron charge (dennsity) in the device once the electron positions have been updated. The algorithm in use here is the well-known *Cloud-in-Cell* method that reduces the spurious oscillations in the electron density due to the random nature of the Monte Carlo method.

### 2.2.6 computecurrents.h

The routine implemented in this file calculates the electron current on the user-defined contacts of the device and print them out on the screen. This routine is invoked only at the end of the simulation, since it has no sense to call it during the simulation of the transient solution.

### 2.2.7 constants.h

In this file, we define all universal physics constants needed for the simulations. In this file, variables like the Boltzmann constant, the reduced Planck constant, the electron charge, the free electron mass, etc, are defined.

## 2.2.8 deviceconfig.h

The routine in this file is called at the beginning of the simulation. It basically set the initial position and pseudo-wave vector of every particle of the simulation, according to the user choice. In particular, the position of the particles will depend on the user specified doping distribution and the pseudo-wave vector depend on the lattice temperature of the device.

## 2.2.9 drift.h

This part is invoked when the position and the pseudo-wave vector is invoked. Given the initial position and pseudo-wave vector of a particle, this routine updates these vectors according to the electrostatic potential and the energy band profile. No scattering or quantum mechanisms are taken into account here.

## 2.2.10 electricfield.h

This part of the code is invoked at every time step of the simulation and updates the electrostatic potential according to the previously calculated electron density (see charge.h). The potential takes also into account the boundary conditions specified by the user.

In this routine the electrostatic potential is calculated by means of pseudo-time dependent Poisson equation which gives the same solution of the classical Poisson equation but at a less demanding memory price.

## 2.2.11 mcparameters.h

This file defines the routine `MCparameters(material)` which input parameter is the name of a semiconductor material. The material can be anything defined in the first part of the file archimedes.c (in the preprocessor definitions) i.e. SILICON, GAAS, GERMANIUM, INSB, ALSB, ALXINXSB, ALXIN1XSB, ALAS, ALP, GAP, GASB, INAS, INP, INXGA1XAS, INXAL1XAS, and INXGAXXAS (which are basically integers).This routine basically defines the material dependent values for the various scattering mechanisms.

To speed up a simulation, Monte Carlo simulators use to calculate several lookup tables before the actual simulation runs. This is done in this file. It is easy to see, from the sources, that the tables calculated in this part of the code refers to the various scattering mechanisms selected by the user.

For example, the code calculates the lookup tables in case the selected material has two valleys and the optical phonons are taken into account.

These calculations, obviously, are done once for all at the beginning of the simulation. There is no need to repeat them once they are done.

### 2.2.12 media.h

This part of the code deals with the calculation of the observables of the simulation, electron density, x and y-velocities, energy, etc.. This part basically averages those variables over several time steps to smooth out the spurious oscillations that can occur due to the randomness of the Monte Carlo method.

### 2.2.13 particlecreation.h

This part of the code is invoked every time a new particle needs to be created. This usually happens during a simulation close to the contacts. The particles are created in such a way that the total charge on the contacts is always neutral. The number of particles to be created in the contacts depends on the number of particles lost through the contact and the doping density in the proximity of the contacts.

### 2.2.14 quantumeffectivepotential.h

This part is invoked at every time step if the user has imposed the use of any particular quantum effective potential model. A model is specified by the user and this routine calculates only the variables related to the model.

### 2.2.15 random.h

This file contains a very short in line function that generates a random number. The first random number is calculated according to a specific seed that can be chosen by the user. If the seed is modified the code has to be re-compiled.

**2.2.16 readinputfile.h**

This file contains the parser for scripts describing the geometry, doping, contacts and applied potentials of the device. For more details, have a look to the source code, since it is easy to understand it and well commented.

**2.2.17 saveoutputfiles.h**

This part of the code is invoked at the end of a simulation or at the end of every time step during the simulation, depending on the user choice. It saves the various observables calculated during the simulation, i.e. density, potential, velocity, energy, etc. The files can be saved in different formats that are selected by the user (for example GNUPlot format).

**2.2.18 scattering.h**

This part simulates the scattering on a particle, which outcome depends on the starting pseudo-wave vector of an electron and on the particular scattering mechanisms the user has selected. The scattering mechanisms that can be chosen are the most common ones, i.e. acoustic, optical and impurities scattering.

**2.2.19 updating.h**

This part of the code is called at every time step and updates the position and pseudo-wave vector of all particles simulated. The updating process depends, obviously, on the methods and models chosen by the user.

**2.3 How to install Archimedes**

Here we give the instructions to compile Archimedes on a local machine. Since the code has been developed in C, and it does not depend on any external library, it can be easily compiled on any kind of Unix machine, even on Windows. We report here the very straightforward steps to compile it.

**2.3.1 The standard way**

The simplest way to compile this package is to follow the instructions below (that are reported in the INSTALL file in Archimedes repository).

1. Go to the directory containing the package's source code
   `# cd archimedes`
    and type
   `# ./configure`
   to configure the package for your system.
2. Type
   `# make`
   to compile the package
3. Optionally, type
   `# make check`
   to run any self-tests that come with the package.
4. Type
   `# make install`
   to install the programs and any data files and documentation.
5. Eventually type
   `# make clean`
   to remove the program binaries and object files. To also remove the files that configure created, type
   `# make distclean`

## 2.3.2 The non-standard way

If, for some reason, the procedure above does not work, the following should still work:

`# cd archimedes/src`

`# gcc -lm archimedes.c -o archimedes`

This will create a binary file whose name is `archimedes`. To run it, type:

`# ./archimedes filename`

Obviously, gcc must be installed in your system.

## References

**[1] www.gnu.org/software/archimedes**

**[2] http://cscope.sourceforge.net/**

[3] Tomizawa

# Chapter 3

# Archimedes Scripts Syntax

Archimedes has a very convenient way to describe a device that has to be simulated. The user can define a device by simply describing it in a script. The script of course cannot be a human description of the device, and it has to follow a proper syntax to be parsed and understood by Archimedes. In this chapter, we report the commands, and the related syntax, to write a script that can be interpreted by Archimedes.

The fact that the user has to write down a script to describe a device could be, at a first glance, considered as an obstacle but, we will see in the following, there is nothing to be scared of. The commands are very intuitive and easy to remember and the syntax is very natural and easy to remember.

## 5.1 An example to start

The best way to understand how to define a device is by first studying an example. The example we will study here is a MESFET, a silicon device well-known by engineers. The MESFET we want to define and simulate is a structure like the one reported below (Figure 1).

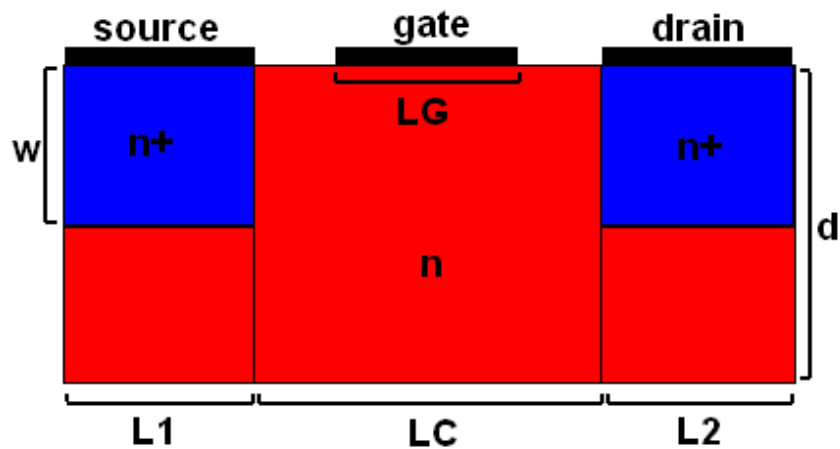

Figure 10 - A Silicon MESFET

A MESFET has three contacts, called the source, the gate and the drain. Two of them, the source and drain, are ohmic contacts, while the gate is a Schottky contact. This device has also two $n^+$ zones and one $n$ zone. The dimensions of the MESFET we want to simulate are $d = 0.2\,\mu m$, $L_1 = 0.1\,\mu m$, $L_C = 0.4\,\mu m$, $L_2 = 0.1\,\mu m$, $L_G = 0.2\,\mu m$ and $W = 0.05\,\mu m$. The densities are $n^+ = 3 \times 10^{23}/m^3$ and $n = 10^{23}/m^3$.

Finally we want to simulate electrons transport using the Monte Carlo method.

In the following, we report a script that set everything we need to simulate electron transport in a MESFET. We will, then, give some details of this script to have a rough idea about how a script works.

```
# Silicon MESFET test-1

MATERIAL SILICON
TRANSPORT MC ELECTRONS

FINALTIME 5.0e-12
TIMESTEP 0.0015e-12

XLENGTH 0.6e-6
YLENGTH 0.2e-6

XSPATIALSTEP 80
YSPATIALSTEP 40

# Definition of the doping concentration
# =============================
DONORDENSITY 0. 0. 0.6e-6 0.2e-6 1.e23
DONORDENSITY 0. 0.15e-6 0.1e-6 0.2e-6 3.e23
DONORDENSITY 0.5e-6 0.15e-6 0.6e-6 0.2e-6 3.e23
ACCEPTORDENSITY 0. 0. 0.6e-6 0.2e-6 1.e20

# Definition of the various contacts
# =========================
CONTACT DOWN 0.0 0.6e-6 INSULATOR 0.0
CONTACT LEFT 0.0 0.2e-6 INSULATOR 0.0
CONTACT RIGHT 0.0 0.2e-6 INSULATOR 0.0
CONTACT UP 0.1e-6 0.2e-6 INSULATOR 0.0
CONTACT UP 0.4e-6 0.5e-6 INSULATOR 0.0
CONTACT UP 0.0 0.1e-6 OHMIC 0.0 3.e23
CONTACT UP 0.2e-6 0.4e-6 SCHOTTKY -1.3
```

```
CONTACT UP 0.5e-6 0.6e-6 OHMIC 1.0 3.e23

NOQUANTUMEFFECTS
MAXIMINI

LATTICETEMPERATURE 300.

STATISTICALWEIGHT 250
MEDIA 500

OUTPUTFORMAT GNUPLOT
```

How to run this example:

> archimedes mesfet.inp

5.2 List of commands

In this section, we report the complete list of the commands available in the Archimedes scripting language. The commands are reported in alphabetical order. Every command has its own syntax so the user should read with attention the descriptions reported below before trying to implemented his/her own script. For every command, a description is also reported along with an example and an explanation of that example. For the sake of completeness, even the commands that were implemented in previous releases of Archimedes are reported even if their use is not encouraged.

Notice: Archimedes scripting language is case sensitive. This means that, for example, two words like *MATERIAL* and *Material* are not considered as equivalent in Archimedes. Every command in Archimedes is always in capitols. Finally when an option is reported in square brackets ([]), it means that this option is needed only in particular cases (reported in the description of the command).

- **ACCEPTORDENSITY**

    *Syntax*:
    **ACCEPTORDENSITY**     *XMIN    YMIN        XMAX    YMAX        DENSITY*

*Description*:
When defining a new device, the user has to specify the acceptor density. This is necessary in order to solve the Poisson equation. This equation needs both the donor and acceptor distribution to give realistic/reliable results. If no acceptor density is specified, Archimedes considers, as the default, that the acceptor density is constant on all the device and it is equal to the intrinsic acceptor density of the material. Furthermore, if the user specify the value of the acceptor density only on a part of the device, the remaining part is considered to be equal to the intrinsic acceptor density.

To specify the acceptor density, one has to specify the area and the density. The area is represented by a rectangle like $[x_{min}, x_{max}] \times [y_{min}, y_{max}]$ where $x_{min}$, $x_{max}$, $y_{min}$, $y_{max}$ are in meters. The density is in $1/m^3$.

*Example*:
ACCEPTORDENSITY    0.0    0.0    1.0e-6    0.1e-6    1.e20

*Meaning*:
The acceptor density on the area $[0,0] \times [1\ \mu m, 0.1\ \mu m]$ is equal to $10^{20}/m^3$.

- **CIMP**

    *Syntax*:
    **CIMP**        *DENSITY*

    *Description*:
    In materials like GaAs, the impurities can be scattering centers of strong nature. The background impurities density in a device is specified for the whole devices. The units are $1/m^3$.

    *Example*:
    CIMP     1.e23

    *Meaning*:
    The impurity density in the device is $10^{23}/m^3$.

- **COMMENTS (#)**

    *Syntax*:

**# comment**

*Description*:
As in every (scripting) language, comments are very important for the clarity of a code. Everything that follow a # symbol is considered as a comment.

*Example*:
```
# This is a comment
# TRANSPORT MC ELECTRONS
```

*Meaning*:
These lines are totally ignored by Archimedes, even in the case one of them contains a valid command. Everything that comes after a # is ignored by Archimedes.

- **CONTACT**

  *Syntax*:
  **CONTACT** *POSITION BEGINNING END TYPE [POTENTIAL] [DENSITY]*

  *Description*:
  This command is used to specify where the contacts/insulator-boundaries of the device are positioned. It is also used to specify which boundaries have to be considered as insulators. First the position of the contact has to be specified. The choices for the position are UP, DOWN, LEFT and RIGHT. Once the position of the contact is defined, the user has to specify where it starts and when it ends. This is done by specifying the values BEGINNING and END (in meters). TYPE is the type of contact the user wants to define. It can be of three types, INSULATOR, OHMIC, SCHOTTKY. Finally, an eventual applied potential is specified (in Volts) and an eventual applied density is specified in $1/m^3$ (for Ohmic contacts only).

  *Example*:
  ```
  CONTACT   UP     0.0    1.0e-6   INSULATOR   0.0
  CONTACT   UP     0.0    0.1e-6   OHMIC       0.0    1.e23
  CONTACT   LEFT   0.2e-6 0.4e-6   SCHOTTKY   -0.8
  ```

  *Meaning*:

The first line describes an insulator boundary on the top edge of the device. The edge starts at $x_{min} = 0.0$ and ends at $x_{max} = 1.0\,\mu m$. The applied voltage on this edge is equal to 0.0 Volts.

The second line describes an ohmic contact on the top edge of the device. The contact starts at $x_{min} = 0.0$ and ends at $x_{max} = 0.1\,\mu m$. The applied potential is equal to 0.0 Volts and the density is equal to $10^{23}/m^3$.

The last line describes a Schottky contact on the left edge of the device that starts at $x_{min} = 0.2\,\mu m$ and ends at $x_{max} = 0.4\,\mu m$. The applied voltage on this edge is equal to $-0.8$ Volts.

- **DONORDENSITY**

  *Syntax*:
  **DONORDENSITY** *XMIN    YMIN      XMAX    YMAX      DENSITY*

  *Description*:
  When defining a device, the user has to specify the donor density. This is necessary in order to solve the Poisson equation correctly. If no donor density is specified, Archimedes considers, as the default, that the donor density is constant on the device and it is equal to the intrinsic donor density of the material. If the user specify the value of the donor density only on a part of the device, the remaining part is considered to be equal to the intrinsic density.

  To specify the donor density, one has to specify the area and the density. The area is represented by a rectangle like $[x_{min}, x_{max}] \times [y_{min}, y_{max}]$ where $x_{min}$, $x_{max}$, $y_{min}$, $y_{max}$ are in meters. The density is in $1/m^3$.

  *Example*:
  ```
  DONORDENSITY   0.0   0.0     1.0e-6   0.1e-6    1.e23
  ```

  *Meaning*:
  The donor density on the area $[0,0] \times [1\,\mu m, 0.1\,\mu m]$ is equal to $10^{23}/m^3$.

- **LEID**

  *Syntax*:
  **LEID**

*Description*:

LEID stands for *Load Electrons Initial Data*. Basically, if the user has a set of results that come from another simulation and wants to use this set as the starting point of a Monte Carlo simulation, this is the command to use. The initial set has to be provided as a set of three files providing the electron density, the electron energy and the potential, and their names have to be respectively:

- *density_start.xyz*
- *energy_start.xyz*
- *potential_start.xyz*

Notice: The number of cells in x and y-directions must be exactly the same as the ones specified in the Monte Carlo simulations.

*Example*:
LEID

*Meaning*:
Load the initial data for the Monte Carlo electrons transport simulation.

- **MATERIAL**

*Syntax*:
**MATERIAL    X**  *XMIN    XMAX*     **Y**   *YMIN    YMAX     ELEMENT   [X]*

*Description*:

This command specify what semiconductor material an area of the device is made of. The area is described by a rectangle $[x_{min}, x_{max}] \times [y_{min}, y_{max}]$ (in meters). The element is a semiconductor element selected among the following list:

- SILICON
- GERMANIUM
- GAAS
- INSB
- ALSB
- ALAS
- ALP

- GAP
- GASB
- INAS
- INP
- ALxINxSB
- ALxIN1-xSB
- INxGA1-xAS
- INxAL1-xAS
- INxGAxAS2
- ALGAAS

For some materials, like , a chemical parameter X needs to be specified. This is done by putting a number between 0 and 1 right after the material chemical name. ALGGAS is the only exception to the rule since no X has to be specified (in this case X is considered to be equal to 0.3).

*Example*:
```
MATERIAL  X 0.0  1.0e-6   Y 0.0  0.1e-6  SILICON
MATERIAL  X 0.0  0.1e-6   Y 0.0  1.0e-6  INxGA1-xAS   0.3
```

*Meaning*:
The first line describes a device that is made of Silicon in the area $[0.0, 1.0 \, \mu m] \times [0.0, 0.1 \, \mu m]$. The second line describes a device that is made of $In_{0.3}Ga_{0.7}As$ in the area $[0.0, 0.1 \, \mu m] \times [0.0, 1.0 \, \mu m]$.

- **TRANSPORT**

   *Syntax*:
   **TRANSPORT**          *TRANSPORT_MODEL*          *CHARGE(S)_TYPE*

   *Description*:
   This command specify the model and the charge to be simulated. The Transport model can be one of the following:

   - MC
   - MEP

   MC stands for Monte Carlo transport, while MEP stands for Maximum Entropy Principle which is an hydrodynamical model (very fast if compared to Monte

Carlo, but quite obsolete now). The MEP model is usually used to generate files for the LEID command (see LEID command for more details) and should not be used for actual simulations (in this version of Archimedes, it could fail for some devices).

Concerning the charges type, the possible choices are:

- ELECTRONS
- HOLES
- BIPOLAR

which have a straightforward meaning.

*Example*:
```
TRANSPORT    MC    ELECTRONS
```

*Meaning*:
This line means that the simulation will be a Monte Carlo method applied to electron charges.

• **FINALTIME**

*Syntax*:
**FINALTIME**      *VALUE*

*Description*:
This command set the final time of the simulation. The unit is the second.

*Example*:
```
FINALTIME    5.0e-12
```

*Meaning*:
The final time of the simulation is set to $5.0 \, picosec$.

• **TAUW**

*Syntax*:
**TAUW**         *VALUE*

*Description*:
This command should be used when the MEP model is invoked otherwise it has no effect. This command set the value of the energy relaxation time (in seconds) of the MEP model.

*Example*:
TAUW    1.0e-12

*Meaning*:
The energy relaxation time is set to $1^{-12}$ seconds.

- **TIMESTEP**

  *Syntax*:
  **TIMESTEP**        *VALUE*

  *Description*:
  This command set the time step of the simulation, whatever the transport model is. The units are the second.

  *Example*:
  TIMESTEP    0.15e-14

  *Meaning*:
  The time step of the simulation is set to $0.15 \times 10^{-14}\ second$.

- **XLENGTH**

  *Syntax*:
  **XLENGTH**        *VALUE*

  *Description*:
  This command specify the total length of the device in the X direction.

  *Example*:
  XLENGTH     100.0e-9

*Meaning*:
This line specify that the total length of the device in the x-direction is equal to $100.0\ nm$.

- **YLENGTH**

  *Syntax*:
  **YLENGTH**       *VALUE*

  *Description*:
  Like XLENGTH but in Y-direction. See XLENGTH for more info.

  *Example*:
  YLENGTH     10.0e-9

  *Meaning*:
  This line specify that the total length of the device in the y-direction is equal to $10.0\ nm$.

- **XSPATIALSTEP**

  *Syntax*:
  **XSPATIALSTEP**       *VALUE*

  *Description*:
  This command specify the number of cells in the x-direction. Any simulated device is represented by a grid that contains a finite number of cells. The value specified must be an integer.

  *Example*:
  XSPATIALSTEP     128

  *Meaning*:
  This line specify the number of cells in the x-direction to be equal to $128$.

- **YSPATIALSTEP**

*Syntax*:
**YSPATIALSTEP**        *VALUE*

*Description*:
Like XSPATIALSTEP, but in y-direction. See XSPATIALSTEP for more details.

*Example*:
YSPATIALSTEP     64

*Meaning*:
This line specify the number of cells in the y-direction equal to 64.

- **QUANTUMEFFECTS (deprecated)**

    *Syntax*:
    **QUANTUMEFFECTS**

    *Description*:
    This command is obsolete/deprecated and is reported only for back compatibility. It should not be used. Use QEP_MODEL and QEP_PARAMETERS instead. When invoked, it includes the Bohm quantum effective potential in the transport calculations to take into account some quantum effects due to the finite size of the electrons.

    *Example*:
    QUANTUMEFFECTS

    *Meaning*:
    This line includes the Bohm quantum effective potential in the transport simulation (deprecated).

- **NOQUANTUMEFFECTS (deprecated)**

    *Syntax*:
    **NOQUANTUMEFFECTS**

    *Description*:

This command is obsolete/deprecated and is reported only for back compatibility. It should not be used. This command turn off any quantum effective potential model in the simulation.

*Example*:
NOQUANTUMEFFECTS

*Meaning*:
Turn off any quantum effective potential in the simulation.

- **MAXIMINI**

  *Syntax*:
  **MAXIMINI**

  *Description*:
  This command turn on the verbosity of Archimedes. Sometimes, it can be very useful to report the maxima and minima of time-dependent variables during a simulation. By invoking this command, Archimedes gives a mini report of the minima and maxima of several variables. This report, usually, look like the following one (extracted from an actual simulation):

  Max. Potential = 1.0
  Min. Potential = 0
  Max. x-elec.field = 2.62846e+06
  Min. x-elec.field = -3.57578e+06
  Max. y-elec.field = -0.35495e+03
  Min. y-elec.field = -0.58273e+03
  Max. Density = 5.14954e+23
  Min. Density = 1.33122e+21

  which meaning is pretty clear.

  *Example*:
  MAXIMINI

  *Meaning*:
  Turn the verbosity of Archimedes to on and print a small report on screen at every time step.

- **QEP_MODEL**

  *Syntax*:
  **QEP_MODEL**        MODEL

  *Description*:
  The user can use this command when a quantum effective model has to be used in a simulation. MODEL is the model that will be used for the simulation, and can be chosen from the following list (list of quantum effective potentials implemented in Archimedes):

  - BOHM
  - CALIBRATED_BOHM
  - FULL
  - DENSITY_GRADIENT

  Every model implements a different set of equations to include quantum effects due to the finite size of electrons in a device. For more details on these models, please read Chapter 1.

  *Example*:
  QEP_MODEL     BOHM

  *Meaning*:
  The simulation will include the Quantum Effective Bohm potential in the electron transport calculations.

- **QEP_PARAMETERS**

  *Syntax*:
  **QEP_PARAMETERS**     ALPHA    [BETA]

  *Description*:

  *Example*:
  QEP_PARAMETERS     0.5    1.0

  *Meaning*:

- **NOMAXIMINI**

  *Syntax*:
  **NOMAXIMINI**

  *Description*:
  Turns the verbosity of Archimedes to off. For more details, see MAXIMINI

  *Example*:
  NOMAXIMINI

  *Meaning*:
  See the description.

- **SAVEEACHSTEP**

  *Syntax*:

  *Description*:

  *Example*:

  *Meaning*:

- **LATTICETEMPERATURE**

  *Syntax*:
  **LATTICETEMPERATURE**          KELVIN

  *Description*:
  Using this command, the user set the lattice temperature of the device to be simulated. The temperature is expressed in Kelvin and has to be a non-negative value.

  *Example*:
  LATTICETEMPERATURE     77
  LATTICETEMPERATURE     300

*Meaning*:
The first line set the temperature to be 77 Kelvin (liquid hydrogen) while the second line set the temperature of the device to be 300 Kelvin (room temperature).

- **STATISTICALWEIGHT**

  *Syntax*:
  **STATISTICALWEIGHT**     INTEGER

  *Description*:
  This command fixes the (initial) number of super-particles in a cell that contains the maximum doping density of the device. The cells that have a lower doping density are filled consequently.

  *Example*:
  STATISTICALWEIGHT   100

  *Meaning*:
  The cells that have the maximum doping will have 100 super-particles.

- **MEDIA**

  *Syntax*:
  **MEDIA**     INTEGER

  *Description*:
  This command fixes the (integer) number of final steps over which the observables are averaged in order to smooth out the natural spurious oscillations intrinsic to the Monte Carlo method.

  *Example*:
  MEDIA   100

  *Meaning*:
  The observables will be averaged over the 100 last final steps of the simulation.

- **OUTPUTFORMAT**

  *Syntax*:
  **OUTPUTFORMAT**        FORMAT

  *Description*:
  The user uses this command to fix the output format of the observables at the end (or during) the simulation. The formats that can be selected are:
  -   GNUPLOT
  -   MESH

  *Example*:
  OUTPUTFORMAT   GNUPLOT

  *Meaning*:
  The output format is compatible with GNUPlot.

## 5.3 A further example

Let us report, in the following, one further example of device to be simulated, along with a short description of the device. This is useful to understand a little bit better the syntax of an Archimedes script.

The device reported here is a simple diode, which is good benchmark for simulations, and which geometry is easy to understand. The structure we want to simulate is the one reported in figure 2.

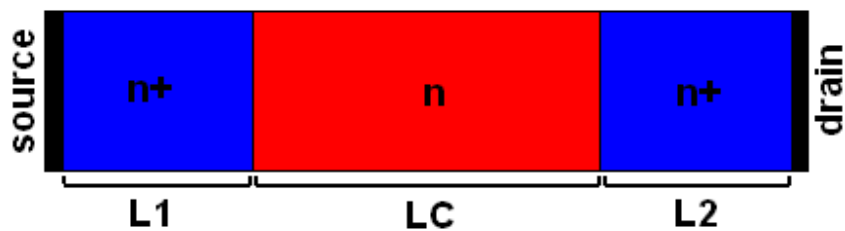

Figure 11 - A Silicon Diode

It is easy to understand the structure and the doping of the device by the figure itself. The diode is a *n+-n-n+* structure, with two ohmic contacts, the source and the drain. A voltage is applied on the drain. Let us report now the script to simulate this device.

```
TRANSPORT MC ELECTRONS

FINALTIME 5.5e-12
TIMESTEP 0.0015e-12

XLENGTH 1.0e-6
YLENGTH 0.1e-6

XSPATIALSTEP 100
YSPATIALSTEP 25

# Phonons scattering
ACOUSTICSCATTERING ON
OPTICALSCATTERING ON
IMPURITYSCATTERING ON

# Parabolic conduction band
CONDUCTIONBAND FULL

# definition of the material (all the device is made of Silicon)
MATERIAL X 0.0 1.0e-6   Y 0.0 0.1e-6  SILICON

# Definition of the doping concentration
# ======================================
DONORDENSITY   0.    0.   1.0e-6   0.1e-6   2.e21
DONORDENSITY   0.    0.   0.3e-6   0.1e-6   5.e23
DONORDENSITY   0.7e-6  0.   1.0e-6   0.1e-6   5.e23
ACCEPTORDENSITY 0.    0.   1.0e-6   0.1e-6   1.e20

# Definition of the various contacts
# ==================================
CONTACT LEFT  0.0   0.1e-6 OHMIC    0.0   5.e23
CONTACT RIGHT 0.0   0.1e-6 OHMIC    2.5   5.e23
CONTACT UP    0.0   1.0e-6 INSULATOR 0.0
CONTACT DOWN  0.0   1.0e-6 INSULATOR 0.0
```

```
NOQUANTUMEFFECTS
MAXIMINI
# SAVEEACHSTEP

LATTICETEMPERATURE 300.

STATISTICALWEIGHT 1500

MEDIA 500
OUTPUTFORMAT GNUPLOT
```

# Chapter 4

# The Graphical User Interface,

# A simpler way to use Archimedes.

A graphical user interface (GUI) for Archimedes has been implemented for those users that do not want to know anything about the scripting language Archimedes can parse and interpret, but still want to run simulations. This GUI has been created using the free tool Rappture, a rapid application infrastructure. The GUI can be run locally (but needs to be installed first) or on-line in a browser (and no installation is needed in this case). In this chapter we will see how to install this GUI on a local machine and how to reach it on-line when a non-installation situation is required, as in a class teaching Monte Carlo simulations, for example. Then we will describe how to use this GUI to simulate simple devices as diodes and MESFETs but also most sofisticated devices.

## 4.1 A short introduction to Rappture

Before we say anything about the GUI, we want to spend a few words about Rappture, that is the infrastructure that has been used to develop it. Rappture is a free software and can be found at the following link:

https://nanohub.org/infrastructure/rappture

As the author of this tool says, "*Rappture is a toolkit for **R**apid **App**lication Infrastruc**ture**, making it quick and easy to develop powerful scientific applications. Rappture has bindings with C/C++, FORTRAN, Matlab, Octave, Perl, Python and Tcl applications. It combines numerical building blocks, such as Poisson equation solvers and iterative matrix solvers, along with a powerful infrastructure for handling user interfaces. Once you describe the input/output for your simulator, Rappture handles the rest, generating a graphical interface automatically based on your description. The resulting application is easy to deploy on the* nanoHUB*, so a large community of users can access it through their web browser.*"

Rappture is a very powerful toolkit that allow developers to implement GUIs just like that. Archimedes GUI strongly relies on Rappture. We report in the following an extract of the introduction given on Rappture website.

### 4.1.1 How does Rappture work?

Instead of inventing your own input/output, you declare the parameters associated with your tool by describing Rappture objects in the Extensible Markup Language (XML). For example, the figure below shows the XML description of a simple plotting tool.

The <input> section contains three elements: a <string> representing the mathematical formula that the user will enter, a <number> for the x-axis minimum, and another <number> for the x-axis maximum. The <output> section for this tool contains a <curve> plotting the value of the function versus x. This is the complete set of parameters for this simple tool. The are many other types of Rappture elements which can be used to describe other input/output data, including simple elements such as <boolean> and <choice>, and more complex elements, such as <structure>, <mesh>, <field>, and <molecule>. For details, see Rappture XML Elements.

Rappture reads the XML description for a tool and generates the GUI automatically. The XML description shown on the left-hand side of the figure above produces the screen shot on the right-hand side. Note that the three input parameters appear on the left side of the screen, and the output curve appears on the right. The user can enter a formula along with the x-axis min/max values, and press the Simulate button to see the graph on the right-hand side. Rappture has a fairly sophisticated analysis environment for viewing results. The user can enter additional formulas and compare the results, or plot them all on the same graph, as shown in the figure.

Describing the inputs and outputs is the first half of the development process. The second half is writing the code within a simulator to access these elements. Rappture has bindings for a variety of programming languages, including C/C++, Fortran, Python, and a variety of other languages. So you can use the Rappture Application Programming Interface (API) naturally within your favorite programming environment.

To continue our example, here is the Python code needed to implement the simple graphical calculator tool shown above:

```
import Rappture
import sys
from math import *

io = Rappture.library(sys.argv[1])

xmin = float(io.get('input.number(min).current'))      ⎫
xmax = float(io.get('input.number(max).current'))      ⎬  instead of read()
formula = io.get('input.string(formula).current')      ⎭
npts = 100

for i in range(npts):
    x = (xmax-xmin)/npts * i + xmin;
    y = eval(formula)
    io.put('output.curve.component.xy', '%g %g\n' % (x,y), append=1)   ⎤ instead of write()
                                                                        ⎦
Rappture.result(io)
```

The lines in bold face emphasize the code needed for Rappture; the rest of the code is needed for the core simulator, regardless of how the input/output is handled. A typical simulator might read values from standard input, compute the results, and write them to standard output. Instead, a Rappture simulator gets inputs from the <input> elements in its tool description, and puts outputs into the <output> elements. In Python, this is accomplished as follows. The import statement loads the Rappture package into the Python interpreter, providing access to elements in the API, such as Rappture.library and Rappture.result. The Rappture.library call loads what we refer to as the driver file, which is normally passed as the first argument to the simulator program. The driver file is the same as the XML tool description shown above, but with <current> values for each of the input elements. The io.get() command gets the current value for each of the input parameters. Each parameter is uniquely identified by a simple path syntax. The path input.number(min).current means "find the <input> tag, then the <number id="min"> within it, then the <current> tag within that."

Once our simulator has all of its inputs, it begins evaluating the formula for each value of x. The io.put() command stores each computed (x,y) point in the output <curve> element; the append=1 flag indicates that each point should be appended to the previous results, instead of overwriting them. When the simulation is complete, the Rappture.result() call reports the results back to the Rappture GUI.

The Rappture GUI drives the whole interaction. There is one generic GUI program called rappture that can be used for all Rappture tools. This program reads the XML description for a tool and produces the interface automatically, on-the-fly. The user interacts with the GUI, entering values, and eventually presses the Simulate button. At that point, Rappture substitutes the current value for each input parameter into the XML description, and launches the simulator with this XML description as the "driver" file. The simulator reads the inputs, computes the outputs, and sends the results back to the Rappture GUI, as described above. The GUI then loads the results into the output analyzer for the user to explore.

It is easy to add new parameters to a simulator. First, update the XML description of the simulator; then, update the code for the simulator itself to access the new parameters. Rappture generates the GUI dynamically, each time you run a tool, based on the information available at that point. So as you make changes to a program, Rappture will detect the changes and adjust the GUI accordingly the very next time you run the tool.

The tool shown here is a fairly simple example. Rappture is capable of supporting real scientific simulators with much more complexity. This demo shows an educational tool with a little more complexity running within a web page on the nanoHUB. If you have a nanoHUB account, you can launch the tool yourself from this page. This tool simulates electronic conduction through a molecule sandwiched between two gold contacts. In addition to the various number parameters, this tool has a <structure> with <field>'s describing the energy levels within it. Rappture renders the structure and its fields, giving the user a graphical overview of the device under test. This graphical input, coupled closely with immediate output, provides an intuitive environment for education and scientific discovery.

## 4.2 Archimedes GUI

In the following we will give a description of how this GUI can be used to set up simple and sofisticated devices simulations, run a simulation and browse through the results obtained. Before that, for the interested users, we report some instructions about how the GUI should be installed on a local machine. The users not interested in that can skip the section 2.2 and directly read about how to use the GUI.

### 4.2.1 How to install Archimedes GUI

In order to install Archimedes on a local machine, several steps have to be followed:

- – Install Rappture
- – Install Archimedes numerical kernel
- – Install Archimedes GUI

If any of these step does not work the GUI cannot be used locally, so special attention have to be given to obtain a working GUI on a machine. Fortunately, these steps are not difficult to implement and, usually, are done using standard installation actions.

### 4.2.2   Rappture installation

Installing Rappture is not difficult but the user needs to follow several steps to make it work on his/her machine. So, we report all the details in Appendix III.

### 4.2.3   Archimedes kernel installation

The installation of Archimedes is done as every standard GNU packages. The best way to install Archimedes is to open the file INSTALL that is provided in any release of the package. We report here an extract of the file. Follow these instructions and the installation of Archimedes will be pretty straightforward.

The simplest way to compile this package is:

1. `cd' to the directory containing the package's source code and type

   `./configure' to configure the package for your system.  If you're

   using `csh' on an old version of System V, you might need to type

   `sh ./configure' instead to prevent `csh' from trying to execute

   `configure' itself.

Running `configure' takes awhile. While running, it prints some messages telling which features it is checking for.

2. Type `make' to compile the package.

3. Optionally, type `make check' to run any self-tests that come with the package.

4. Type `make install' to install the programs and any data files and documentation.

5. You can remove the program binaries and object files from the source code directory by typing `make clean'. To also remove the files that `configure' created (so you can compile the package for a different kind of computer), type `make distclean'. There is also a `make maintainer-clean' target, but that is intended mainly for the package's developers. If you use it, you may have to get all sorts of other programs in order to regenerate files that came with the distribution.

There is nothing more than that to be done to install Archimedes, it's simple like that!

### 4.2.4  Archimedes GUI installation

There is actually nothing to install at this point. Just go to your Archimedes rappture directory :

`# cd archimedes/rappture`

and type

`# rappture`

You will see get something like in figure 1. Let us see now how to run simulations using this GUI.

## 4.3  How to use Archimedes GUI

## 4.4  Examples

We report here some examples of semiconductor devices that can be simulated with Archimedes GUI. The examples we report are usually used as benchmarks for newly implemented devices simulators, i.e. a diode and a MESFET.

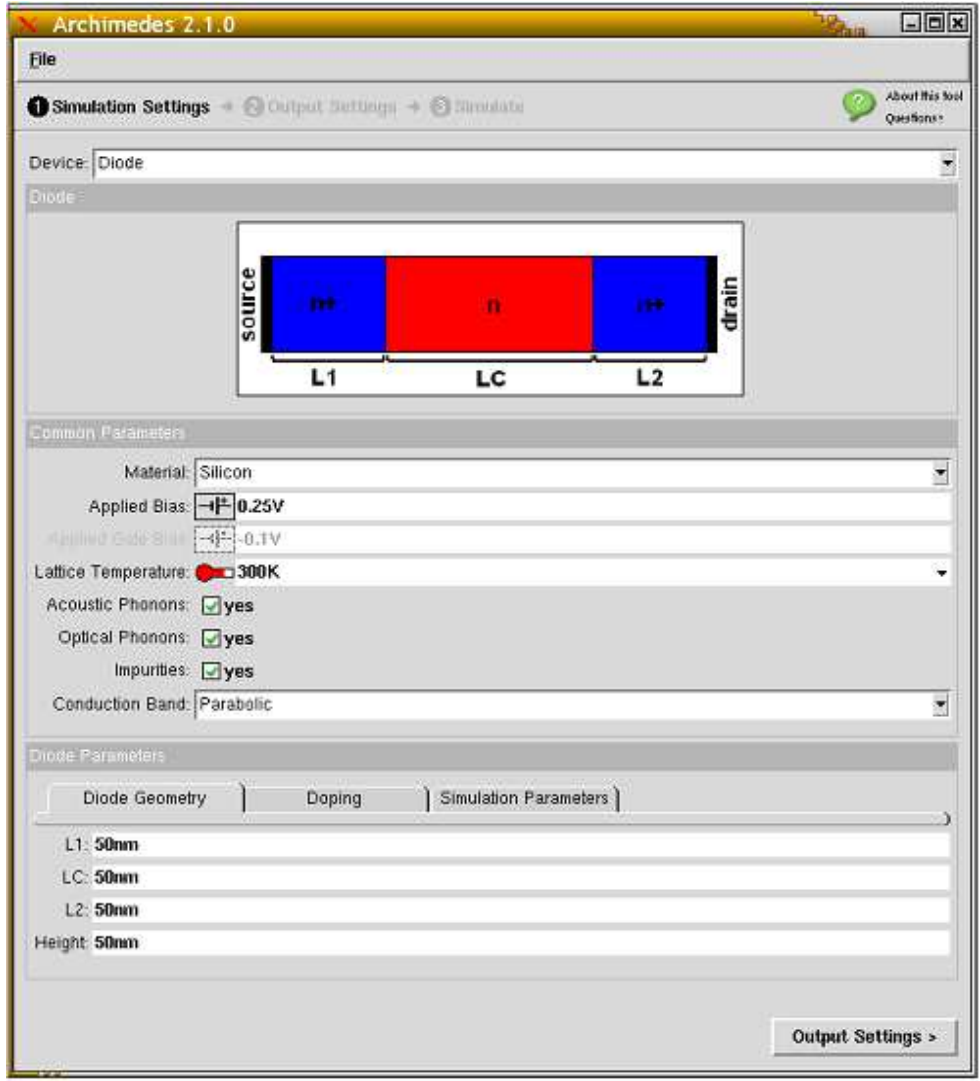

Figure 12 - Archimedes GUI, initial view

When the GUI is started the first time, it look like the one reported in Figure 1. It is easy to understand the meaning of the various buttons and labels, even at the very first glance. For example, there is a *Device* menu the user can use to select a predefined device. When the user select a device from this menu, the device show up on the screen. The user can select the material the device is made of using the *Material* menu.

If a diode is selected, an applied potential can be specified. This is done by inputting a value in the corresponding box. In this particular case, the bias is applied to the drain of the device, while the source is fixed to zero volt.

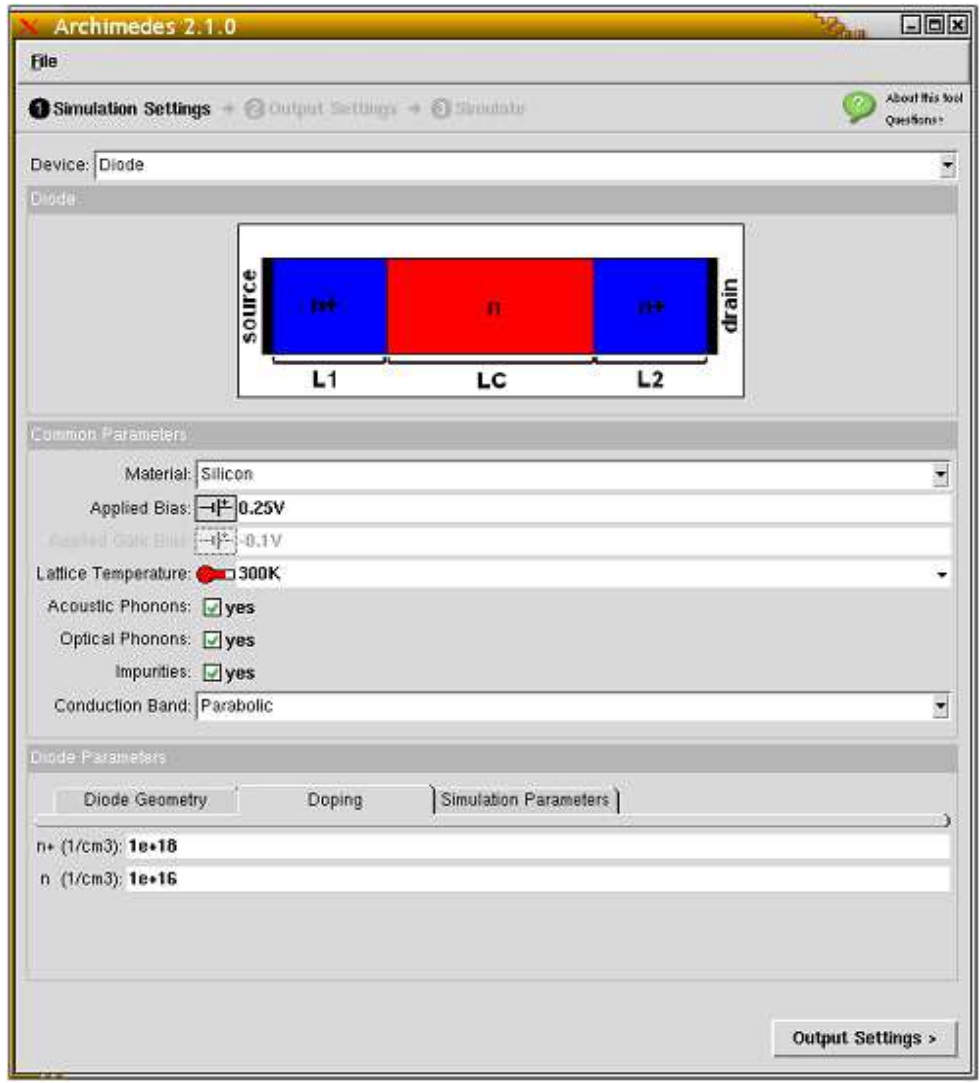

Figure 13 - Archimedes GUI, doping selection

The user can also select the temperature of the device. The temperature is given by the user and is applied as an homogeneous temperature on the whole device. One mechanism that depends on the temperature is the scattering. The user can specify what kind of scattering has to be included.

The energy band of electrons is specified by using the menu labeled *Conduction Band*.

The lowest part of the GUI is labeled *Diode Parameters*. In this part of the GUI, there are three tabs that can be selected, i.e. *Diode Geometry*, *Doping*, *Simulation Parameters*. There are shown in Figure 1, Figure 2 and Figure 3 respectively.

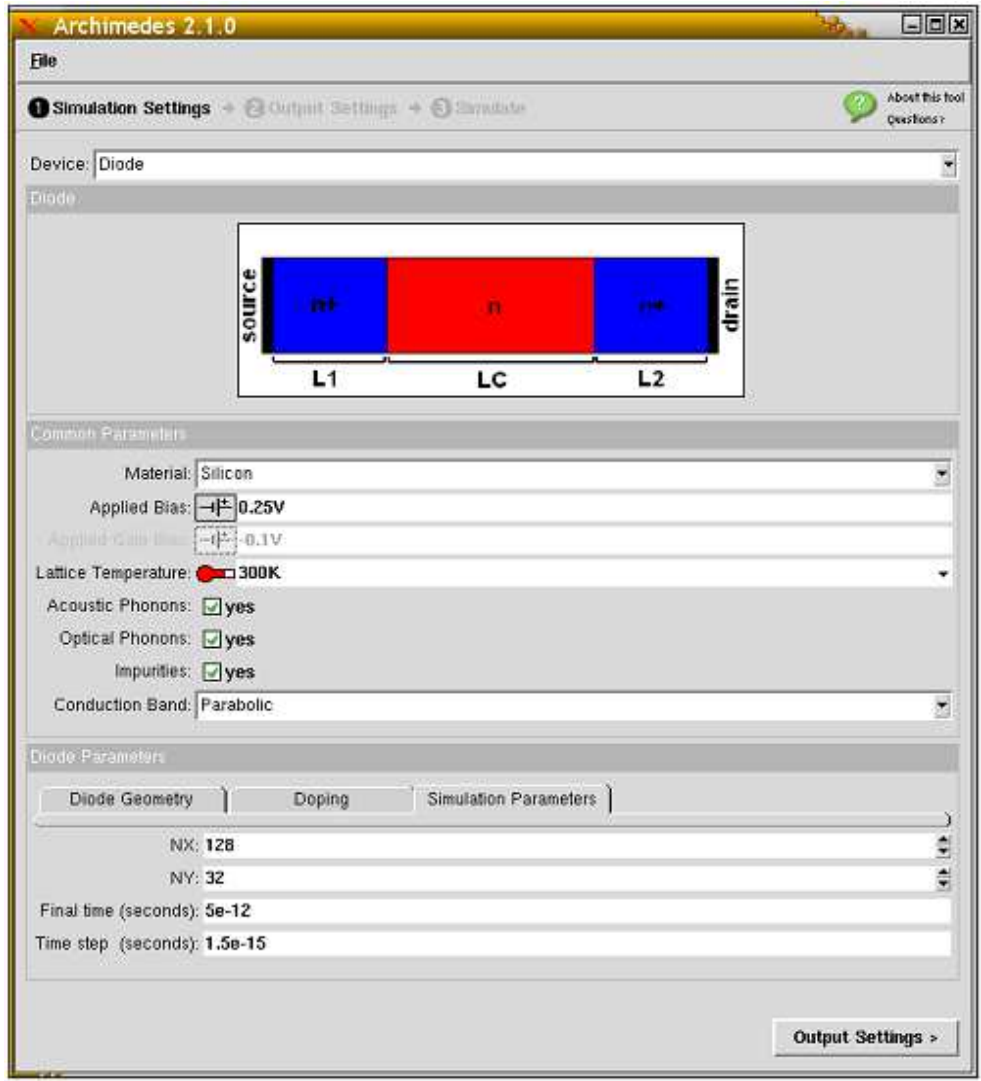

Figure 14 - Archimedes GUI, Simulation Parameters

In the *Diode Geometry* tab (Figure 1), the user can specify several variables for the simulation, i.e. *L1*, *L2*, *LC* and the height of the device. The meaning of those variables is easy to understand if one have a look to the device drawn in the GUI. *L1* is the length of the *n+* doping area attached to the source. *L2* is the length of the *n+* doping area attached to the drain. *LC* is the length of the channel. Finally, *height* is obviously the height of the diode.

Clicking on the *Doping* tab (Figure 2), the user can then specify the doping density *n* and *n+*. They have to be specified in 1/cm3.

Finally, by clicking on the *Simulation Parameters* (Figure 3) the user can specify the number of cells in the x and y-direction, *NX* and *NY* respectively, the final time of the simulation (in seconds) and the simulation time step (or time increment).

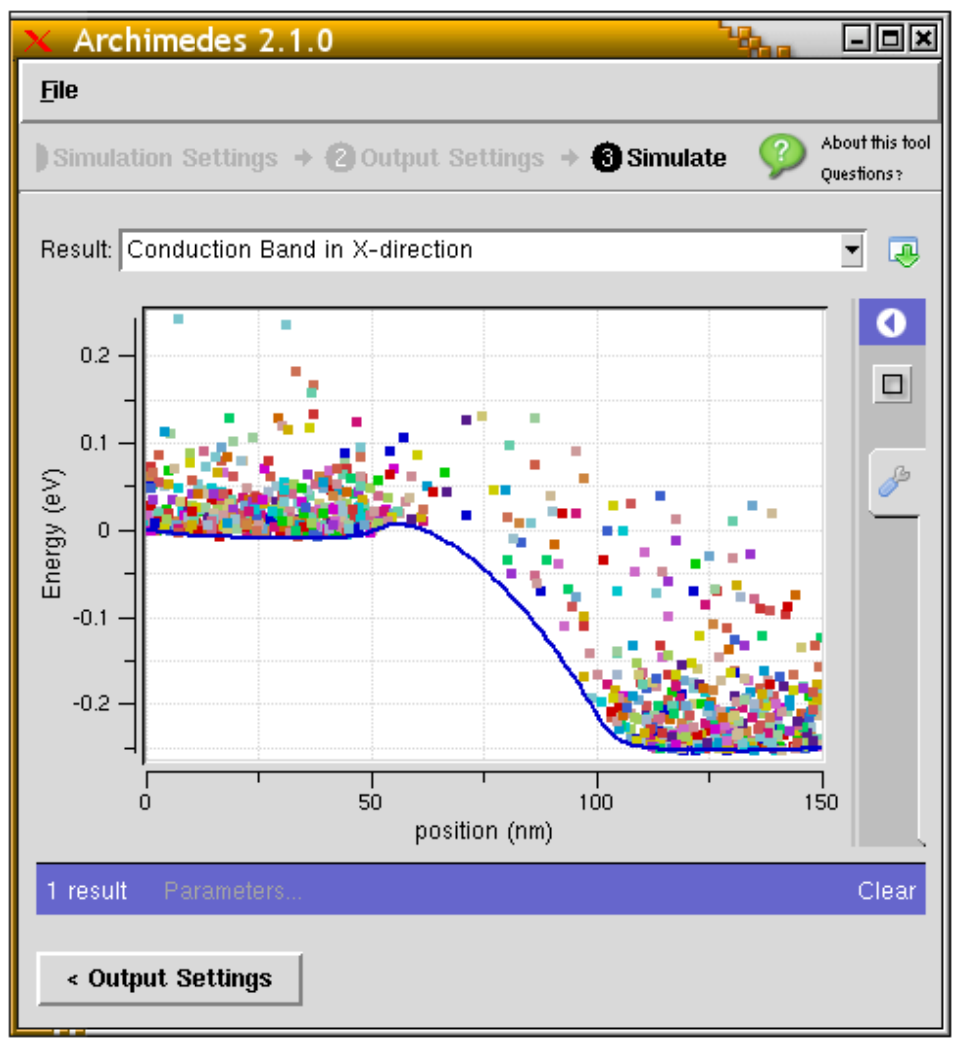

Figure 15 - Electrons in the Diode conduction band

If the user clicks on the *Simulate* button (in the right top part of the GUI), then the simulation starts taking into account the parameters that have been specified.

After waiting for the calculations being done, the results show up on the screen. Be aware that once you run the simulation, obtaining the results can take a very long time. Do not expect to get the results in a few minutes. Usually, Monte Carlo simulations can take evern days, depending on the device to be simulated and the CPU used to carry out the simulation.

The results for the default diode of Archimedes GUI are reported from Figure 4 to Figure 11.

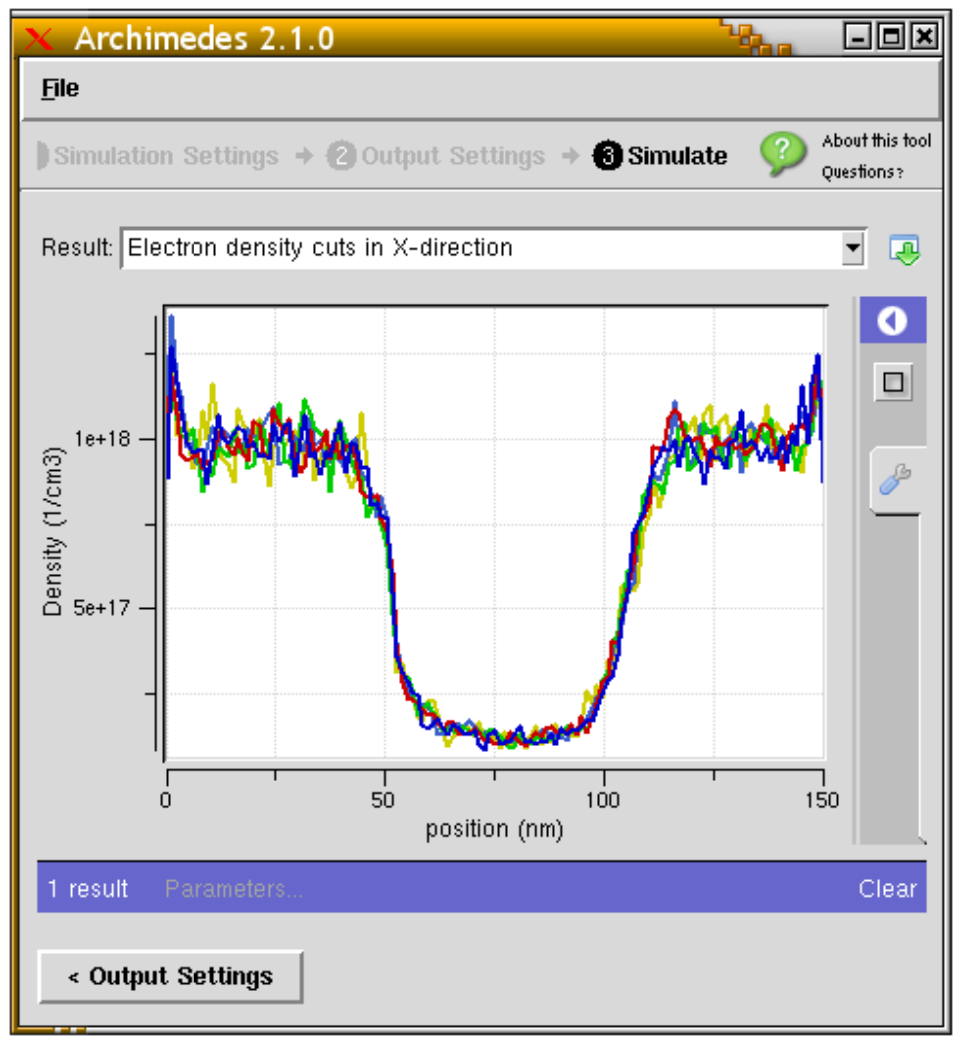

Figure 16 - Density cut over the diode

There is a variety of results than can be visualized once the calculations are done. In Figure 4, electrons are reported in the conduction band of the diode. The conduction band is taken as a cut over the diode in the x-direction.

Figure 5 reports several electron density cuts over the diode in the x-direction. Since this device has a 1D symmetry the cuts overlap and it is possible to see, basically, only one density cut.

Figure 6 reports several electron energy cuts over the diode, in the x-direction. Even in this case, it is possible to see only one cut since the device has a 1D symmetry.

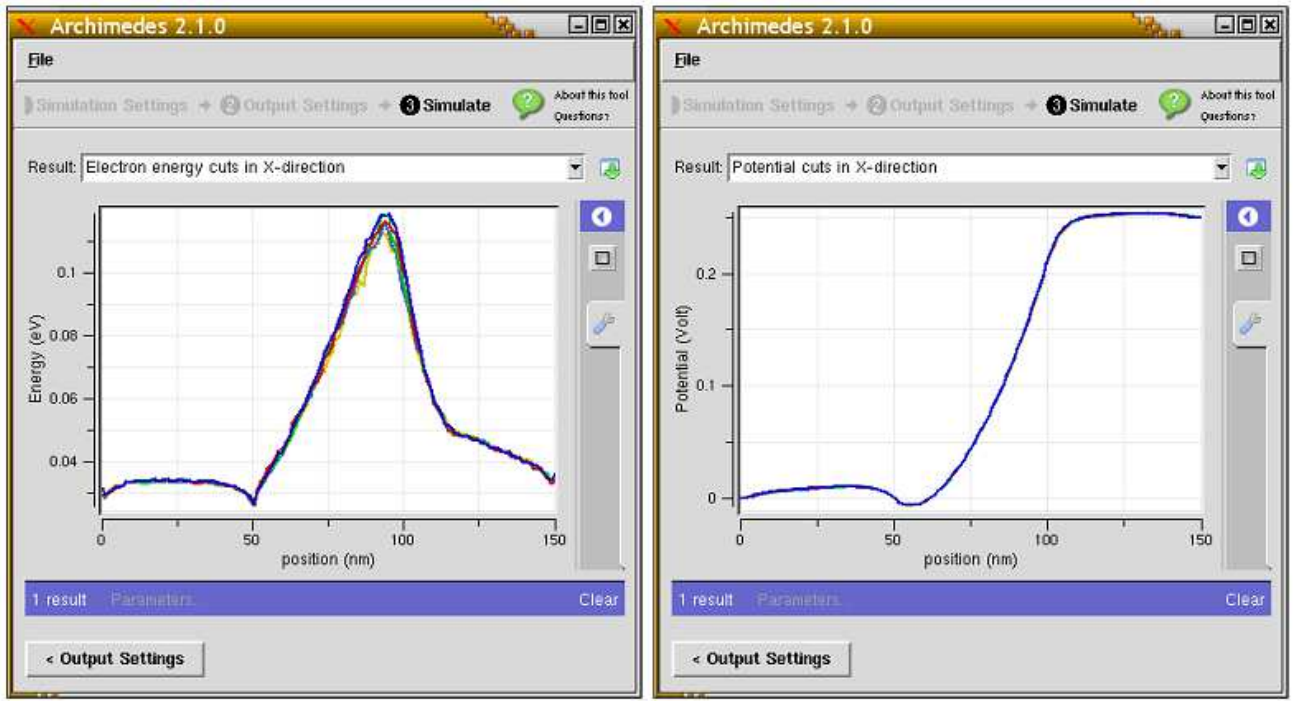

Figure 7 shows the cuts of the electrostatic potential over the simulated diode in the x-direction.

Figure 8 shows the 2D plots that can be obtained using the Archimedes GUI. The top left plot shows the electron density, the top right shows the electron energy, the bottom left shows the electrostatic potential and the bottom right shows the Boltzmann distribution function. All figures show the stationary solutions at 5 picoseconds.

Finally in figure 9 and 10 we report the results of a MESFET simulation done with the Archimede GUI, with a final time equal to 5 picoseconds.

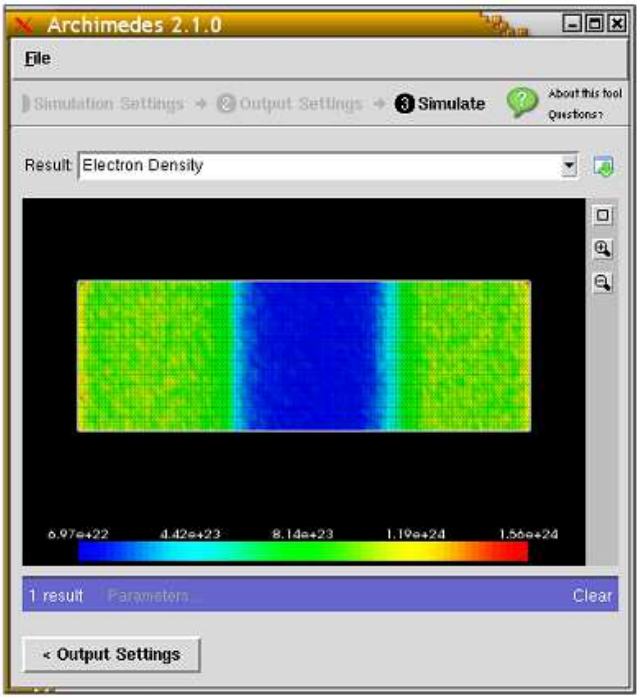
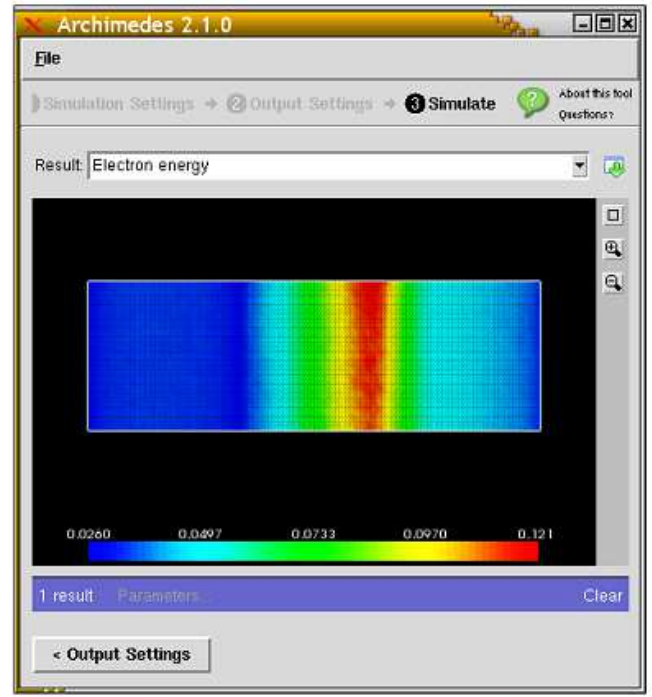
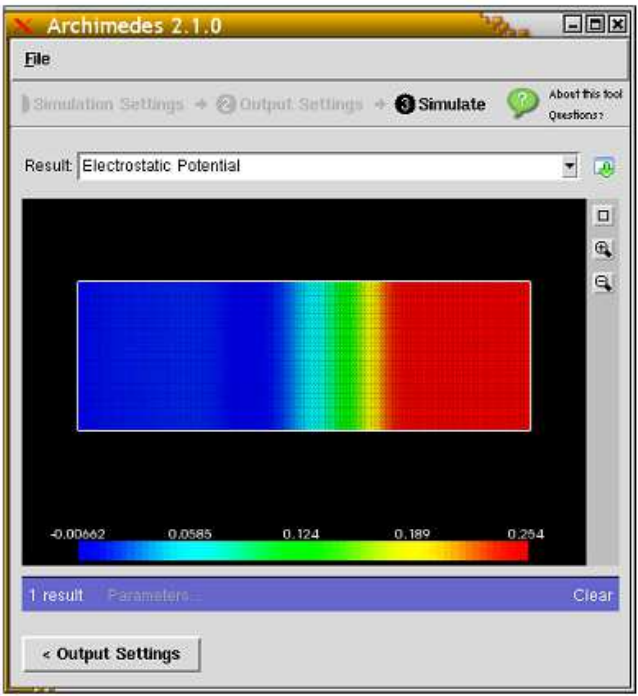
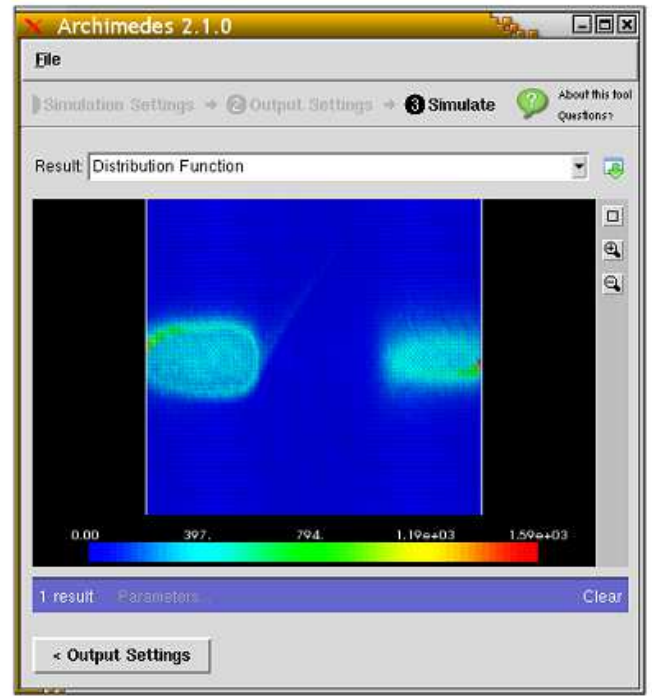

Figure 8 - Diode simulation. Top left: Electron density, Top right: Electron energy, Bottom left: Electrostatic potential, Bottom right: Boltzmann distribution function

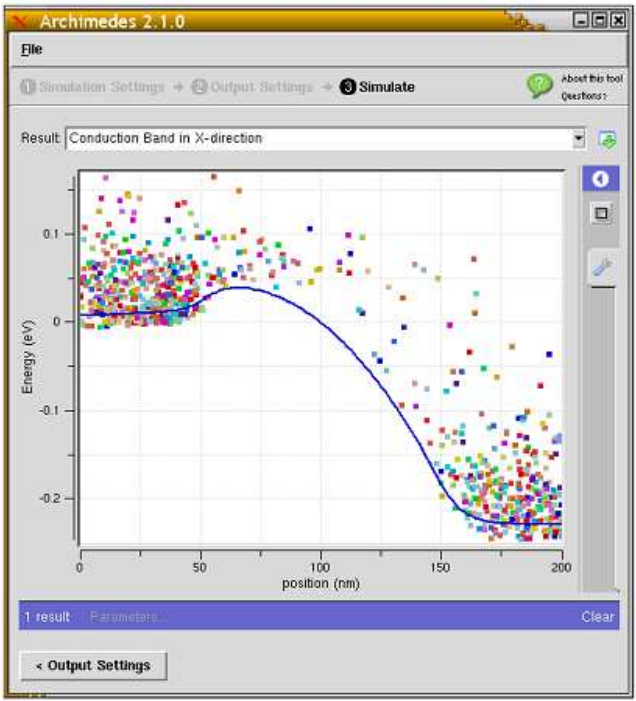
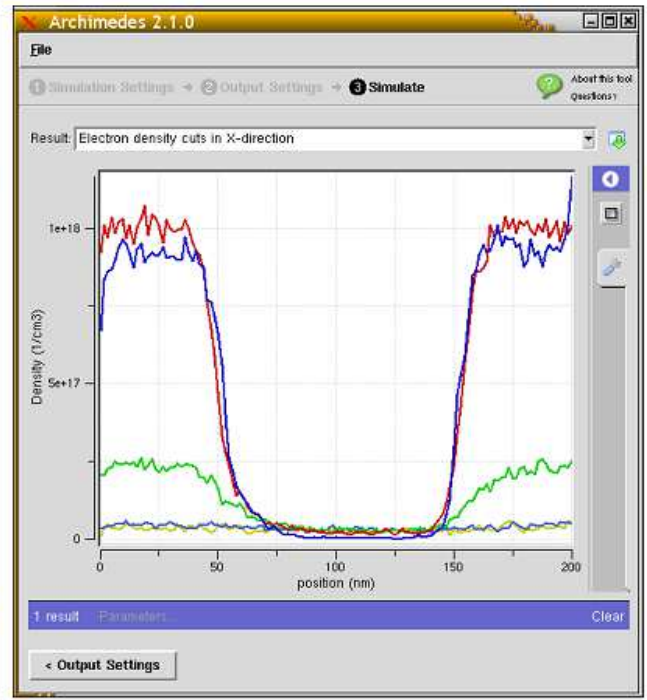
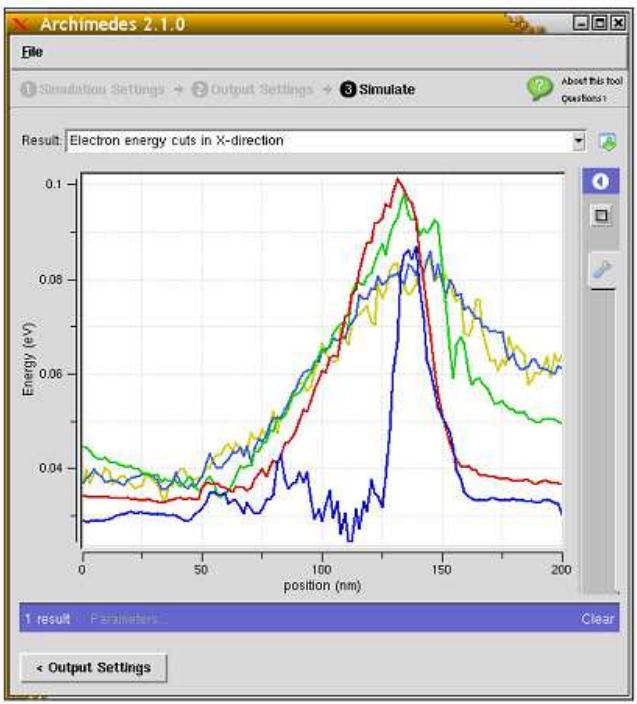
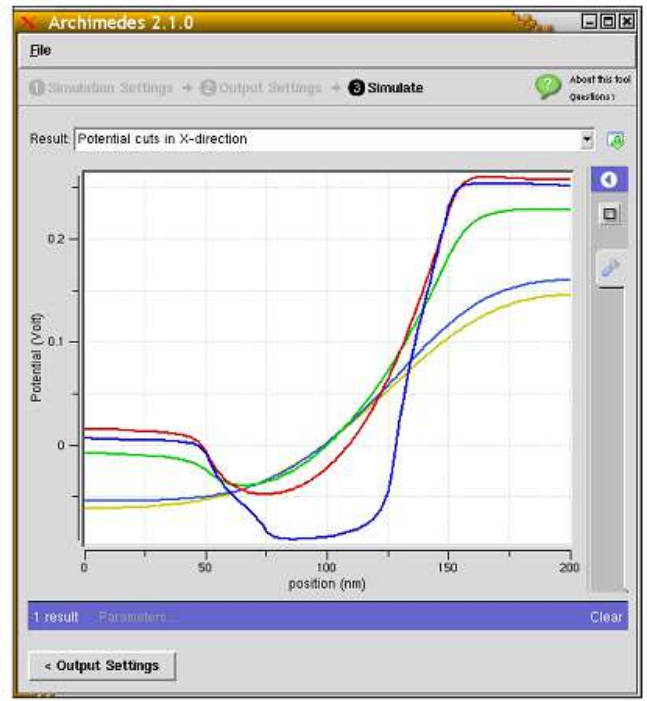

Figure 9 - MESFET - Top left: Electrons in conduction band, Top right: Electron density cuts, Bottom left: Electron energy cuts, Bottom right: Electrostatic potential cuts

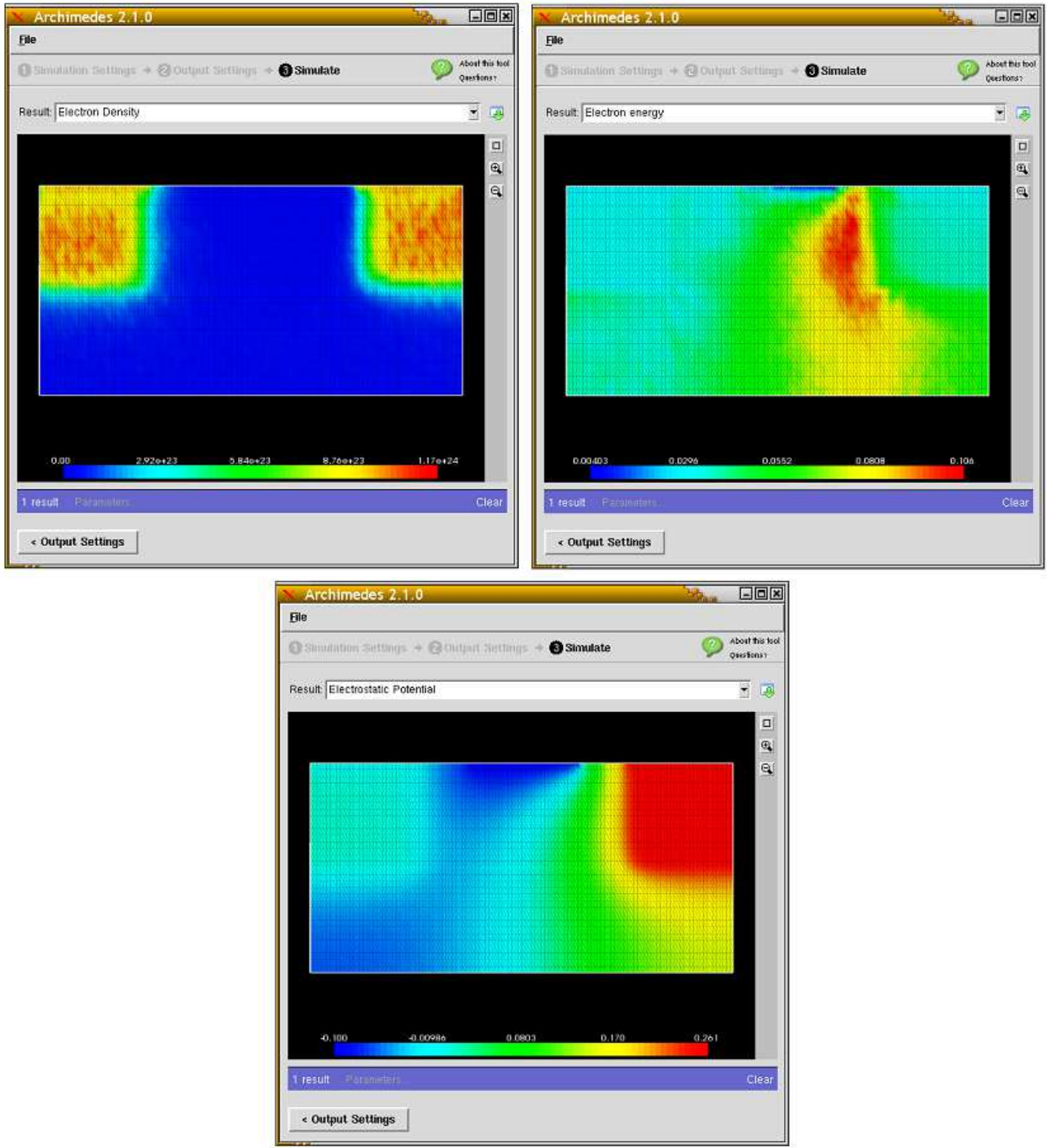

Figure 10 - MESFET - Top left: 2D Electron density, Top right: 2D Electron density, Bottom: Electrostatic potential

# Appendix I

# -

# Some useful Parameters

We report the file **constants.h** of Archimedes. This file, basically, contains the constants used in Archimedes. Some useful definitions are also present in **archimedes.c** and an extract is reported where some values are defined for the different semiconductor materials implemented (in particular, the Virtual Crystal Approximation is shown here).

## constants.h

```
// Universal Physical Constants in M.K.S.C. system
// Boltzmann constant (Joule/Kelvin)
 const real KB=1.380658e-23;
// Electron charge in absolute value (Coulomb)
 const real Q=1.60217733e-19;
// Reduced Planck constant (Joule*sec)
 const real HBAR=1.05457266e-34;
// Permittivity of free space (F/m)
 const real EPS0=8.854187817e-12;
// Electron Mass (Kg)
 const real M=9.1093897e-31;
// Silicon intrinsic density for room temperature
 const real NI=1.45e16;
// Pi number
 const real PI=3.141592654;
// electron energy step (eV) for the MC method
 const real DE=0.002;
// Silicon low field mobility (m^2/(V*sec))
 real MIU0=1400.e-4;
// Silicon saturation velocity (m/sec)
 real VS=1.e5;
// Silicon heavy hole effective mass
 real mstarhole=0.57;
// Silicon heavy hole low field mobility (m^2/(V*sec))
 real MIU0hole=0.0471;
// Silicon heavy hole saturation velocity (m/sec)
 real VShole=1.e5;
// Silicon Schottky contact density (1/m^3)
```

```c
 real NGATE=3.9e11;
```

## archimedes.c

```c
// =======================
// Material constants here!
// =======================
// Silicon electrons in the X-valley
// Germanium electrons in the Gemma-valley
// GaAs electrons in the Gamma- and L-valley
// For the others, see comments below
    NOVALLEY[SILICON]=1;    // X-valley
    NOVALLEY[GERMANIUM]=1;  // G-valley
    NOVALLEY[GAAS]=2;       // G and L-valleys
    NOVALLEY[INSB]=1;       // G valley
    NOVALLEY[ALSB]=1;       // G-valley
    NOVALLEY[ALXINXSB]=1;   // G-valley
    NOVALLEY[ALXIN1XSB]=1;  // G-valley
    NOVALLEY[ALAS]=1;       // G-valley
    NOVALLEY[ALP]=1;        // G-valley
    NOVALLEY[GAP]=1;        // G-valley
    NOVALLEY[GASB]=1;       // G-valley
    NOVALLEY[INAS]=1;       // G-valley
    NOVALLEY[INP]=1;        // G-valley
    NOVALLEY[INXGA1XAS]=1;  // only G-valley
    NOVALLEY[INXAL1XAS]=1;  // G-valley
    NOVALLEY[INXGAXXAS]=1;  // only G-valley
// Dielectric constant for Silicon Oxide SiO2
    EPSRSIO2=3.9*EPS0;         // see http://en.wikipedia.org/wiki/Relative_permittivity
// Dielectric constant for Semiconducting materials
// STATIC
// ======
    EPSR[SILICON]=11.68;      // see http://en.wikipedia.org/wiki/Relative_permittivity
    EPSR[GERMANIUM]=16.2;       // see http://www.ioffe.ru/SVA/NSM/Semicond/Ge/basic.html
    EPSR[GAAS]=12.90;         // see http://www.ioffe.ru/SVA/NSM/Semicond/GaAs/basic.html
    EPSR[INSB]=16.8;          // see http://www.ioffe.ru/SVA/NSM/Semicond/InSb/basic.html
    EPSR[ALSB]=12.04;         // Fischetti conversations
    EPSR[ALAS]=12.90-2.84;     // see http://www.ioffe.ru/SVA/NSM/Semicond/AlGaAs/basic.html
(x=1.0)
    EPSR[ALP]=9.80;           // Fischetti conversations
    EPSR[GAP]=11.10;          // see http://www.ioffe.ru/SVA/NSM/Semicond/GaP/basic.html
    EPSR[GASB]=15.69;          // see http://www.ioffe.ru/SVA/NSM/Semicond/GaSb/basic.html
    EPSR[INAS]=15.15;          // see http://www.ioffe.ru/SVA/NSM/Semicond/InAs/basic.html
    EPSR[INP]=12.50;           // see http://www.ioffe.ru/SVA/NSM/Semicond/InP/basic.html
// III-V semiconductor compounds high frequency dielectric constant
// HIGH FREQUENCY
// ==============
    EPF[GAAS]=10.89;           // see http://www.ioffe.ru/SVA/NSM/Semicond/GaAs/basic.html
    EPF[INSB]=15.68;           // Fischetti conversations
    EPF[ALSB]=9.88;            // Fiscehtti conversations
```

```cpp
    EPF[ALAS]=10.89-2.73;      // see http://www.ioffe.ru/SVA/NSM/Semicond/AlGaAs/basic.html
(x=1.0)
    EPF[ALP]=7.54;              // Fischetti conversations
    EPF[GAP]=9.11;              // see http://www.ioffe.ru/SVA/NSM/Semicond/GaP/basic.html
    EPF[GASB]=14.44;            // see http://www.ioffe.ru/SVA/NSM/Semicond/GaSb/basic.html
    EPF[INAS]=12.3;             // see http://www.ioffe.ru/SVA/NSM/Semicond/InAs/basic.html
    EPF[INP]=9.61;              // see http://www.ioffe.ru/SVA/NSM/Semicond/InP/basic.html
    int ii;
    for(ii=0;ii<NOAMTIA;ii++){
      int i;
      for(i=0;i<6;i++){
        HWO[ii][i]=0.;
        DTK[ii][i]=0.;
        ZF[ii][i]=0.;
      }
    }
// Optical phonon scattering energy (eV)
    HWO[SILICON][0]=0.0120;    // Sellier, Tomizawa
    HWO[SILICON][1]=0.0185;    // Sellier, Tomizawa
    HWO[SILICON][2]=0.0190;    // Sellier, Tomizawa
    HWO[SILICON][3]=0.0474;    // Sellier, Tomizawa
    HWO[SILICON][4]=0.0612;    // Sellier, Tomizawa
    HWO[SILICON][5]=0.0590;    // Sellier, Tomizawa
    HWO[GERMANIUM][0]=0.03704; // Fischetti
    HWO[GAAS][0]=0.03536;      // Fischetti
    HWO[INSB][0]=0.02404;      // Fischetti
    HWO[ALSB][0]=0.0360;       // Fischetti
    HWO[ALAS][0]=0.05009;      // Fischetti
    HWO[ALP][0]=0.06211;       // Fischetti
    HWO[GAP][0]=0.04523;       // Fischetti
    HWO[GASB][0]=0.02529;      // Fischetti
    HWO[INAS][0]=0.03008;      // Fischetti
    HWO[INP][0]=0.04240;       // Fischetti
// Optical coupling constants (eV/m)
    DTK[SILICON][0]=0.05e11;    // Jacoboni Reggiani
    DTK[SILICON][1]=0.08e11;    // Jacoboni Reggiani
    DTK[SILICON][2]=0.03e11;    // Jacoboni Reggiani
    DTK[SILICON][3]=0.20e11;    // Jacoboni Reggiani
    DTK[SILICON][4]=1.14e11;    // Jacoboni Reggiani
    DTK[SILICON][5]=0.20e11;    // Jacoboni Reggiani
    DTK[GERMANIUM][0]=0.0;      // Jacoboni Reggiani
    DTK[GERMANIUM][1]=0.079e11; // Jacoboni Reggiani
    DTK[GERMANIUM][2]=0.0;      // Jacoboni Reggiani
    DTK[GERMANIUM][3]=0.0;      // Jacoboni Reggiani
    DTK[GERMANIUM][4]=0.95e11;  // Jacoboni Reggiani
    DTK[GERMANIUM][5]=0.0;      // Jacoboni Reggiani
    DTK[GAAS][0]=1.11e11;       // Sellier, Tomizawa
    DTK[INSB][0]=0.47e11; // see ???
    DTK[ALSB][0]=0.55e11; // see ???
    DTK[ALAS][0]=3.0e11;  // see ???
    DTK[ALP][0]=0.95e11;  // see ???
```

```cpp
    DTK[GAP][0]=5.33e11;    // see ???
    DTK[GASB][0]=0.94e11;   // see ???
    DTK[INAS][0]=3.59e11;   // see ???
    DTK[INP][0]=2.46e11;    // see ???
// Optical phonon Z-factor
    ZF[SILICON][0]=1.;      // Sellier
    ZF[SILICON][1]=1.;      // Sellier
    ZF[SILICON][2]=4.;      // Sellier
    ZF[SILICON][3]=4.;      // Sellier
    ZF[SILICON][4]=1.;      // Sellier
    ZF[SILICON][5]=4.;      // Sellier
    ZF[GERMANIUM][0]=1.;    // see ???
    ZF[GAAS][0]=1.;         // Sellier
    ZF[INSB][0]=1.;         // see ???
    ZF[ALSB][0]=1.;         // see ???
    ZF[ALAS][0]=1.;         // see ???
    ZF[ALP][0]=1.;          // see ???
    ZF[GAP][0]=1.;          // see ???
    ZF[GASB][0]=1.;         // see ???
    ZF[INAS][0]=1.;         // see ???
    ZF[INP][0]=1.;          // see ???
// Crystal Density (Kg/m^3)
    RHO[SILICON]=2.33e3;    // Fischetti conversations
    RHO[GERMANIUM]=5.32e3;  // Fischetti conversations
    RHO[GAAS]=5.36e3;       // Fischetti conversations
    RHO[INSB]=5.78e3;       // Fischetti conversations
    RHO[ALSB]=4.26e3;       // Fischetti conversations
    RHO[ALAS]=3.76e3;       // Fischetti conversations
    RHO[ALP]=2.40e3;        // Fischetti conversations
    RHO[GAP]=4.14e3;        // Fischetti conversations
    RHO[GASB]=5.61e3;       // Fischetti conversations
    RHO[INAS]=5.67e3;       // Fischetti conversations
    RHO[INP]=4.81e3;        // Fischetti conversations
// Acoustic deformation potential (Joule)
    DA[SILICON]=9.*Q;       // Fischetti -- Jacoboni Reggiani
    DA[GERMANIUM]=9.*Q;     // Fischetti -- Jacoboni Reggiani
    DA[GAAS]=7.*Q;          // Fischetti - Gamma valley
    DA[INSB]=7.*Q;          // Fischetti
    DA[ALSB]=4.6*Q;         // Fischetti
    DA[ALAS]=9.3*Q;         // Fischetti
    DA[ALP]=9.3*Q;          // Fischetti
    DA[GAP]=7.4*Q;          // Fischetti
    DA[GASB]=9.*Q;          // Fischetti
    DA[INAS]=8.2*Q;         // Fischetti
    DA[INP]=6.2*Q;          // Fischetti
// Longitudinal sound velocity (m/sec)
    UL[SILICON]=9.18e3;     // Fischetti
    UL[GERMANIUM]=5.4e3;    // Fischetti
    UL[GAAS]=5.24e3;        // Fischetti
    UL[INSB]=3.41e3;        // Fischetti
    UL[ALSB]=4.25e3;        // Fischetti
```

```
        UL[ALAS]=5.65e3;      // Fischetti
        UL[ALP]=7.41e3;       // Fischetti
        UL[GAP]=5.85e3;       // Fischetti
        UL[GASB]=3.97e3;      // Fischetti
        UL[INAS]=4.28e3;      // Fischetti
        UL[INP]=5.13e3;       // Fischetti
// Band minimum energy
// first valley
        EMIN[SILICON][1]=0.0;      // Sellier, Fischetti, etc.
        EMIN[GERMANIUM][1]=0.173;
        EMIN[GAAS][1]=0.0;         // Tomizawa
        EMIN[INSB][1]=0.0;
        EMIN[ALSB][1]=0.507;
        EMIN[ALAS][1]=0.767;
        EMIN[ALP][1]=1.237;
        EMIN[GAP][1]=0.496;
        EMIN[GASB][1]=0.0;
        EMIN[INAS][1]=0.0;
        EMIN[INP][1]=0.0;
// eventual second valley
        EMIN[GAAS][2]=0.323;

// Definition of effective mass for all materials in all valleys
        MSTAR[SILICON][1]=0.32;      // see Sellier, Tomizawa, etc.
        MSTAR[GAAS][1]=0.067;        // Gamma-valley -- see Tomizawa
        MSTAR[GAAS][2]=0.350;        // L-valley    -- see Tomizawa
        MSTAR[GERMANIUM][1]=0.12;    // Gamma valley -- see
http://ecee.colorado.edu/~bart/book/effmass.htm#long
        MSTAR[INSB][1]=0.0135;       // Gamma-valley -- see Ram-Mohan
        MSTAR[ALSB][1]=0.14;         // Gamma-valley -- See Ram-Mohan
        MSTAR[ALAS][1]=0.149;        // Gamma-valley -- see Ram-Mohan J.App.Phys. Vol.89, Num.11
        MSTAR[ALP][1]=0.22;          // Gamma-valley -- see Ram-Mohan
        MSTAR[GAP][1]=0.13;          // Gamma-valley -- see Ram-Mohan J.App.Phys. Vol.89, Num.11
        MSTAR[GASB][1]=0.039;        // Gamma-valley -- see Ram-Mohan
        MSTAR[INAS][1]=0.026;        // Gamma-valley -- see Ram-Mohan J.App.Phys. Vol.89, Num.11
        MSTAR[INP][1]=0.0795;        // Gamma-valley -- see Ram-Mohan
// non-parabolicity coefficients
        alphaK[SILICON][1]=0.5;      // see Sellier, Tomizawa
        alphaK[GERMANIUM][1]=0.3;    // Gamma valley - Jacoboni Reggiani
// Lattice constants
        LATTCONST[GAAS]=565.35e-12;      // CODATA
        LATTCONST[SILICON]=543.102e-12;  // CODATA
        LATTCONST[GERMANIUM]=564.613e-12; // CODATA
        LATTCONST[ALP]=545.10e-12;       // CODATA
        LATTCONST[ALAS]=565.05e-12;      // CODATA
        LATTCONST[ALSB]=613.55e-12;      // CODATA
        LATTCONST[GAP]=545.12e-12;       // CODATA
        LATTCONST[GASB]=609.59e-12;      // CODATA
        LATTCONST[INP]=586.87e-12;       // CODATA
        LATTCONST[INAS]=605.83e-12;      // CODATA
        LATTCONST[INSB]=647.9e-12;       // CODATA
```

...

```c
// III-V Semiconductor materials energy gap (depending on the lattice temperature)
    printf("\n");
    EG[SILICON]=1.21-3.333e-4*TL;
    printf("EG[SILICON]    = %g\n",EG[SILICON]);
    EG[GERMANIUM]=0.747-3.587e-4*TL;
    printf("EG[GERMANIUM]  = %g\n",EG[GERMANIUM]);
    EG[GAAS]=1.54-4.036e-4*TL;
    printf("EG[GAAS]       = %g\n",EG[GAAS]);
    EG[INSB]=0.2446-2.153e-4*TL;
    printf("EG[INSB]       = %g\n",EG[INSB]);
    EG[ALSB]=1.696-2.20e-4*TL;
    printf("EG[ALSB]       = %g\n",EG[ALSB]);
    EG[ALAS]=2.314-3.0e-4*TL;
    printf("EG[ALAS]       = %g\n",EG[ALAS]);
    EG[ALP]=2.51-3.333e-4*TL;
    printf("EG[ALP]        = %g\n",EG[ALP]);
    EG[GAP]=2.35-2.667e-4*TL;
    printf("EG[GAP]        = %g\n",EG[GAP]);
    EG[GASB]=0.81-3.667e-4*TL;
    printf("EG[GASB]       = %g\n",EG[GASB]);
    EG[INAS]=0.434-2.601e-4*TL;
    printf("EG[INAS]       = %g\n",EG[INAS]);
    EG[INP]=1.445-3.296e-4*TL;
    printf("EG[INP]        = %g\n",EG[INP]);
    printf("\n");

    if(CONDUCTION_BAND==KANE || CONDUCTION_BAND==PARABOLIC ||
CONDUCTION_BAND==FULL){
        // USED WHATEVER IS THE CONDUCTION BAND FOR THE INITIAL PSUEDO WAVE VECTOR
        // OF THE PSEUDO PARTICLES
// all the following non-parabolicity coefficients depend on lattice temperature
// non-parabolicity coefficient for GaAs in the GAMMA-valley
        alphaK[GAAS][1]=pow(1.-
MSTAR[GAAS][1],2.)/(EG[GAAS]+EMIN[GAAS][1]);//expected value = 0.611
        printf("alphaK_gamma[GaAs] = %g\n",alphaK[GAAS][1]);
// non-parabolicity coefficient for GaAs in the L-valley
        alphaK[GAAS][2]=pow(1.-
MSTAR[GAAS][2],2.)/(EG[GAAS]+EMIN[GAAS][2]);//expected value = 0.242;
        printf("alphaK_L[GaAs]     = %g\n",alphaK[GAAS][2]);
// non-parabolicity coefficient for InSb in the GAMMA-valley
        alphaK[INSB][1]=pow(1.-MSTAR[INSB][1],2.)/(EG[INSB]+EMIN[INSB][1]);//5.59;
        printf("alphaK_gamma[InSb] = %g\n",alphaK[INSB][1]);
// non-parabolicity coefficient for AlSb in the GAMMA-valley
        alphaK[ALSB][1]=pow(1.-MSTAR[ALSB][1],2.)/(EG[ALSB]+EMIN[ALSB][1]);//0.321;
        printf("alphaK_gamma[AlSb] = %g\n",alphaK[ALSB][1]);
// non-parabolicity coefficient for AlAs in the GAMMA-valley
        alphaK[ALAS][1]=pow(1.-MSTAR[ALAS][1],2.)/(EG[ALAS]+EMIN[ALAS][1]);
        printf("alphaK_gamma[AlAs] = %g\n",alphaK[ALAS][1]);
// non-parabolicity coefficient for AlP in the GAMMA-valley
        alphaK[ALP][1]=pow(1.-MSTAR[ALP][1],2.)/(EG[ALP]+EMIN[ALP][1]);
```

```c
    printf("alphaK_gamma[AlP] = %g\n",alphaK[ALP][1]);
// non-parabolicity coefficient for GaP in the GAMMA-valley
    alphaK[GAP][1]=pow(1.-MSTAR[GAP][1],2.)/(EG[GAP]+EMIN[GAP][1]);
    printf("alphaK_gamma[GaP] = %g\n",alphaK[GAP][1]);
// non-parabolicity coefficient for GaSb in the GAMMA-valley
    alphaK[GASB][1]=pow(1.-MSTAR[GASB][1],2.)/(EG[GASB]+EMIN[GASB][1]);
    printf("alphaK_gamma[GaSb] = %g\n",alphaK[GASB][1]);
// non-parabolicity coefficient for InAs in the GAMMA-valley
    alphaK[INAS][1]=pow(1.-MSTAR[INAS][1],2.)/(EG[INAS]+EMIN[INAS][1]);
    printf("alphaK_gamma[InAs] = %g\n",alphaK[INAS][1]);
// non-parabolicity coefficient for InP in the GAMMA-valley
    alphaK[INP][1]=pow(1.-MSTAR[INP][1],2.)/(EG[INP]+EMIN[INP][1]);
    printf("alphaK_gamma[InP] = %g\n",alphaK[INP][1]);
    }

// Semiconductor compounds
// ***
// Relative dielectric constant for semiconductor compounds
    EPSR[ALXINXSB]=XVAL[ALXINXSB]*EPSR[ALSB]+XVAL[ALXINXSB]*EPSR[INSB];
    EPSR[ALXIN1XSB]=XVAL[ALXIN1XSB]*EPSR[ALSB]+(1.-
XVAL[ALXIN1XSB])*EPSR[INSB];
    EPSR[INXGA1XAS]=XVAL[INXGA1XAS]*EPSR[INAS]+(1.-
XVAL[INXGA1XAS])*EPSR[GAAS];
    EPSR[INXAL1XAS]=XVAL[INXAL1XAS]*EPSR[INAS]+(1.-
XVAL[INXAL1XAS])*EPSR[ALAS];

EPSR[INXGAXXAS]=XVAL[INXGAXXAS]*EPSR[INAS]+XVAL[INXGAXXAS]*EPSR[GAAS];
// semiconductor compounds high frequency dieletric constant
    EPF[ALXINXSB]=XVAL[ALXINXSB]*(EPF[ALSB]+EPF[INSB]);
    EPF[ALXIN1XSB]=XVAL[ALXIN1XSB]*EPF[ALSB]+(1.-XVAL[ALXIN1XSB])*EPF[INSB];
    EPF[INXGA1XAS]=XVAL[INXGA1XAS]*EPF[INAS]+(1.-
XVAL[INXGA1XAS])*EPF[GAAS];
    EPF[INXAL1XAS]=XVAL[INXAL1XAS]*EPF[INAS]+(1.-XVAL[INXAL1XAS])*EPF[ALAS];
    EPF[INXGAXXAS]=XVAL[INXGAXXAS]*EPF[INAS]+XVAL[INXGAXXAS]*EPF[GAAS];
// semiconductor compounds optical phonon scattering energy (eV)
    HWO[ALXINXSB][0]=XVAL[ALXINXSB]*(HWO[ALSB][0]+HWO[INSB][0]);
    HWO[ALXIN1XSB][0]=XVAL[ALXIN1XSB]*HWO[ALSB][0]+(1.-
XVAL[ALXIN1XSB])*HWO[INSB][0];
    HWO[INXGA1XAS][0]=XVAL[INXGA1XAS]*HWO[INAS][0]+(1.-
XVAL[INXGA1XAS])*HWO[GAAS][0];
    HWO[INXAL1XAS][0]=XVAL[INXAL1XAS]*HWO[INAS][0]+(1.-
XVAL[INXAL1XAS])*HWO[ALAS][0];

HWO[INXGAXXAS][0]=XVAL[INXGAXXAS]*HWO[INAS][0]+XVAL[INXGAXXAS]*HWO[GA
AS][0];
// semiconductor compounds optical coupling constants (eV/m)
    DTK[ALXINXSB][0]=XVAL[ALXINXSB]*(DTK[ALSB][0]+DTK[INSB][0]);
    DTK[ALXIN1XSB][0]=XVAL[ALXIN1XSB]*DTK[ALSB][0]+(1.-
XVAL[ALXIN1XSB])*DTK[INSB][0];
    DTK[INXGA1XAS][0]=XVAL[INXGA1XAS]*DTK[INAS][0]+(1.-
XVAL[INXGA1XAS])*DTK[GAAS][0];
```

```
    DTK[INXAL1XAS][0]=XVAL[INXAL1XAS]*DTK[INAS][0]+(1.-
XVAL[INXAL1XAS])*DTK[ALAS][0];

DTK[INXGAXXAS][0]=XVAL[INXGAXXAS]*DTK[INAS][0]+XVAL[INXGAXXAS]*DTK[GAAS
][0];
```
// semiconductor compounds optical phonon Z-factor
```
    ZF[ALXINXSB][0]=XVAL[ALXINXSB]*(ZF[ALSB][0]+ZF[INSB][0]);
    ZF[ALXIN1XSB][0]=XVAL[ALXIN1XSB]*ZF[ALSB][0]+(1.-
XVAL[ALXIN1XSB])*ZF[INSB][0];
    ZF[INXGA1XAS][0]=XVAL[INXGA1XAS]*ZF[INAS][0]+(1.-
XVAL[INXGA1XAS])*ZF[GAAS][0];
    ZF[INXAL1XAS][0]=XVAL[INXAL1XAS]*ZF[INAS][0]+(1.-
XVAL[INXAL1XAS])*ZF[ALAS][0];

ZF[INXGAXXAS][0]=XVAL[INXGAXXAS]*ZF[INAS][0]+XVAL[INXGAXXAS]*ZF[GAAS][0];
```
// semiconductor compounds Crystal Density (Kg/m^3)
```
    RHO[ALXINXSB]=XVAL[ALXINXSB]*(RHO[ALSB]+RHO[INSB]);
    RHO[ALXIN1XSB]=XVAL[ALXIN1XSB]*RHO[ALSB]+(1.-
XVAL[ALXIN1XSB])*RHO[INSB];
    RHO[INXGA1XAS]=XVAL[INXGA1XAS]*RHO[INAS]+(1.-
XVAL[INXGA1XAS])*RHO[GAAS];
    RHO[INXAL1XAS]=XVAL[INXAL1XAS]*RHO[INAS]+(1.-
XVAL[INXAL1XAS])*RHO[ALAS];
    RHO[INXGAXXAS]=XVAL[INXGAXXAS]*RHO[INAS]+XVAL[INXGAXXAS]*RHO[GAAS];
```
// semiconductor compounds acoustic deformation potential (Joule)
```
    DA[ALXINXSB]=XVAL[ALXINXSB]*(DA[ALSB]+DA[INSB]);
    DA[ALXIN1XSB]=XVAL[ALXIN1XSB]*DA[ALSB]+(1.-XVAL[ALXIN1XSB])*DA[INSB];
    DA[INXGA1XAS]=XVAL[INXGA1XAS]*DA[INAS]+(1.-XVAL[INXGA1XAS])*DA[GAAS];
    DA[INXAL1XAS]=XVAL[INXAL1XAS]*DA[INAS]+(1.-XVAL[INXAL1XAS])*DA[ALAS];
    DA[INXGAXXAS]=XVAL[INXGAXXAS]*DA[INAS]+XVAL[INXGAXXAS]*DA[GAAS];
```
// semiconductor compounds longitudinal sound velocity (m/sec)
```
    UL[ALXINXSB]=XVAL[ALXINXSB]*(UL[ALSB]+UL[INSB]);
    UL[ALXIN1XSB]=XVAL[ALXIN1XSB]*UL[ALSB]+(1.-XVAL[ALXIN1XSB])*UL[INSB];
    UL[INXGA1XAS]=XVAL[INXGA1XAS]*UL[INAS]+(1.-XVAL[INXGA1XAS])*UL[GAAS];
    UL[INXAL1XAS]=XVAL[INXAL1XAS]*UL[INAS]+(1.-XVAL[INXAL1XAS])*UL[ALAS];
    UL[INXGAXXAS]=XVAL[INXGAXXAS]*UL[INAS]+XVAL[INXGAXXAS]*UL[GAAS];
```
// semiconductor compounds energy gap
```
    EG[ALXINXSB]=XVAL[ALXINXSB]*(EG[ALSB]+EG[INSB]);
    EG[ALXIN1XSB]=XVAL[ALXIN1XSB]*EG[ALSB]+(1.-XVAL[ALXIN1XSB])*EG[INSB];
    EG[INXGA1XAS]=XVAL[INXGA1XAS]*EG[INAS]+(1.-XVAL[INXGA1XAS])*EG[GAAS];
    EG[INXAL1XAS]=XVAL[INXAL1XAS]*EG[INAS]+(1.-XVAL[INXAL1XAS])*EG[ALAS];
    EG[INXGAXXAS]=XVAL[INXGAXXAS]*EG[INAS]+XVAL[INXGAXXAS]*EG[GAAS];
```
// semiconductor compounds energy minimum of GAMMA-valley
```
    EMIN[ALXINXSB][1]=XVAL[ALXINXSB]*(EMIN[ALSB][1]+EMIN[INSB][1]);
    EMIN[ALXIN1XSB][1]=XVAL[ALXIN1XSB]*EMIN[ALSB][1]+(1.-
XVAL[ALXIN1XSB])*EMIN[INSB][1];
    EMIN[INXGA1XAS][1]=XVAL[INXGA1XAS]*EMIN[INAS][1]+(1.-
XVAL[INXGA1XAS])*EMIN[GAAS][1];
    EMIN[INXAL1XAS][1]=XVAL[INXAL1XAS]*EMIN[INAS][1]+(1.-
XVAL[INXAL1XAS])*EMIN[ALAS][1];

EMIN[INXGAXXAS][1]=XVAL[INXGAXXAS]*EMIN[INAS][1]+XVAL[INXGAXXAS]*EMIN[GA
AS][1];
```

// semiconductor compounds energy minimum 0f L-valley
    EMIN[ALXINXSB][2]=XVAL[ALXINXSB]*(EMIN[ALSB][2]+EMIN[INSB][2]);
    EMIN[ALXIN1XSB][2]=XVAL[ALXIN1XSB]*EMIN[ALSB][2]+(1.-
XVAL[ALXIN1XSB])*EMIN[INSB][2];
    EMIN[INXGA1XAS][2]=XVAL[INXGA1XAS]*EMIN[INAS][2]+(1.-
XVAL[INXGA1XAS])*EMIN[GAAS][2];
    EMIN[INXAL1XAS][2]=XVAL[INXAL1XAS]*EMIN[INAS][2]+(1.-
XVAL[INXAL1XAS])*EMIN[ALAS][2];

EMIN[INXGAXXAS][2]=XVAL[INXGAXXAS]*EMIN[INAS][2]+XVAL[INXGAXXAS]*EMIN[GA
AS][2];
// GAMMA-valley effective mass
    MSTAR[ALXINXSB][1]=XVAL[ALXINXSB]*(MSTAR[ALSB][1]+MSTAR[INSB][1]);
    MSTAR[ALXIN1XSB][1]=XVAL[ALXIN1XSB]*MSTAR[ALSB][1]+(1.-
XVAL[ALXIN1XSB])*MSTAR[INSB][1];
    MSTAR[INXGA1XAS][1]=XVAL[INXGA1XAS]*MSTAR[INAS][1]+(1.-
XVAL[INXGA1XAS])*MSTAR[GAAS][1];
    MSTAR[INXAL1XAS][1]=XVAL[INXAL1XAS]*MSTAR[INAS][1]+(1.-
XVAL[INXAL1XAS])*MSTAR[ALAS][1];

MSTAR[INXGAXXAS][1]=XVAL[INXGAXXAS]*MSTAR[INAS][1]+XVAL[INXGAXXAS]*MSTA
R[GAAS][1];
// L-valley effective mass
    MSTAR[ALXINXSB][2]=XVAL[ALXINXSB]*(MSTAR[ALSB][2]+MSTAR[INSB][2]);
    MSTAR[ALXIN1XSB][2]=XVAL[ALXIN1XSB]*MSTAR[ALSB][2]+(1.-
XVAL[ALXIN1XSB])*MSTAR[INSB][2];
    MSTAR[INXGA1XAS][2]=XVAL[INXGA1XAS]*MSTAR[INAS][2]+(1.-
XVAL[INXGA1XAS])*MSTAR[GAAS][2];
    MSTAR[INXAL1XAS][2]=XVAL[INXAL1XAS]*MSTAR[INAS][2]+(1.-
XVAL[INXAL1XAS])*MSTAR[ALAS][2];

MSTAR[INXGAXXAS][2]=XVAL[INXGAXXAS]*MSTAR[INAS][2]+XVAL[INXGAXXAS]*MSTA
R[GAAS][2];
// non-parabolicity coefficient for semiconductor compounds in the GAMMA-valley
    alphaK[ALXINXSB][1]=XVAL[ALXINXSB]*(alphaK[ALSB][1]+alphaK[INSB][1]);
    alphaK[ALXIN1XSB][1]=XVAL[ALXIN1XSB]*alphaK[ALSB][1]+(1.-
XVAL[ALXIN1XSB])*alphaK[INSB][1];
    alphaK[INXGA1XAS][1]=XVAL[INXGA1XAS]*alphaK[INAS][1]+(1.-
XVAL[INXGA1XAS])*alphaK[GAAS][1];
    alphaK[INXAL1XAS][1]=XVAL[INXAL1XAS]*alphaK[INAS][1]+(1.-
XVAL[INXAL1XAS])*alphaK[ALAS][1];

alphaK[INXGAXXAS][1]=XVAL[INXGAXXAS]*alphaK[INAS][1]+XVAL[INXGAXXAS]*alpha
K[GAAS][1];
// non-parabolicity coefficient for Al_x In_(1-x) Sb in the L-valley
    alphaK[ALXINXSB][2]=XVAL[ALXINXSB]*(alphaK[ALSB][2]+alphaK[INSB][2]);
    alphaK[ALXIN1XSB][2]=XVAL[ALXIN1XSB]*alphaK[ALSB][2]+(1.-
XVAL[ALXIN1XSB])*alphaK[INSB][2];
    alphaK[INXGA1XAS][2]=XVAL[INXGA1XAS]*alphaK[INAS][2]+(1.-
XVAL[INXGA1XAS])*alphaK[GAAS][2];
    alphaK[INXAL1XAS][2]=XVAL[INXAL1XAS]*alphaK[INAS][2]+(1.-
XVAL[INXAL1XAS])*alphaK[ALAS][2];

```
alphaK[INXGAXXAS][2]=XVAL[INXGAXXAS]*alphaK[INAS][2]+XVAL[INXGAXXAS]*alpha
K[GAAS][2];
// ***
```

# Appendix II

# -

# GPL v3

Since Archimedes is a GNU package released under GNU Public License (GPL) v3, we report, in the following, the text of this license. Any future version of Archimedes will always be released under GPL v3 or later. If you plan to use Archimedes on a regular base or if you plan to use it as starting point to your simulation program, please have read the license below.

## GNU GENERAL PUBLIC LICENSE

Version 3, 29 June 2007

Copyright © 2007 Free Software Foundation, Inc. <http://fsf.org/>

Everyone is permitted to copy and distribute verbatim copies of this license document, but changing it is not allowed.

**Preamble**

The GNU General Public License is a free, copyleft license for software and other kinds of works.

The licenses for most software and other practical works are designed to take away your freedom to share and change the works. By contrast, the GNU General Public License is intended to guarantee your freedom to share and change all versions of a program--to make sure it remains free software for all its users. We, the Free Software Foundation, use the GNU General Public License for most of our software; it applies also to any other work released this way by its authors. You can apply it to your programs, too.

When we speak of free software, we are referring to freedom, not price. Our General Public Licenses are designed to make sure that you have the freedom to distribute copies of free software (and charge for them if you wish), that you receive source code or can get it if you want it, that you can change the software or use pieces of it in new free programs, and that you know you can do these things.

To protect your rights, we need to prevent others from denying you these rights or asking you to surrender the rights. Therefore, you have certain responsibilities if you distribute copies of the software, or if you modify it: responsibilities to respect the freedom of others.

For example, if you distribute copies of such a program, whether gratis or for a fee, you must pass on to the recipients the same freedoms that you received. You must make sure that they, too, receive or can get the source code. And you must show them these terms so they know their rights.

Developers that use the GNU GPL protect your rights with two steps: (1) assert copyright on the software, and (2) offer you this License giving you legal permission to copy, distribute and/or modify it.

For the developers' and authors' protection, the GPL clearly explains that there is no warranty for this free software. For both users' and authors' sake, the GPL requires that modified versions be marked as changed, so that their problems will not be attributed erroneously to authors of previous versions.

Some devices are designed to deny users access to install or run modified versions of the software inside them, although the manufacturer can do so. This is fundamentally incompatible with the aim of protecting users' freedom to change the software. The systematic pattern of such abuse occurs in the area of products for individuals to use, which is precisely where it is most unacceptable. Therefore, we have designed this version of the GPL to prohibit the practice for those products. If such problems arise substantially in other domains, we stand ready to extend this provision to those domains in future versions of the GPL, as needed to protect the freedom of users.

Finally, every program is threatened constantly by software patents. States should not allow patents to restrict development and use of software on general-purpose computers, but in those that do, we wish to avoid the special danger that patents applied to a free program could make it effectively proprietary. To prevent this, the GPL assures that patents cannot be used to render the program non-free.

The precise terms and conditions for copying, distribution and modification follow.

**TERMS AND CONDITIONS**

**0. Definitions.**
"This License" refers to version 3 of the GNU General Public License.

"Copyright" also means copyright-like laws that apply to other kinds of works, such as semiconductor masks.

"The Program" refers to any copyrightable work licensed under this License. Each licensee is addressed as "you". "Licensees" and "recipients" may be individuals or organizations.

To "modify" a work means to copy from or adapt all or part of the work in a fashion requiring copyright permission, other than the making of an exact copy. The resulting work is called a "modified version" of the earlier work or a work "based on" the earlier work.

A "covered work" means either the unmodified Program or a work based on the Program.

To "propagate" a work means to do anything with it that, without permission, would make you directly or secondarily liable for infringement under applicable copyright law, except executing it on a computer or modifying a private copy. Propagation includes copying, distribution (with or without modification), making available to the public, and in some countries other activities as well.

To "convey" a work means any kind of propagation that enables other parties to make or receive copies. Mere interaction with a user through a computer network, with no transfer of a copy, is not conveying.

An interactive user interface displays "Appropriate Legal Notices" to the extent that it includes a convenient and prominently visible feature that (1) displays an appropriate copyright notice, and (2) tells the user that there is no warranty for the work (except to the extent that warranties are provided), that licensees may convey the work under this License, and how to view a copy of this License. If the interface presents a list of user commands or options, such as a menu, a prominent item in the list meets this criterion.

## 1. Source Code.
The "source code" for a work means the preferred form of the work for making modifications to it. "Object code" means any non-source form of a work.

A "Standard Interface" means an interface that either is an official standard defined by a recognized standards body, or, in the case of interfaces specified for a particular programming language, one that is widely used among developers working in that language.

The "System Libraries" of an executable work include anything, other than the work as a whole, that (a) is included in the normal form of packaging a Major Component, but which is not part of that Major Component, and (b) serves only to enable use of the work with that Major Component, or to implement a Standard Interface for which an implementation is available to the public in source code form. A "Major Component", in this context, means a major essential component (kernel, window system, and so on) of the specific operating system (if any) on which the executable work runs, or a compiler used to produce the work, or an object code interpreter used to run it.

The "Corresponding Source" for a work in object code form means all the source code needed to generate, install, and (for an executable work) run the object code and to modify the work, including scripts to control those activities. However, it does not include the work's System Libraries, or general-purpose tools or generally available free programs which are used unmodified in performing those activities but which are not part of the work. For example, Corresponding Source includes interface definition files associated with source files for the work, and the source code for shared libraries and dynamically linked subprograms that the work is specifically designed to require, such as by intimate data communication or control flow between those subprograms and other parts of the work.

The Corresponding Source need not include anything that users can regenerate automatically from other parts of the Corresponding Source.

The Corresponding Source for a work in source code form is that same work.

**2. Basic Permissions.**
All rights granted under this License are granted for the term of copyright on the Program, and are irrevocable provided the stated conditions are met. This License explicitly affirms your unlimited permission to run the unmodified Program. The output from running a covered work is covered by this License only if the output, given its content, constitutes a covered work. This License acknowledges your rights of fair use or other equivalent, as provided by copyright law.

You may make, run and propagate covered works that you do not convey, without conditions so long as your license otherwise remains in force. You may convey covered works to others for the sole purpose of having them make modifications exclusively for you, or provide you with facilities for running those works, provided that you comply with the terms of this License in conveying all material for which you do not control copyright. Those thus making or running the covered works for you must do so exclusively on your behalf, under your direction and control, on terms that prohibit them from making any copies of your copyrighted material outside their relationship with you.

Conveying under any other circumstances is permitted solely under the conditions stated below. Sublicensing is not allowed; section 10 makes it unnecessary.

**3. Protecting Users' Legal Rights From Anti-Circumvention Law.**

No covered work shall be deemed part of an effective technological measure under any applicable law fulfilling obligations under article 11 of the WIPO copyright treaty adopted on 20 December 1996, or similar laws prohibiting or restricting circumvention of such measures.

When you convey a covered work, you waive any legal power to forbid circumvention of technological measures to the extent such circumvention is effected by exercising rights under this License with respect to the covered work, and you disclaim any intention to limit operation or modification of the work as a means of enforcing, against the work's users, your or third parties' legal rights to forbid circumvention of technological measures.

### 4. Conveying Verbatim Copies.

You may convey verbatim copies of the Program's source code as you receive it, in any medium, provided that you conspicuously and appropriately publish on each copy an appropriate copyright notice; keep intact all notices stating that this License and any non-permissive terms added in accord with section 7 apply to the code; keep intact all notices of the absence of any warranty; and give all recipients a copy of this License along with the Program.

You may charge any price or no price for each copy that you convey, and you may offer support or warranty protection for a fee.

### 5. Conveying Modified Source Versions.

You may convey a work based on the Program, or the modifications to produce it from the Program, in the form of source code under the terms of section 4, provided that you also meet all of these conditions:

- a) The work must carry prominent notices stating that you modified it, and giving a relevant date.
- b) The work must carry prominent notices stating that it is released under this License and any conditions added under section 7. This requirement modifies the requirement in section 4 to "keep intact all notices".
- c) You must license the entire work, as a whole, under this License to anyone who comes into possession of a copy. This License will therefore apply, along with any applicable section 7 additional terms, to the whole of the work, and all its parts, regardless of how they are packaged. This License gives no permission to license the work in any other way, but it does not invalidate such permission if you have separately received it.
- d) If the work has interactive user interfaces, each must display Appropriate Legal Notices; however, if the Program has interactive interfaces that do not display Appropriate Legal Notices, your work need not make them do so.

A compilation of a covered work with other separate and independent works, which are not by their nature extensions of the covered work, and which are not combined with it such as to form a larger program, in or on a volume of a storage or distribution medium, is called an "aggregate" if the compilation and its resulting copyright are not used to limit the access or legal rights of the compilation's users beyond what the individual works permit. Inclusion of a covered work in an aggregate does not cause this License to apply to the other parts of the aggregate.

**6. Conveying Non-Source Forms.**
You may convey a covered work in object code form under the terms of sections 4 and 5, provided that you also convey the machine-readable Corresponding Source under the terms of this License, in one of these ways:

- a) Convey the object code in, or embodied in, a physical product (including a physical distribution medium), accompanied by the Corresponding Source fixed on a durable physical medium customarily used for software interchange.
- b) Convey the object code in, or embodied in, a physical product (including a physical distribution medium), accompanied by a written offer, valid for at least three years and valid for as long as you offer spare parts or customer support for that product model, to give anyone who possesses the object code either (1) a copy of the Corresponding Source for all the software in the product that is covered by this License, on a durable physical medium customarily used for software interchange, for a price no more than your reasonable cost of physically performing this conveying of source, or (2) access to copy the Corresponding Source from a network server at no charge.
- c) Convey individual copies of the object code with a copy of the written offer to provide the Corresponding Source. This alternative is allowed only occasionally and noncommercially, and only if you received the object code with such an offer, in accord with subsection 6b.
- d) Convey the object code by offering access from a designated place (gratis or for a charge), and offer equivalent access to the Corresponding Source in the same way through the same place at no further charge. You need not require recipients to copy the Corresponding Source along with the object code. If the place to copy the object code is a network server, the Corresponding Source may be on a different server (operated by you or a third party) that supports equivalent copying facilities, provided you maintain clear directions next to the object code saying where to find the Corresponding Source. Regardless of what server hosts the Corresponding Source, you remain obligated to ensure that it is available for as long as needed to satisfy these requirements.

- e) Convey the object code using peer-to-peer transmission, provided you inform other peers where the object code and Corresponding Source of the work are being offered to the general public at no charge under subsection 6d.

A separable portion of the object code, whose source code is excluded from the Corresponding Source as a System Library, need not be included in conveying the object code work.

A "User Product" is either (1) a "consumer product", which means any tangible personal property which is normally used for personal, family, or household purposes, or (2) anything designed or sold for incorporation into a dwelling. In determining whether a product is a consumer product, doubtful cases shall be resolved in favor of coverage. For a particular product received by a particular user, "normally used" refers to a typical or common use of that class of product, regardless of the status of the particular user or of the way in which the particular user actually uses, or expects or is expected to use, the product. A product is a consumer product regardless of whether the product has substantial commercial, industrial or non-consumer uses, unless such uses represent the only significant mode of use of the product.

"Installation Information" for a User Product means any methods, procedures, authorization keys, or other information required to install and execute modified versions of a covered work in that User Product from a modified version of its Corresponding Source. The information must suffice to ensure that the continued functioning of the modified object code is in no case prevented or interfered with solely because modification has been made.

If you convey an object code work under this section in, or with, or specifically for use in, a User Product, and the conveying occurs as part of a transaction in which the right of possession and use of the User Product is transferred to the recipient in perpetuity or for a fixed term (regardless of how the transaction is characterized), the Corresponding Source conveyed under this section must be accompanied by the Installation Information. But this requirement does not apply if neither you nor any third party retains the ability to install modified object code on the User Product (for example, the work has been installed in ROM).

The requirement to provide Installation Information does not include a requirement to continue to provide support service, warranty, or updates for a work that has been modified or installed by the recipient, or for the User Product in which it has been modified or installed. Access to a network may be denied when the modification itself materially and adversely affects the operation of the network or violates the rules and protocols for communication across the network.

Corresponding Source conveyed, and Installation Information provided, in accord with this section must be in a format that is publicly documented (and with an implementation available to the public in source code form), and must require no special password or key for unpacking, reading or copying.

**7. Additional Terms.**
"Additional permissions" are terms that supplement the terms of this License by making exceptions from one or more of its conditions. Additional permissions that are applicable to the entire Program shall be treated as though they were included in this License, to the extent that they are valid under applicable law. If additional permissions apply only to part of the Program, that part may be used separately under those permissions, but the entire Program remains governed by this License without regard to the additional permissions.

When you convey a copy of a covered work, you may at your option remove any additional permissions from that copy, or from any part of it. (Additional permissions may be written to require their own removal in certain cases when you modify the work.) You may place additional permissions on material, added by you to a covered work, for which you have or can give appropriate copyright permission.

Notwithstanding any other provision of this License, for material you add to a covered work, you may (if authorized by the copyright holders of that material) supplement the terms of this License with terms:

- a) Disclaiming warranty or limiting liability differently from the terms of sections 15 and 16 of this License; or
- b) Requiring preservation of specified reasonable legal notices or author attributions in that material or in the Appropriate Legal Notices displayed by works containing it; or
- c) Prohibiting misrepresentation of the origin of that material, or requiring that modified versions of such material be marked in reasonable ways as different from the original version; or
- d) Limiting the use for publicity purposes of names of licensors or authors of the material; or
- e) Declining to grant rights under trademark law for use of some trade names, trademarks, or service marks; or
- f) Requiring indemnification of licensors and authors of that material by anyone who conveys the material (or modified versions of it) with contractual assumptions of liability to the recipient, for any liability that these contractual assumptions directly impose on those licensors and authors.

All other non-permissive additional terms are considered "further restrictions" within the meaning of section 10. If the Program as you received it, or any part of it, contains a notice stating that it is governed by this License along with a term that is a further restriction, you may remove that term. If a license document contains a further restriction but permits relicensing or conveying under this License, you may add to a covered work material governed by the terms of that license document, provided that the further restriction does not survive such relicensing or conveying.

If you add terms to a covered work in accord with this section, you must place, in the relevant source files, a statement of the additional terms that apply to those files, or a notice indicating where to find the applicable terms.

Additional terms, permissive or non-permissive, may be stated in the form of a separately written license, or stated as exceptions; the above requirements apply either way.

**8. Termination.**
You may not propagate or modify a covered work except as expressly provided under this License. Any attempt otherwise to propagate or modify it is void, and will automatically terminate your rights under this License (including any patent licenses granted under the third paragraph of section 11).

However, if you cease all violation of this License, then your license from a particular copyright holder is reinstated (a) provisionally, unless and until the copyright holder explicitly and finally terminates your license, and (b) permanently, if the copyright holder fails to notify you of the violation by some reasonable means prior to 60 days after the cessation.

Moreover, your license from a particular copyright holder is reinstated permanently if the copyright holder notifies you of the violation by some reasonable means, this is the first time you have received notice of violation of this License (for any work) from that copyright holder, and you cure the violation prior to 30 days after your receipt of the notice.

Termination of your rights under this section does not terminate the licenses of parties who have received copies or rights from you under this License. If your rights have been terminated and not permanently reinstated, you do not qualify to receive new licenses for the same material under section 10.

**9. Acceptance Not Required for Having Copies.**
You are not required to accept this License in order to receive or run a copy of the Program. Ancillary propagation of a covered work occurring solely as a consequence of using peer-to-peer transmission to receive a copy likewise does not require acceptance. However, nothing other than this License grants you permission to propagate or modify any covered work. These actions infringe copyright if you do not accept this License. Therefore, by modifying or propagating a covered work, you indicate your acceptance of this License to do so.

## 10. Automatic Licensing of Downstream Recipients.

Each time you convey a covered work, the recipient automatically receives a license from the original licensors, to run, modify and propagate that work, subject to this License. You are not responsible for enforcing compliance by third parties with this License.

An "entity transaction" is a transaction transferring control of an organization, or substantially all assets of one, or subdividing an organization, or merging organizations. If propagation of a covered work results from an entity transaction, each party to that transaction who receives a copy of the work also receives whatever licenses to the work the party's predecessor in interest had or could give under the previous paragraph, plus a right to possession of the Corresponding Source of the work from the predecessor in interest, if the predecessor has it or can get it with reasonable efforts.

You may not impose any further restrictions on the exercise of the rights granted or affirmed under this License. For example, you may not impose a license fee, royalty, or other charge for exercise of rights granted under this License, and you may not initiate litigation (including a cross-claim or counterclaim in a lawsuit) alleging that any patent claim is infringed by making, using, selling, offering for sale, or importing the Program or any portion of it.

## 11. Patents.

A "contributor" is a copyright holder who authorizes use under this License of the Program or a work on which the Program is based. The work thus licensed is called the contributor's "contributor version".

A contributor's "essential patent claims" are all patent claims owned or controlled by the contributor, whether already acquired or hereafter acquired, that would be infringed by some manner, permitted by this License, of making, using, or selling its contributor version, but do not include claims that would be infringed only as a consequence of further modification of the contributor version. For purposes of this definition, "control" includes the right to grant patent sublicenses in a manner consistent with the requirements of this License.

Each contributor grants you a non-exclusive, worldwide, royalty-free patent license under the contributor's essential patent claims, to make, use, sell, offer for sale, import and otherwise run, modify and propagate the contents of its contributor version.

In the following three paragraphs, a "patent license" is any express agreement or commitment, however denominated, not to enforce a patent (such as an express permission to practice a patent or covenant not to sue for patent infringement). To "grant" such a patent license to a party means to make such an agreement or commitment not to enforce a patent against the party.

If you convey a covered work, knowingly relying on a patent license, and the Corresponding Source of the work is not available for anyone to copy, free of charge and under the terms of this License, through a publicly available network server or other readily accessible means, then you must either (1) cause the Corresponding Source to be so available, or (2) arrange to deprive yourself of the benefit of the patent license for this particular work, or (3) arrange, in a manner consistent with the requirements of this License, to extend the patent license to downstream recipients. "Knowingly relying" means you have actual knowledge that, but for the patent license, your conveying the covered work in a country, or your recipient's use of the covered work in a country, would infringe one or more identifiable patents in that country that you have reason to believe are valid.

If, pursuant to or in connection with a single transaction or arrangement, you convey, or propagate by procuring conveyance of, a covered work, and grant a patent license to some of the parties receiving the covered work authorizing them to use, propagate, modify or convey a specific copy of the covered work, then the patent license you grant is automatically extended to all recipients of the covered work and works based on it.

A patent license is "discriminatory" if it does not include within the scope of its coverage, prohibits the exercise of, or is conditioned on the non-exercise of one or more of the rights that are specifically granted under this License. You may not convey a covered work if you are a party to an arrangement with a third party that is in the business of distributing software, under which you make payment to the third party based on the extent of your activity of conveying the work, and under which the third party grants, to any of the parties who would receive the covered work from you, a discriminatory patent license (a) in connection with copies of the covered work conveyed by you (or copies made from those copies), or (b) primarily for and in connection with specific products or compilations that contain the covered work, unless you entered into that arrangement, or that patent license was granted, prior to 28 March 2007.

Nothing in this License shall be construed as excluding or limiting any implied license or other defenses to infringement that may otherwise be available to you under applicable patent law.

**12. No Surrender of Others' Freedom.**

If conditions are imposed on you (whether by court order, agreement or otherwise) that contradict the conditions of this License, they do not excuse you from the conditions of this License. If you cannot convey a covered work so as to satisfy simultaneously your obligations under this License and any other pertinent obligations, then as a consequence you may not convey it at all. For example, if you agree to terms that obligate you to collect a royalty for further conveying from those to whom you convey the Program, the only way you could satisfy both those terms and this License would be to refrain entirely from conveying the Program.

### 13. Use with the GNU Affero General Public License.

Notwithstanding any other provision of this License, you have permission to link or combine any covered work with a work licensed under version 3 of the GNU Affero General Public License into a single combined work, and to convey the resulting work. The terms of this License will continue to apply to the part which is the covered work, but the special requirements of the GNU Affero General Public License, section 13, concerning interaction through a network will apply to the combination as such.

### 14. Revised Versions of this License.

The Free Software Foundation may publish revised and/or new versions of the GNU General Public License from time to time. Such new versions will be similar in spirit to the present version, but may differ in detail to address new problems or concerns.

Each version is given a distinguishing version number. If the Program specifies that a certain numbered version of the GNU General Public License "or any later version" applies to it, you have the option of following the terms and conditions either of that numbered version or of any later version published by the Free Software Foundation. If the Program does not specify a version number of the GNU General Public License, you may choose any version ever published by the Free Software Foundation.

If the Program specifies that a proxy can decide which future versions of the GNU General Public License can be used, that proxy's public statement of acceptance of a version permanently authorizes you to choose that version for the Program.

Later license versions may give you additional or different permissions. However, no additional obligations are imposed on any author or copyright holder as a result of your choosing to follow a later version.

### 15. Disclaimer of Warranty.

THERE IS NO WARRANTY FOR THE PROGRAM, TO THE EXTENT PERMITTED BY APPLICABLE LAW. EXCEPT WHEN OTHERWISE STATED IN WRITING THE COPYRIGHT HOLDERS AND/OR OTHER PARTIES PROVIDE THE PROGRAM "AS IS" WITHOUT WARRANTY OF ANY KIND, EITHER EXPRESSED OR IMPLIED,

INCLUDING, BUT NOT LIMITED TO, THE IMPLIED WARRANTIES OF MERCHANTABILITY AND FITNESS FOR A PARTICULAR PURPOSE. THE ENTIRE RISK AS TO THE QUALITY AND PERFORMANCE OF THE PROGRAM IS WITH YOU. SHOULD THE PROGRAM PROVE DEFECTIVE, YOU ASSUME THE COST OF ALL NECESSARY SERVICING, REPAIR OR CORRECTION.

**16. Limitation of Liability.**
IN NO EVENT UNLESS REQUIRED BY APPLICABLE LAW OR AGREED TO IN WRITING WILL ANY COPYRIGHT HOLDER, OR ANY OTHER PARTY WHO MODIFIES AND/OR CONVEYS THE PROGRAM AS PERMITTED ABOVE, BE LIABLE TO YOU FOR DAMAGES, INCLUDING ANY GENERAL, SPECIAL, INCIDENTAL OR CONSEQUENTIAL DAMAGES ARISING OUT OF THE USE OR INABILITY TO USE THE PROGRAM (INCLUDING BUT NOT LIMITED TO LOSS OF DATA OR DATA BEING RENDERED INACCURATE OR LOSSES SUSTAINED BY YOU OR THIRD PARTIES OR A FAILURE OF THE PROGRAM TO OPERATE WITH ANY OTHER PROGRAMS), EVEN IF SUCH HOLDER OR OTHER PARTY HAS BEEN ADVISED OF THE POSSIBILITY OF SUCH DAMAGES.

**17. Interpretation of Sections 15 and 16.**
If the disclaimer of warranty and limitation of liability provided above cannot be given local legal effect according to their terms, reviewing courts shall apply local law that most closely approximates an absolute waiver of all civil liability in connection with the Program, unless a warranty or assumption of liability accompanies a copy of the Program in return for a fee.

END OF TERMS AND CONDITIONS

**How to Apply These Terms to Your New Programs**

If you develop a new program, and you want it to be of the greatest possible use to the public, the best way to achieve this is to make it free software which everyone can redistribute and change under these terms.

To do so, attach the following notices to the program. It is safest to attach them to the start of each source file to most effectively state the exclusion of warranty; and each file should have at least the "copyright" line and a pointer to where the full notice is found.

```
<one line to give the program's name and a brief idea of what it
does.>

Copyright (C) <year>  <name of author>
```

```
This  program  is  free  software:  you  can  redistribute  it  and/or
modify  it  under  the  terms  of  the  GNU  General  Public  License
as  published  by  the  Free  Software  Foundation,  either  version  3  of
the  License,  or  (at  your  option)  any  later  version.

This  program  is  distributed  in  the  hope  that  it  will  be  useful,
but  WITHOUT  ANY  WARRANTY;  without  even  the  implied  warranty  of
MERCHANTABILITY  or  FITNESS  FOR  A  PARTICULAR  PURPOSE.   See  the
GNU  General  Public  License  for  more  details.

You  should  have  received  a  copy  of  the  GNU  General  Public
License  along  with  this  program.   If  not,  see
<http://www.gnu.org/licenses/>.
```

Also add information on how to contact you by electronic and paper mail.

If the program does terminal interaction, make it output a short notice like this when it starts in an interactive mode:

```
<program>  Copyright (C) <year>  <name of author>

This  program  comes  with  ABSOLUTELY  NO  WARRANTY;  for  details  type
`show  w'.  This  is  free  software,  and  you  are  welcome  to
redistribute  it  under  certain  conditions;  type  `show  c'  for
details.
```

The hypothetical commands `show w' and `show c' should show the appropriate parts of the General Public License. Of course, your program's commands might be different; for a GUI interface, you would use an "about box".

You should also get your employer (if you work as a programmer) or school, if any, to sign a "copyright disclaimer" for the program, if necessary. For more information on this, and how to apply and follow the GNU GPL, see <http://www.gnu.org/licenses/>.

The GNU General Public License does not permit incorporating your program into proprietary programs. If your program is a subroutine library, you may consider it more useful to permit linking proprietary applications with the library. If this is what you want to do, use the GNU Lesser General Public License instead of this License. But first, please read <http://www.gnu.org/philosophy/why-not-lgpl.html>.

# Appendix III

# -

# How to install Rappture

To install Rappture, one first must have the sources. They can be found at the following address:

[https://nanohub.org/infrastructure/rappture/wiki/Downloads](https://nanohub.org/infrastructure/rappture/wiki/Downloads)

and they are released under the following license (which gives freedom to the user to do anything with the code as long as the same freedom to use modified copy of Rappture is provided). Here is an extract:

Permission is hereby granted, free of charge, to any person obtaining a copy of this software and associated documentation files (the "Software"), to deal with the Software without restriction, including without limitation the rights to use, copy, modify, merge, publish, distribute, sublicense, and/or sell copies of the Software, and to permit persons to whom the Software is furnished to do so, subject to the following conditions:

   * Redistributions of source code must retain the above copyright
     notice, this list of conditions and the following disclaimers.

   * Redistributions in binary form must reproduce the above copyright
     notice, this list of conditions and the following disclaimers
     in the documentation and/or other materials provided with the
     distribution.

\* Neither the names of the Network for Computational Nanotechnology, Purdue University, nor the names of its contributors may be used to endorse or promote products derived from this Software without specific prior written permission.

Details on how to install Rappture are given at:

https://nanohub.org/infrastructure/rappture/wiki/rappture_installation_hardway

We report here some guidelines, just to have an idea on what it takes to install it. As one will see, there is nothing special to install it and the installation is pretty straightforward.

First of all, the system requirements are the following:

You will need a C and C++ compiler. The following is a list of packages for debian distributions that should be installed.

- gcc
- g++
- libssl-dev
- make
- patch
- subversion
- libx11-dev
- libxext-dev
- libfreetype6-dev
- libxft-dev
- libxrandr-dev
- libpng12-dev
- libjpeg62-dev
- libtiff4-dev
- libxpm-dev

(Other distribution's package names may differ slightly. The version numbers don't matter. For example, libpng12-dev or libpng14-dev can be used. )

Bindings will be built for whatever languages are found installed on your system. What this means is that if you want to use Rappture with let's say python, you need to have python development package installed before building Rappture.

Here is a list of optional debian packages

- gfortran
- perl
- python
- python-dev
- ruby
- ruby-dev

- libperl-dev
- sun-java6-jdk
- sun-java6-jre

The following optional debian packages are useful too.

- gdb
- libavcodec-dev
- libavformat-dev
- libavutil-dev

Once one has these packages installed on his/her machine he/she can start the steps of the actual installation of Rappture.

First of all, let us extract the source files from the downloaded tarballs.

```
# tar xpvf rappture-src-*.tar.gz
# tar xpvf rappture-runtime-src-*.tar.gz
```

This will extract the Rappture and the Rappture run-time sources.

# APPENDIX IV

**-**

0. PREAMBLE

The purpose of this License is to make a manual, textbook, or other
functional and useful document "free" in the sense of freedom: to
assure everyone the effective freedom to copy and redistribute it,
with or without modifying it, either commercially or noncommercially.
Secondarily, this License preserves for the author and publisher a way
to get credit for their work, while not being considered responsible
for modifications made by others.

This License is a kind of "copyleft", which means that derivative
works of the document must themselves be free in the same sense.  It
complements the GNU General Public License, which is a copyleft
license designed for free software.

We have designed this License in order to use it for manuals for free
software, because free software needs free documentation: a free
program should come with manuals providing the same freedoms that the
software does.  But this License is not limited to software manuals;
it can be used for any textual work, regardless of subject matter or
whether it is published as a printed book.  We recommend this License
principally for works whose purpose is instruction or reference.

1. APPLICABILITY AND DEFINITIONS

This License applies to any manual or other work, in any medium, that
contains a notice placed by the copyright holder saying it can be
distributed under the terms of this License.  Such a notice grants a
world-wide, royalty-free license, unlimited in duration, to use that
work under the conditions stated herein.  The "Document", below,
refers to any such manual or work.  Any member of the public is a
licensee, and is addressed as "you".  You accept the license if you
copy, modify or distribute the work in a way requiring permission
under copyright law.

A "Modified Version" of the Document means any work containing the
Document or a portion of it, either copied verbatim, or with
modifications and/or translated into another language.

A "Secondary Section" is a named appendix or a front-matter section of
the Document that deals exclusively with the relationship of the
publishers or authors of the Document to the Document's overall
subject (or to related matters) and contains nothing that could fall
directly within that overall subject.  (Thus, if the Document is in
part a textbook of mathematics, a Secondary Section may not explain
any mathematics.)  The relationship could be a matter of historical
connection with the subject or with related matters, or of legal,
commercial, philosophical, ethical or political position regarding
them.

The "Invariant Sections" are certain Secondary Sections whose titles
are designated, as being those of Invariant Sections, in the notice
that says that the Document is released under this License.  If a
section does not fit the above definition of Secondary then it is not
allowed to be designated as Invariant.  The Document may contain zero
Invariant Sections.  If the Document does not identify any Invariant
Sections then there are none.

The "Cover Texts" are certain short passages of text that are listed,
as Front-Cover Texts or Back-Cover Texts, in the notice that says that
the Document is released under this License.  A Front-Cover Text may
be at most 5 words, and a Back-Cover Text may be at most 25 words.

A "Transparent" copy of the Document means a machine-readable copy,
represented in a format whose specification is available to the
general public, that is suitable for revising the document
straightforwardly with generic text editors or (for images composed of
pixels) generic paint programs or (for drawings) some widely available
drawing editor, and that is suitable for input to text formatters or
for automatic translation to a variety of formats suitable for input
to text formatters.  A copy made in an otherwise Transparent file
format whose markup, or absence of markup, has been arranged to thwart
or discourage subsequent modification by readers is not Transparent.
An image format is not Transparent if used for any substantial amount
of text.  A copy that is not "Transparent" is called "Opaque".

Examples of suitable formats for Transparent copies include plain
ASCII without markup, Texinfo input format, LaTeX input format, SGML
or XML using a publicly available DTD, and standard-conforming simple
HTML, PostScript or PDF designed for human modification.  Examples of
transparent image formats include PNG, XCF and JPG.  Opaque formats
include proprietary formats that can be read and edited only by
proprietary word processors, SGML or XML for which the DTD and/or
processing tools are not generally available, and the
machine-generated HTML, PostScript or PDF produced by some word
processors for output purposes only.

The "Title Page" means, for a printed book, the title page itself,
plus such following pages as are needed to hold, legibly, the material
this License requires to appear in the title page.  For works in
formats which do not have any title page as such, "Title Page" means
the text near the most prominent appearance of the work's title,
preceding the beginning of the body of the text.

The "publisher" means any person or entity that distributes copies of the Document to the public.

A section "Entitled XYZ" means a named subunit of the Document whose title either is precisely XYZ or contains XYZ in parentheses following text that translates XYZ in another language. (Here XYZ stands for a specific section name mentioned below, such as "Acknowledgements", "Dedications", "Endorsements", or "History".) To "Preserve the Title" of such a section when you modify the Document means that it remains a section "Entitled XYZ" according to this definition.

The Document may include Warranty Disclaimers next to the notice which states that this License applies to the Document. These Warranty Disclaimers are considered to be included by reference in this License, but only as regards disclaiming warranties: any other implication that these Warranty Disclaimers may have is void and has no effect on the meaning of this License.

2. VERBATIM COPYING

You may copy and distribute the Document in any medium, either commercially or noncommercially, provided that this License, the copyright notices, and the license notice saying this License applies to the Document are reproduced in all copies, and that you add no other conditions whatsoever to those of this License. You may not use technical measures to obstruct or control the reading or further copying of the copies you make or distribute. However, you may accept compensation in exchange for copies. If you distribute a large enough number of copies you must also follow the conditions in section 3.

You may also lend copies, under the same conditions stated above, and you may publicly display copies.

3. COPYING IN QUANTITY

If you publish printed copies (or copies in media that commonly have printed covers) of the Document, numbering more than 100, and the Document's license notice requires Cover Texts, you must enclose the copies in covers that carry, clearly and legibly, all these Cover Texts: Front-Cover Texts on the front cover, and Back-Cover Texts on the back cover. Both covers must also clearly and legibly identify you as the publisher of these copies. The front cover must present the full title with all words of the title equally prominent and visible. You may add other material on the covers in addition. Copying with changes limited to the covers, as long as they preserve the title of the Document and satisfy these conditions, can be treated as verbatim copying in other respects.

If the required texts for either cover are too voluminous to fit legibly, you should put the first ones listed (as many as fit reasonably) on the actual cover, and continue the rest onto adjacent pages.

If you publish or distribute Opaque copies of the Document numbering more than 100, you must either include a machine-readable Transparent copy along with each Opaque copy, or state in or with each Opaque copy a computer-network location from which the general network-using public has access to download using public-standard network protocols
a complete Transparent copy of the Document, free of added material.
If you use the latter option, you must take reasonably prudent steps,
when you begin distribution of Opaque copies in quantity, to ensure
that this Transparent copy will remain thus accessible at the stated
location until at least one year after the last time you distribute an
Opaque copy (directly or through your agents or retailers) of that
edition to the public.

It is requested, but not required, that you contact the authors of the
Document well before redistributing any large number of copies, to
give them a chance to provide you with an updated version of the
Document.

4. MODIFICATIONS

You may copy and distribute a Modified Version of the Document under
the conditions of sections 2 and 3 above, provided that you release
the Modified Version under precisely this License, with the Modified
Version filling the role of the Document, thus licensing distribution
and modification of the Modified Version to whoever possesses a copy
of it.  In addition, you must do these things in the Modified Version:

A. Use in the Title Page (and on the covers, if any) a title distinct
   from that of the Document, and from those of previous versions
   (which should, if there were any, be listed in the History section
   of the Document).  You may use the same title as a previous version
   if the original publisher of that version gives permission.
B. List on the Title Page, as authors, one or more persons or entities
   responsible for authorship of the modifications in the Modified
   Version, together with at least five of the principal authors of the
   Document (all of its principal authors, if it has fewer than five),
   unless they release you from this requirement.
C. State on the Title page the name of the publisher of the
   Modified Version, as the publisher.
D. Preserve all the copyright notices of the Document.
E. Add an appropriate copyright notice for your modifications
   adjacent to the other copyright notices.
F. Include, immediately after the copyright notices, a license notice
   giving the public permission to use the Modified Version under the
   terms of this License, in the form shown in the Addendum below.
G. Preserve in that license notice the full lists of Invariant Sections
   and required Cover Texts given in the Document's license notice.
H. Include an unaltered copy of this License.
I. Preserve the section Entitled "History", Preserve its Title, and add
   to it an item stating at least the title, year, new authors, and
   publisher of the Modified Version as given on the Title Page.  If
   there is no section Entitled "History" in the Document, create one
   stating the title, year, authors, and publisher of the Document as
   given on its Title Page, then add an item describing the Modified
   Version as stated in the previous sentence.
J. Preserve the network location, if any, given in the Document for
   public access to a Transparent copy of the Document, and likewise
   the network locations given in the Document for previous versions
   it was based on.  These may be placed in the "History" section.
   You may omit a network location for a work that was published at least four years before the Document itself, or if the original
publisher of the version it refers to gives permission.
K. For any section Entitled "Acknowledgements" or "Dedications",
   Preserve the Title of the section, and preserve in the section all
   the substance and tone of each of the contributor acknowledgements
   and/or dedications given therein.
L. Preserve all the Invariant Sections of the Document,
   unaltered in their text and in their titles.  Section numbers
   or the equivalent are not considered part of the section titles.
M. Delete any section Entitled "Endorsements".  Such a section
   may not be included in the Modified Version.
N. Do not retitle any existing section to be Entitled "Endorsements"
   or to conflict in title with any Invariant Section.
O. Preserve any Warranty Disclaimers.

If the Modified Version includes new front-matter sections or
appendices that qualify as Secondary Sections and contain no material
copied from the Document, you may at your option designate some or all
of these sections as invariant.  To do this, add their titles to the
list of Invariant Sections in the Modified Version's license notice.
These titles must be distinct from any other section titles.

You may add a section Entitled "Endorsements", provided it contains
nothing but endorsements of your Modified Version by various
parties--for example, statements of peer review or that the text has
been approved by an organization as the authoritative definition of a
standard.

You may add a passage of up to five words as a Front-Cover Text, and a
passage of up to 25 words as a Back-Cover Text, to the end of the list
of Cover Texts in the Modified Version.  Only one passage of
Front-Cover Text and one of Back-Cover Text may be added by (or
through arrangements made by) any one entity.  If the Document already
includes a cover text for the same cover, previously added by you or
by arrangement made by the same entity you are acting on behalf of,
you may not add another; but you may replace the old one, on explicit
permission from the previous publisher that added the old one.

The author(s) and publisher(s) of the Document do not by this License
give permission to use their names for publicity for or to assert or
imply endorsement of any Modified Version.

5. COMBINING DOCUMENTS

You may combine the Document with other documents released under this
License, under the terms defined in section 4 above for modified
versions, provided that you include in the combination all of the
Invariant Sections of all of the original documents, unmodified, and
list them all as Invariant Sections of your combined work in its
license notice, and that you preserve all their Warranty Disclaimers.

The combined work need only contain one copy of this License, and
multiple identical Invariant Sections may be replaced with a single
copy.  If there are multiple Invariant Sections with the same name but
different contents, make the title of each such section unique by
adding at the end of it, in parentheses, the name of the original author or publisher of that section if known, or else a unique number.
Make the same adjustment to the section titles in the list of
Invariant Sections in the license notice of the combined work.

In the combination, you must combine any sections Entitled "History"
in the various original documents, forming one section Entitled
"History"; likewise combine any sections Entitled "Acknowledgements",
and any sections Entitled "Dedications".  You must delete all sections
Entitled "Endorsements".

6. COLLECTIONS OF DOCUMENTS

You may make a collection consisting of the Document and other
documents released under this License, and replace the individual
copies of this License in the various documents with a single copy
that is included in the collection, provided that you follow the rules
of this License for verbatim copying of each of the documents in all
other respects.

You may extract a single document from such a collection, and
distribute it individually under this License, provided you insert a
copy of this License into the extracted document, and follow this
License in all other respects regarding verbatim copying of that
document.

7. AGGREGATION WITH INDEPENDENT WORKS

A compilation of the Document or its derivatives with other separate
and independent documents or works, in or on a volume of a storage or
distribution medium, is called an "aggregate" if the copyright
resulting from the compilation is not used to limit the legal rights
of the compilation's users beyond what the individual works permit.
When the Document is included in an aggregate, this License does not
apply to the other works in the aggregate which are not themselves
derivative works of the Document.

If the Cover Text requirement of section 3 is applicable to these
copies of the Document, then if the Document is less than one half of
the entire aggregate, the Document's Cover Texts may be placed on
covers that bracket the Document within the aggregate, or the
electronic equivalent of covers if the Document is in electronic form.
Otherwise they must appear on printed covers that bracket the whole
aggregate.

8. TRANSLATION

Translation is considered a kind of modification, so you may
distribute translations of the Document under the terms of section 4.
Replacing Invariant Sections with translations requires special
permission from their copyright holders, but you may include
translations of some or all Invariant Sections in addition to the
original versions of these Invariant Sections.  You may include a
translation of this License, and all the license notices in the
Document, and any Warranty Disclaimers, provided that you also include the original English version of this License and the original versions of those notices and disclaimers.  In case of a disagreement between the translation and the original version of this License or a notice or disclaimer, the original version will prevail.

If a section in the Document is Entitled "Acknowledgements", "Dedications", or "History", the requirement (section 4) to Preserve its Title (section 1) will typically require changing the actual title.

9. TERMINATION

You may not copy, modify, sublicense, or distribute the Document except as expressly provided under this License.  Any attempt otherwise to copy, modify, sublicense, or distribute it is void, and will automatically terminate your rights under this License.

However, if you cease all violation of this License, then your license from a particular copyright holder is reinstated (a) provisionally, unless and until the copyright holder explicitly and finally terminates your license, and (b) permanently, if the copyright holder fails to notify you of the violation by some reasonable means prior to 60 days after the cessation.

Moreover, your license from a particular copyright holder is reinstated permanently if the copyright holder notifies you of the violation by some reasonable means, this is the first time you have received notice of violation of this License (for any work) from that copyright holder, and you cure the violation prior to 30 days after your receipt of the notice.

Termination of your rights under this section does not terminate the licenses of parties who have received copies or rights from you under this License.  If your rights have been terminated and not permanently reinstated, receipt of a copy of some or all of the same material does not give you any rights to use it.

10. FUTURE REVISIONS OF THIS LICENSE

The Free Software Foundation may publish new, revised versions of the GNU Free Documentation License from time to time.  Such new versions will be similar in spirit to the present version, but may differ in detail to address new problems or concerns.  See http://www.gnu.org/copyleft/.

Each version of the License is given a distinguishing version number. If the Document specifies that a particular numbered version of this License "or any later version" applies to it, you have the option of following the terms and conditions either of that specified version or of any later version that has been published (not as a draft) by the Free Software Foundation.  If the Document does not specify a version number of this License, you may choose any version ever published (not as a draft) by the Free Software Foundation.  If the Document specifies that a proxy can decide which future versions of this License can be used, that proxy's public statement of acceptance of a version permanently authorizes you to choose that version for the Document.

11. RELICENSING

"Massive Multiauthor Collaboration Site" (or "MMC Site") means any World Wide Web server that publishes copyrightable works and also provides prominent facilities for anybody to edit those works.  A public wiki that anybody can edit is an example of such a server.  A "Massive Multiauthor Collaboration" (or "MMC") contained in the site means any set of copyrightable works thus published on the MMC site.

"CC-BY-SA" means the Creative Commons Attribution-Share Alike 3.0 license published by Creative Commons Corporation, a not-for-profit corporation with a principal place of business in San Francisco, California, as well as future copyleft versions of that license published by that same organization.

"Incorporate" means to publish or republish a Document, in whole or in part, as part of another Document.

An MMC is "eligible for relicensing" if it is licensed under this License, and if all works that were first published under this License somewhere other than this MMC, and subsequently incorporated in whole or in part into the MMC, (1) had no cover texts or invariant sections, and (2) were thus incorporated prior to November 1, 2008.

The operator of an MMC Site may republish an MMC contained in the site under CC-BY-SA on the same site at any time before August 1, 2009, provided the MMC is eligible for relicensing.

ADDENDUM: How to use this License for your documents

To use this License in a document you have written, include a copy of the License in the document and put the following copyright and license notices just after the title page:

    Copyright (c)  YEAR  YOUR NAME.
    Permission is granted to copy, distribute and/or modify this document
    under the terms of the GNU Free Documentation License, Version 1.3
    or any later version published by the Free Software Foundation;
    with no Invariant Sections, no Front-Cover Texts, and no Back-Cover
Texts.
    A copy of the license is included in the section entitled "GNU
    Free Documentation License".

If you have Invariant Sections, Front-Cover Texts and Back-Cover Texts, replace the "with...Texts." line with this:

    with the Invariant Sections being LIST THEIR TITLES, with the
    Front-Cover Texts being LIST, and with the Back-Cover Texts being LIST.

If you have Invariant Sections without Cover Texts, or some other combination of the three, merge those two alternatives to suit the situation.

If your document contains nontrivial examples of program code, we
recommend releasing these examples in parallel under your choice of
free software license, such as the GNU General Public License,
to permit their use in free software.

.